\documentclass[twocolumn,prx,hyperref,footinbib,longbibliography]{revtex4-1}

\usepackage{bbm,amssymb}
\usepackage{times}
\usepackage{graphicx}
\usepackage{dcolumn}
\usepackage{bm}
\usepackage{float}
\usepackage{mathrsfs}
\usepackage{color}
\usepackage{hyperref} 
\hypersetup{
    colorlinks=true,
    linkcolor=red,
    filecolor=blue,      
    urlcolor=blue,
    citecolor=blue
}
 
\usepackage{amsmath}
\usepackage{braket} 
\usepackage{soul}

\usepackage{tabularx}
\usepackage{multirow}
\usepackage{booktabs}
\usepackage{makecell}

\begin{document}

\title{Doubly geometric quantum control}

\author{Wenzheng Dong}
\email{dongwz@vt.edu}
\author{Fei Zhuang}
\author{Sophia  E.  Economou}
\author{Edwin Barnes}
\email{efbarnes@vt.edu}
\affiliation{ Department of Physics, Virginia Tech, Blacksburg, Virginia 24061, USA}

\begin{abstract}
In holonomic quantum computation, \textcolor{black}{quantum} gates are performed using driving protocols that trace out closed loops on the  \textcolor{black}{Bloch} sphere, making them robust to certain pulse errors. However, dephasing noise that is transverse to the drive, which is significant in many qubit platforms, lies outside the family of correctable errors. Here, we present a general procedure that combines two types of geometry---holonomy loops on the Bloch sphere and geometric space curves in three dimensions---to design gates that simultaneously suppress pulse errors and transverse noise errors. We demonstrate this doubly geometric control technique by designing explicit  examples of \textcolor{black}{single-qubit and two-qubit} dynamically corrected holonomic gates.
\end{abstract}

\maketitle

\section{Introduction}
Quantum information processing demands unprecedented precision in the control of qubits. This is made challenging by the ubiquitous noise from the environment and unavoidable control imperfections, which can substantially reduce the control fidelity. \textcolor{black}{Tremendous progress has been made in the development of quantum optimal control techniques~\cite{Glaser.TEPJ.2015,Stefanatos.EPL.2020,Vandersypen.RMP.2005,johansson2012qutip,Jr-Shin.Li.PRA.2006,Ruschhaupt.NJP.2012,Daems.PRL.2013,Dridi.PRL.2020,Dridi.PRA.2020,Tian.PRA.2020,Lapert.PRL.2010,YuanHaidong.PRA.2012,Lampert.SciRep.2012,Doria.PRL.2011,Sarandy.PRL.2005,Goerz.NJP.2014,Marx.PRA.2010,schulte2012control,Martinis_and_Geller.PRA.2014,Stefanatos.PRA.2019,Stefanatos.JPAMT.2020,khaneja.JMR.2005,Machnes.PRA.2011,Reich.JCP.2012,Palao.PRL.2002,Palao.PRA.2003,Tesch.PRL.2002}.}
Holonomic quantum computation \cite{ZANARDI199994}, where the gates are based on geometric phases \cite{ANANDAN1988171,Wilczek.Zee,Berry_2009,SolinasPRA2004,Sjoqvist.IJQC.2015,Xu.PRL.2012,Utkan.JPSJ.2014}, is one approach to boosting gate fidelities in the presence of noise. 
Using geometric rather than dynamical phases to implement quantum gates can mitigate the effect of noise that leaves holonomy loops in the control space unperturbed. Geometric phases can be accrued using either adiabatic \cite{Berry.Geometric.Phase,DeChiara.PRL.2003,Yale2016} or non-adiabatic driving~\cite{PhysRevLett.58.1593,Sj_qvist_2012,Sjoqvist,XueZhengyuan.PRA.2018,Ribeiro.PRA.2019,LiuBoajie.PRL.2019,YingZuJian.PRR.2020,Shkolnikov.PRB.2020,li2020dynamically,Ji2021noncyclic,Zhao.holonomicDD.PRL.2021}; the latter alleviates decoherence by reducing the operation time. Non-adiabatic holonomic (geometric) gates have been successfully realized in superconducting systems \cite{YanTongxing.PRL.2019,SunLuYan.PRL.2020}, trapped ions \cite{Duan.Science.2001,AiMingzhong.PRApp.2020}, and NV centers in diamond \cite{BrianZhou.PRL.2017,Zhou2016,Sekiguchi.2017}. An advantage of the holonomic approach is that it affords substantial flexibility in choosing experimentally friendly pulse shapes to generate the gates. However, while holonomic gates are resistant to errors along the holonomy loop, they remain susceptible to noise that is transverse to it, a type of noise that is common in many qubit platforms.  

Dynamically corrected gates constitute another approach to designing controls that are robust against various types of noise~\cite{Goelman_JMR89,Viola.PRL.1998,Biercuk_Nature09,Khodjasteh.PRL.2010,Barnes.SciRep.2015,Wang.NatCom.2012,Kestner.PRL.2013,vanderSar_Nature12,Green_NJP13,Merrill_Wiley14,CalderonVargasPRL2017,ButerakosPRB2018,GungorduPRB2018,Zeng.PRA.2019,Kanaar.PRB.2021}. Analytical approaches in particular enable one to search for globally optimal solutions to a given control task. However, these approaches often rely on fixed pulse waveforms that can be  experimentally infeasible or can necessitate the use of longer sequences that expose the system to adverse effects that arise on longer time scales. Finding shorter sequences is made challenging by the need to solve highly nonlinear equations that enforce the noise cancellation conditions~\cite{Wang.NatCom.2012,Barnes.SciRep.2015}. A recent approach to dynamically corrected gates sidesteps many of these issues by mapping qubit evolution to geometric space curves in three dimensions~\cite{Zeng.NJP.2018,ZengPRA2018,Zeng.PRA.2019,Buterakos.PRXQ.2021}. Control fields that cancel transverse noise errors can be obtained from closed space curves. While some progress has been made in extending this method to the cancellation of control field errors, this often comes at the expense of reduced flexibility in waveform shaping~\cite{Barnes.SciRep.2015,Throckmorton2019,Gungordu2019}.

In this paper, we present a general technique for constructing non-adiabatic holonomic gates that dynamically correct transverse noise errors. This is achieved by bringing together two types of geometry: holonomic trajectories on the Bloch sphere and geometric space curves that quantify the error accrued by transverse noise. We show how to systematically design qubit evolutions that exhibit both closed holonomy loops and closed error curves, achieving the simultaneous cancellation of both control field errors and transverse noise errors that are beyond the reach of purely holonomic methods. We refer to these evolutions as Doubly Geometric (DoG) gates. We demonstrate our approach by constructing several families of DoG gates based on smooth pulses.

The paper is organized as follows. In Sec.~\ref{sec:2geometries} we discuss the two types of geometry that characterize the evolution of a noisy quantum system: the holonomy of quantum states and geometric space curves that quantify the response of the system to transverse noise. In Sec.~\ref{sec:DoGgates}, we show how to combine space curves with holonomy to arrive at a general procedure for finding control fields that suppress both control errors and transverse noise errors. We present several explicit examples of robust gates generated by this procedure, \textcolor{black}{including single-qubit and two-qubit entangling gates. Quantum speed limits and experimental implementations are discussed.} We conclude in Sec.~\ref{sec:conslusion}. Several appendices contain technical details about our general technique and our explicit examples.

\section{Two geometries of quantum evolution}\label{sec:2geometries}

We begin by describing the two types of geometry that come into play when a qubit evolves in the presence of noise. The first is the notion of holonomy on the Bloch sphere, which facilitates the design of gate operations that are based on geometric phases. The second is a geometric space curve formalism that allows one to view the accumulation of errors due to transverse noise as a three-dimensional path in the space of traceless Hermitian operators that we refer to as the ``error curve". After we describe these two types of geometry, we show how they can be combined together to form a framework for designing dynamically corrected holonomic gates in the following section. 

\subsection{Geometric phases and holonomy}

\begin{figure}[h!]
    \centering
    \includegraphics[width=0.5\textwidth]{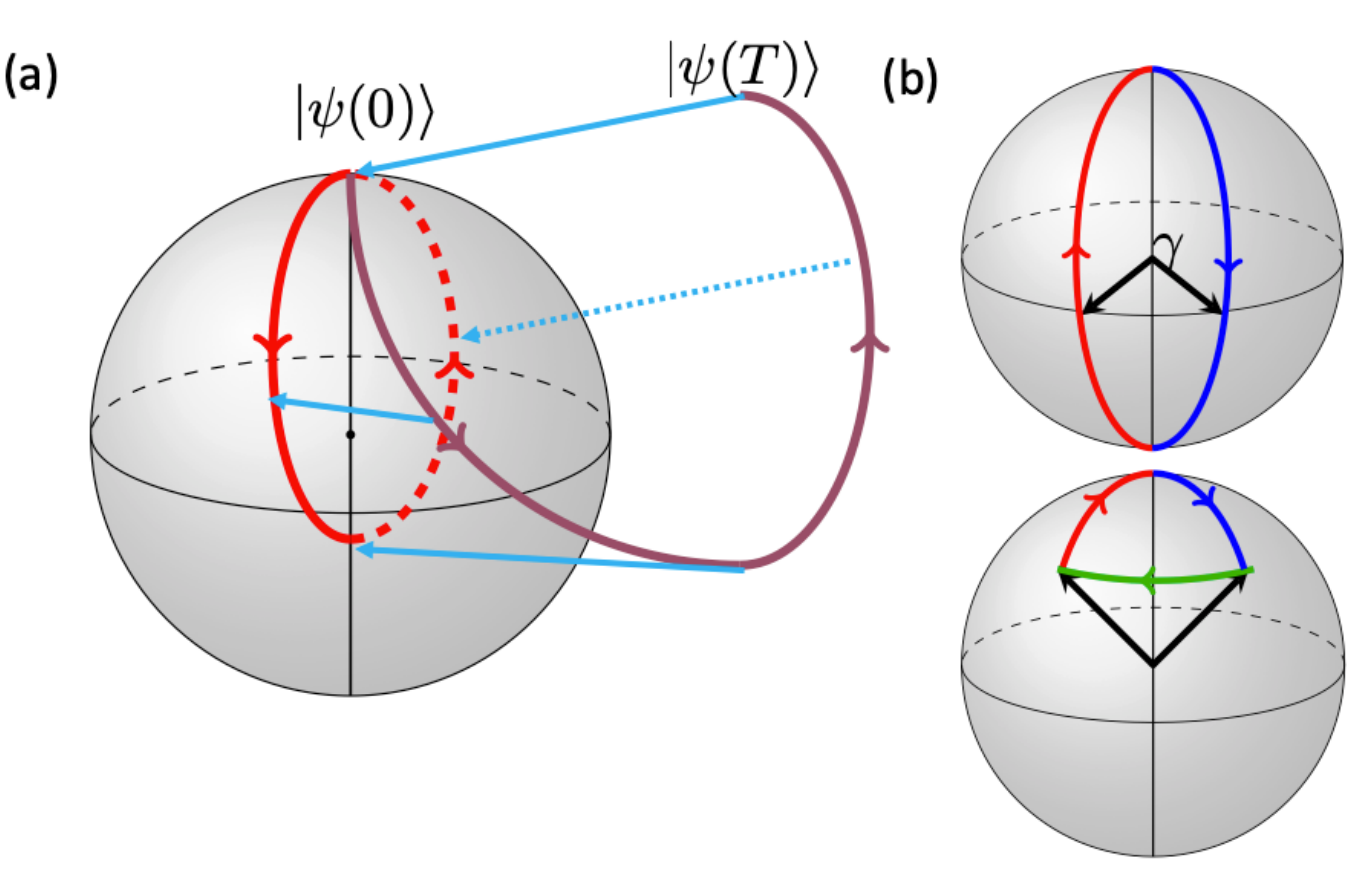}
    \caption{(a) Schematic depiction of how the Aharonov-Anandan geometric phase is defined for a two-level system. States in the full Hilbert space (open dark magenta curve) are projected onto the Bloch sphere (blue arrows), and the projected states (closed red curve) enclose an area that determines the geometric phase.
    (b) Different Bloch sphere paths that realize holonomic gates, including the orange-slice model (top) and the triangle-cap model (bottom).}
    \label{fig:holonomic_paths}
\end{figure}

Aharonov and Anandan~\cite{PhysRevLett.58.1593} showed that there is a geometric phase associated with any cyclic evolution that starts from state $\ket{\psi(0)}$ and ends in the parallel state $\ket{\psi(T)}$ at time $T$. This phase is given by
\begin{equation}\label{eq:def_of_geometric_phase}
\begin{aligned}
    \beta_g(T)=&\arg \langle\psi(0) | \psi(T)\rangle \textcolor{black}{+} i \int_{0}^{T}\langle\psi(t)| \dot{\psi}(t)\rangle d t   \\
    =& i\int^{T}_0\langle \nu(t)|\partial_t| \nu(t)\rangle d t.
\end{aligned}
\end{equation}
The right-hand side of the first line of Eq.~\eqref{eq:def_of_geometric_phase} amounts to subtracting the dynamical phase from the total phase, $\beta_g(T)=\beta_{\text{total}}(T)-\beta_d(T)$. The dynamical phase is often written as an integral of the expectation value of the Hamiltonian, $\bra{\psi(t)} H(t) \ket{\psi(t)}$. Note however that this expression is frame-dependent, as we discuss in Appendix~\ref{appendix:def_DP}. The second line of Eq.~\eqref{eq:def_of_geometric_phase} comes from mapping $\ket{\psi(t)}$ onto state $\ket{\nu(t)}$ in a projective space defined such that all parallel $\ket{\psi(t)}$ map to the same $\ket{\nu(t)}$. In the case of a two-level system, which is the focus of this work, $\ket{\nu(t)}$ lives on the Bloch sphere. This mapping is illustrated in Fig.~\ref{fig:holonomic_paths}(a). The area enclosed by the trajectory of $\ket{\nu(t)}$ determines the geometric phase. It is worth noting that unlike in the case of the Berry phase \cite{Berry.Geometric.Phase}, the Aharonov-Anandan phase is defined without reference to an adiabatic condition; it is exact for any cyclic evolution. Hereafter, we refer to the Aharonov-Anandan phase as the geometric phase. 

The defining feature of the geometric phase is that it depends only on the path traced out by $\ket{\nu(t)}$ on the Bloch sphere~\cite{Samuel.Bhandari.PRL.1988,Pachos.IJMPB.2001}. 
As a consequence, the geometric phase is insensitive to ``parallel" errors, i.e., errors that only affect the rate at which the path is traversed and that have no effect on the shape of the path~\cite{ZhuShiLiang.PRA.2005,DeChiara.PRL.2003}.
As we show in Appendix~\ref{appendix:robustness}, this property of the geometric phase makes it robust against first- and second-order parallel errors, including certain driving field errors. Note that in contrast, the  dynamical phase is generally susceptible to these errors even at first order.
The robustness of geometric phases against noise errors is the central idea behind holonomic quantum computation~\cite{ZANARDI199994}. However, it is important to note that the geometric phase is still sensitive to noise that is ``perpendicular" to the path, meaning that the noise deforms the path. Our work concerns the simultaneous correction of these additional errors.

To design an abelian holonomic gate on a qubit based on the geometric phase, we start from a cyclic evolution of a two-dimensional system (the same strategy also applies in higher dimensions):
\begin{equation}
\begin{aligned}
\label{eq:holonomic_evolution}
    \ket{\psi_k(T)}= e^{i\alpha_k(T)} \ket{\psi_k(0)},
 \end{aligned}
\end{equation}
where $k\in \{0,1\}$ and where $\alpha_k(T)$ is the total phase accumulated during the evolution. 
The parallel transport condition, $\bra{\psi_k(t)} H(t) \ket{\psi_k(t)}=0$,
ensures that the dynamical phase vanishes, so that the total phase is equal to the geometric phase and is therefore robust. In the following, we use the term holonomic evolution to refer to cyclic evolution that satisfies the parallel transport condition to distinguish it from evolution that is merely cyclic. This abelian holonomic evolution is cyclic for both initial states $\ket{\psi_k(0)}$ with $k=0,1$. This is in contrast to the non-abelian case where it is the manifold, instead of each basis state, that is cyclic \cite{ANANDAN1988171,Wilczek.Zee}. Non-abelian gates require auxiliary levels to satisfy the parallel transport condition in the computational manifold \cite{Xu.PRL.2012,Sjoqvist,Zhao.PZ.PRA.2020}.
In this work, we only consider abelian holonomic evolution since it is sufficient to construct arbitrary single-qubit  gates.

The explicit construction of a holonomic single-qubit gate is as follows. First, parameterize the evolution operator as
\begin{equation}
\label{eq:evolution_operator}
    U_c(t)=\left(
\begin{array}{cc}
 e^{i \alpha (t)} \cos \frac{\theta (t)}{2} & -e^{-i (\alpha (t)+\phi (t))} \sin \frac{\theta (t)}{2} \\
 e^{i (\alpha (t)+\phi (t))} \sin \frac{\theta (t)}{2} & e^{-i \alpha (t)} \cos \frac{\theta (t)}{2}\\
\end{array}
\right). 
\end{equation}
The columns are the time-evolved computational states: $\ket{\psi_0(t)}$= $e^{i\alpha(t)}$ $\left(\begin{array}{cc}
    \cos \frac{\theta(t)}{2} ,    \sin\frac{\theta(t)}{2}e^{i\phi(t)} 
\end{array}\right)^\intercal$
and $\ket{\psi_1(t)}$=$e^{-i\alpha(t)}$ $\left(\begin{array}{cc}
    -\sin \frac{\theta(t)}{2}e^{-i\phi(t)},
   \cos\frac{\theta(t)}{2}
\end{array}\right)^\intercal,$
where $\intercal$ is the transpose operator.
It is important to note that these two states pick up opposite geometric phases after cyclic evolution ($\beta_{g,0}(T)=-\beta_{g,1}(T):= \beta_{g}(T)$). The evolution path $\theta(t),\phi(t)$ on the Bloch sphere determines the total accumulated phase, provided the parallel transport condition, $\dot{\alpha}(t)=-\frac{1}{2}\big(1-\cos\theta(t)\big)\dot{\phi}(t)$, is satisfied. When this is the case, the resulting holonomic evolution operator is then given by
\begin{equation}
\label{eq:holonomic_gate_two_base}
    U_c(T)= e^{i\beta_g(T)}\ket{\psi_0(0)}\bra{\psi_0(0)}+e^{-i\beta_g(T)}\ket{\psi_1(0)}\bra{\psi_1(0)},
\end{equation}
where the geometric phase is $\beta_g(T)=-\frac{1}{2}\int^T_0(1-\cos\theta)d\phi$.
Although this construction yields a single-axis rotation, arbitrary holonomic gates can be implemented by changing the initial state and Bloch sphere trajectory. In this work, we focus on the initial state $\ket{\psi_0(0)}=(1,0)^\intercal$ without loss of generality.

It remains to determine how the qubit should be driven such that it undergoes holonomic evolution. Consider a general three-field Hamiltonian:
\begin{equation}
\label{eq:3_field_hamiltonian}
    H_c(t)=\frac{\Omega(t)}{2}
\left(\begin{array}{cc}
 0 & e^{-i \Phi(t)} \\
 e^{i \Phi(t)} & 0 \\
\end{array} \right) + \frac{\Delta(t)}{2}
\left(\begin{array}{cc}
 1 & 0 \\
0 & -1 \\
\end{array} \right),
\end{equation}
which includes the Rabi frequency $\Omega(t)$, the phase field $\Phi(t)$ and the detuning field $\Delta(t)$. 
Interestingly, there is a one-to-one mapping between a holonomic trajectory on the Bloch sphere and the control fields that generate it:
\begin{equation}
\label{eq:field_path_map_TDM}
        \begin{aligned}
            \Omega(t)=&\sqrt{\dot{\theta}(t)^2+\sin ^2\theta (t) \cos ^2\theta (t) \dot{\phi}(t)^2}, \\
             \Phi(t)=&  \arg \bigg(-\big[\dot{\theta}(t) \sin \phi(t)+\dot{\phi}(t) \sin \theta(t) \cos \theta(t) \cos \phi(t)\big] \\
             & \quad +i\big[\dot{\theta}(t) \cos \phi(t)-\dot{\phi}(t) \sin \theta(t) \cos \theta(t) \sin \phi(t)\big] \bigg) , \\
              \Delta(t)=& \sin^2 \theta(t) \dot{\phi}(t).
        \end{aligned}
\end{equation}
The algebraic steps that lead to this result can be found in Appendix~\ref{appendix:holonomy_derive}. The same result was also derived by Li \textit{et al.} in Ref.~\cite{LiKZ.PRR.2020}. Given a holonomic path, we can determine the control fields that produce it using Eq.~\eqref{eq:field_path_map_TDM}.

The so-called orange-slice model holonomic gate \cite{KwiatPRL1991,Mousolou.PRA.2014} can be implemented using a two-field control Hamiltonian ($\Delta(t)\equiv0$), where $\Phi(t)$ is a step function. The corresponding Bloch sphere trajectory is illustrated in Fig.~\ref{fig:holonomic_paths}(b). The resulting geometric phase is robust to systematic noise in the driving field $\Omega(t)$.

\subsection{Geometric error curves}\label{sec:space_curves}

In addition to errors in control fields, many qubit platforms also suffer from noise fluctuations in the detuning~\cite{BluhmPRL2010,BylanderNaturePhysics2011,MartinsPRL2016,KlimovPRL2018,Burnett.npjQI.2019,CHYang.NatEle.2019}. Because this noise is transverse to the driving field $\Omega(t)$, it is not correctable using the holonomic approach described above. In systems such as semiconductor spin qubits and superconducting qubits, these fluctuations are often slow compared to typical gate times, allowing one to describe them using a quasistatic noise model~\cite{BluhmPRL2010,BylanderNaturePhysics2011,MartinsPRL2016,KlimovPRL2018,BurnettnpjQI2019,YangNatureElectronics2019}. In this case, the Hamiltonian becomes
\begin{equation}\label{eq:noisy_3field_ham}
    H(t)=H_c(t)+\delta_z\sigma_z,
\end{equation}
where $\delta_z$ is an unknown constant that gives rise to errors in the target gate operation.
If $\delta_z$ is sufficiently small, then its detrimental effects can be ameliorated by designing $H_c(t)$ appropriately.

Zeng \textit{et al.} presented a general approach to finding control fields that suppress quasistatic noise errors in the detuning~\cite{Zeng.NJP.2018,Zeng.PRA.2019}. This method utilizes a mapping between qubit evolution and space curves in three dimensions. Here, we refer to these as ``error curves" because they quantify the extent to which the qubit deviates from its ideal evolution. In this section, we briefly review this error-curve formalism and adapt it to the noisy three-field Hamiltonian of Eq.~\eqref{eq:noisy_3field_ham}. In the next section, we show how this technique can be combined with holonomy to produce gates that are insensitive to noise in both the driving field and detuning.

The error-curve formalism begins by transforming to the interaction picture, in which the Hamiltonian becomes $H_I(t)=\delta_zU_c^{\dagger}(t)\sigma_zU_c(t)$, where $U_c(t)$ is the evolution operator generated by $H_c(t)$. Note that in the ideal case where $\delta_z=0$, the evolution operator $U_I(t)$ corresponding to $H_I(t)$ would equal the identity at all times. The main idea behind dynamically corrected gates is to engineer $H_c(t)$ such that $U_I(T)$ is as close to the identity as possible. Enforcing this condition is made difficult by the fact that it is generally hard to calculate $U_I(t)$ analytically. However, if $\delta_z$ is sufficiently small, we can employ a Magnus expansion, which to first order gives $U_I(t)=\exp[-i\delta_zA_1(t)]$, with $A_1(t)=\int^t_0U_c^{\dagger}(t')\sigma_zU_c(t')dt'$. This function can be expanded in a basis of Pauli matrices:
\begin{equation}
     A_1(t)=\bm{r}(t)\cdot \bm{\sigma} =  x(t)\sigma_x+y(t)\sigma_y+z(t)\sigma_z.
\end{equation}
The vector $\bm{r}(t)$ traces out a three-dimensional space curve as time evolves. Because the length of $\bm{r}(t)$ quantifies the deviation of $U_I(t)$ away from the identity, we call this the error curve. Clearly, we have $\bm{r}(0)=0$. If the error curve closes on itself at the final time $T$, then $U_I(T)=\mathbbm{1}+\mathcal{O}(\delta_z^2)$, and the resulting gate is first-order insensitive to detuning noise. Thus, there is a direct relationship between robust gates and closed error curves.

We can write down an explicit parameterization of the error curve using our general form for $U_c(t)$ from Eq.~\eqref{eq:evolution_operator}:
\begin{equation}
\label{eq:error_curve}
\begin{aligned}
   &\bm{r}(t)=-\hat x\int^t_0\sin\theta(t')\cos\big(2\alpha(t')+\phi(t')\big)dt' \\
    &-\hat y\int^t_0\sin\theta(t') \sin\big(2\alpha(t')+\phi(t')\big)dt'+\hat z\int^t_0\cos\theta(t')dt'.
\end{aligned}
\end{equation}
The derivative of the error curve yields another curve called the tangent indicatrix (or tantrix for short): 
\begin{equation}
\label{eq:tantrix}
\begin{aligned}
    \dot{\bm{r}}(t)=&-\sin\theta(t)\cos\big(2\alpha(t)+\phi(t)\big)\hat x  \\
   & -\sin\theta(t) \sin\big(2\alpha(t)+\phi(t)\big)\hat y+\cos\theta(t)\hat z.
\end{aligned}
\end{equation}
At each time $t$, $\dot{\bm{r}}(t)$ is oriented in the direction that is tangent to the error curve at that point.
Note that the tantrix is a unit vector: $\|\dot{\bm{r}}(t)\|^2=1$, so it lives on a unit sphere. One should be careful to distinguish the tantrix from the projected evolution path on the Bloch sphere that is used to define the geometric phase. The relationship between these two curves plays an important role in the next section. Also notice that because the tantrix is normalized, time coincides with the arc length of the curve. Therefore, the length of the error curve is equal to the gate time $T$.

\begin{figure}
    \centering
    \includegraphics[width=0.45\textwidth]{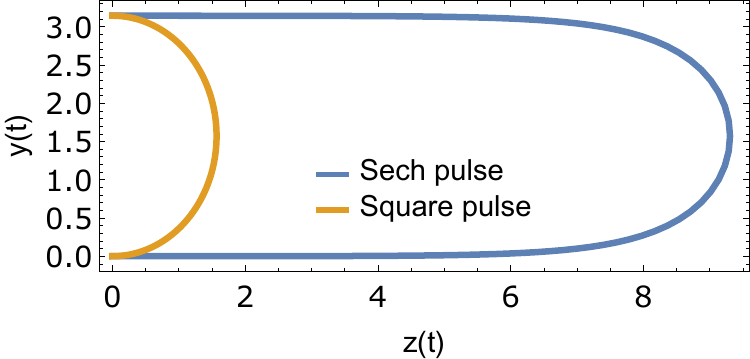}
    \caption{Error curve $\bm{r}(t)$ for $H_c(t)=\Omega(t)\sigma_x$ where $\Omega(t)$ is a square pulse (orange) or a hyperbolic secant pulse (blue). In both cases, the pulses have area $\pi$. The sech pulse is given by $\Omega(t)=\text{sech}(t)$, where $t\in[-10,10]$.}
    \label{fig:fig2_error_curve_2D}
\end{figure}

Three-dimensional space curves are characterized by two real functions: the curvature $\kappa(t)$ and the torsion $\tau(t)$. Given these two functions, the space curve can be generated by integrating the Frenet-Serret equations~\cite{Zeng.PRA.2019}. Conversely, the curvature and torsion can be computed from the curve. Here, we can use the explicit parameterization in Eq.~\eqref{eq:error_curve} together with the fact that $U_c$ satisfies the Schr\"odinger equation involving $H_c$ to relate these quantities to the control fields (see Appendix~\ref{appendix:geometric_formalism}):
\begin{equation}
\begin{aligned}\label{eq:curvature_torsion}
    \kappa(t)&= \|\ddot{\bm{r}}(t)\| =\Omega(t),\\
 \tau(t) &=  \frac{(\dot{\bm{r}}\times \ddot{\bm{r}})\cdot\dddot{\bm{r}} }{\big\| \dot{\bm{r}}\times \ddot{\bm{r}}  \big\|^2}  = \dot{\Phi}(t)-\Delta(t).  
\end{aligned}
\end{equation}
We see that the curvature and torsion of the error curve are given by the driving amplitude, $\Omega(t)$, and by the difference of the derivative of the phase field and the detuning, $\dot{\Phi}(t)-\Delta(t)$, respectively. As an example, in Fig.~\ref{fig:fig2_error_curve_2D} we show two error curves with $\Delta=0=\Phi$, where in one case $\Omega(t)$ is a square pulse, and in the other a hyperbolic secant (sech) pulse. Notice that the error curve for the square pulse is a semicircle, reflecting the fact that the curvature is constant in this case. The long straight segments of the sech error curve correspond to the long tails of the sech pulse. 

\begin{figure} []
    \centering
    \includegraphics[width=0.48\textwidth]{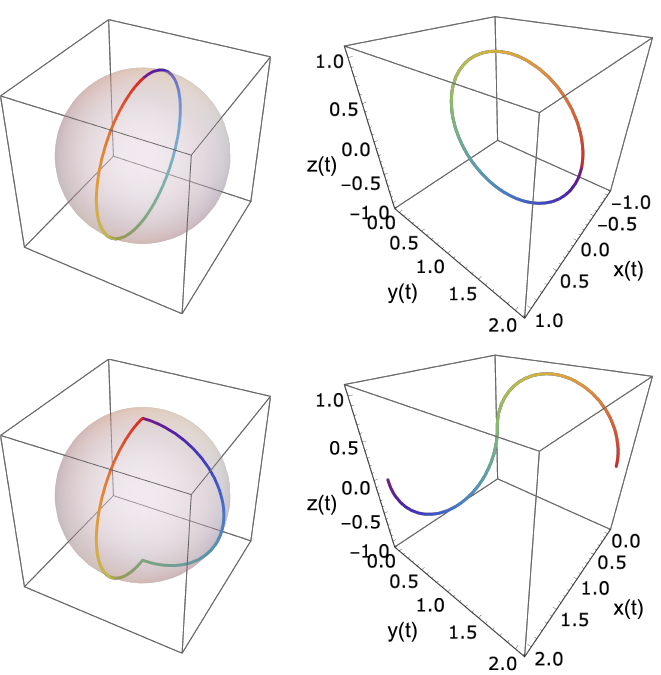}
    \caption{The Bloch sphere representations (left) and error curves $\bm{r}(t)$ (right) of the standard orange-slice model holonomic gate, generated by the control Hamiltonian $H_{\text{o-s}}(t)$ (Eq.~\eqref{eq:orange_slice_ham}). The upper (lower) plots  correspond to $\Phi_0=0$ ($\Phi_0=\pi/2$). The curves are colored to indicate which part of the Bloch sphere trajectory corresponds to which part of the error curve, starting from red at $t=0$ and ending with violet at $t=T$.}
    \label{fig:fig3_simple_orange_slice_model}
\end{figure}


The form of the error curve in the square pulse case allows us to understand pictorially the sensitivity of the orange-slice model holonomic gate \cite{Mousolou.PRA.2014,SunLuYan.PRL.2020,AiMingzhong.PRApp.2020} to detuning noise. This gate is implemented using a two-field control Hamiltonian that evolves states sequentially along two geodesic lines that join together at the two poles of the Bloch sphere:
\begin{equation}\label{eq:orange_slice_ham}
    H_{\text{o-s}}(t)= 
    \begin{cases}
     \frac{\Omega_1(t)}{2} \left(\begin{array}{cc}
         0 &1  \\
          1 & 0 
     \end{array} \right) \quad  0\leq t\leq T_1 \\
     \frac{\Omega_2(t)}{2} \left(\begin{array}{cc}
         0 & e^{-i \Phi_0}  \\
          e^{i \Phi_0} & 0 
     \end{array} \right) \quad  T_1<t \leq  T, \\
    \end{cases}
\end{equation}
where $\int^{T_1}_0\Omega_1(t)dt=\int^{T}_{T_1}\Omega_2(t)dt=\pi$ gives the two pole-to-pole evolution paths. Switching $\Phi$ from 0 to $\Phi_0$ when the state reaches the south pole at time $T_1$ causes the system to return to the north pole along a different geodesic, yielding a closed ``orange-slice" trajectory overall. This is depicted in Fig.~\ref{fig:fig3_simple_orange_slice_model} for $\Phi_0=0$ and $\Phi_0=\pi/2$. The resulting rotation angle of the gate is determined by the solid angle enclosed by the slice, which in turn depends on $\Phi_0$. These gates tend to be robust to systematic errors in the pulse $\Omega(t)$ that are parallel to the path~\cite{AiMingzhong.PRApp.2020,SunLuYan.PRL.2020}. However, the improvements afforded by this approach are generally sensitive to the details of how the gate is implemented. These subtleties are discussed in Appendix~\ref{appendix:HG_pulse_error}.

Fig.~\ref{fig:fig3_simple_orange_slice_model} also shows the corresponding error curves for both values of $\Phi_0$. We see that for $\Phi_0=0$, the error curve is closed, indicating that first-order detuning noise errors are cancelled. On the other hand, for $\Phi_0=\pi/2$, the error curve is not closed, and so detuning errors are not corrected in this case. More generally, the error curve does not close for any nonzero value of $\Phi_0$. In these cases, the error curve consists of two semicircles that lie in planes that are rotated relative to each other by angle $\Phi_0$. The intersection point at $t=T_1$ is a singular point with infinite torsion ($\dot{\Phi}(T_1)=\infty$). Note that replacing the square pulses with other single-pulse waveforms (e.g., hyperbolic secant or Gaussian pulses) cannot make the error curve close as long as $\Phi_0\ne0$.

Finally, we note that different Hamiltonians can generate the same error curve. To see this consider transforming to a frame defined by the the detuning field (with transformation matrix $e^{-i\int^t_0\frac{\Delta(t')}{2}\sigma_z dt'}$), in which case we obtain a {two-field control} Hamiltonian $\tilde{H}_c(t)=\frac{\Omega (t )}{2}
\left( \begin{array}{cc}
 0 &   e^{-i \tilde{\Phi}(t)} \\
    e^{i \tilde{\Phi}(t)} & 0 \\
\end{array} \right), $ where $\dot{\tilde{\Phi}}(t)-\dot{\Phi}(t)=-\Delta(t)$. $\tilde{H}_c(t)$ and $H_c(t)$ map to the same error curve, because the transformation between them commutes with the noise term in Eq.~\eqref{eq:noisy_3field_ham}.

\section{Doubly geometric robust gates}\label{sec:DoGgates}

\subsection{\textcolor{black}{Single-qubit DoG gates}}

We saw in the case of the orange-slice model that there exist qubit evolutions (albeit trivial) which are both holonomic and have closed error curves (see Fig.~\ref{fig:fig3_simple_orange_slice_model}). However, we also saw that holonomic trajectories $\bm{h}(t)$ on the Bloch sphere do not generically translate to closed error curves. This leads to the question: Is there a systematic way to find nontrivial holonomic evolutions that have closed error curves? If so, this would enable the design of dynamically corrected gates that correct both pulse and detuning errors. We now show that this is indeed possible, and we present an explicit procedure for constructing such gates.

We can always map a holonomic trajectory parameterized by $\theta(t)$ and $\phi(t)$ onto an error curve by imposing the parallel transport condition in Eq.~\eqref{eq:tantrix}, which yields a restricted form for the tantrix:
\begin{equation}
\label{eq:holonomic_tantrix}
\begin{aligned}
        \dot{\bm{r}}(t)=&-\sin\theta(t)\cos\big(\int^t_0\cos\theta(t')\dot{\phi}(t')dt'\big)\hat x \\& -\sin\theta(t) \sin\big(\int^t_0\cos\theta(t')\dot{\phi}(t')dt'\big)\hat y+\cos\theta(t)\hat z.
\end{aligned} 
\end{equation}
If we consider a closed trajectory for $\bm{h}(t)$ that starts and ends at the north pole of the Bloch sphere at times $t=0$ and $t=T$, respectively, then this implies $\theta(0)=\theta(T)=0$. This in turn means that the tantrix must be closed, with $\dot{\bm{r}}(T)=\dot{\bm{r}}(0)=\hat z$. This condition can always be satisfied if the error curve is smooth and closed, because in this case the tantrix will be continuous, and we can always perform a rigid rotation to align the initial and final tangent vectors with $\hat z$. However, it is not easy to design a holonomic evolution $\bm{h}(t)$ such that the corresponding error curve obtained by integrating Eq.~\eqref{eq:holonomic_tantrix} is closed. 

To circumvent this difficulty, we can instead start by choosing a smooth, closed error curve $\bm{r}(t)$. Eq.~\eqref{eq:holonomic_tantrix} implies that this maps to a {\it unique} holonomic trajectory $\bm{h}(t)$, which can be obtained by differentiating $\bm{r}(t)$ and then extracting $\theta(t)$ and $\phi(t)$ from the result. The control fields can then be obtained using Eq.~\eqref{eq:field_path_map_TDM}. These equations can be recast in terms of quantities obtained directly from the error curve. The general procedure can then be summarized in terms of the following four steps:
\begin{enumerate}
    \item Design a smooth error curve $\bm{r}(t)=x(t)\hat x+y(t)\hat y+z(t)\hat z$ such that $\bm{r}(T)=\bm{r}(0)$ and $\dot{\bm{r}}(T)=\dot{\bm{r}}(0)=\hat z$, and such that the tantrix is normalized, $\|\dot{\bm{r}}(t)\|=1$.
    \item Compute the holonomic control fields using
    \begin{align}
        \Omega(t)&=\|\ddot{\bm{r}}(t)\|,\\
        \Delta(t)&=\frac{\dot x\ddot y-\dot y\ddot x}{\dot z},\\
        \Phi(t)&=\int_0^t[\tau(t')+\Delta(t')]dt',
    \end{align}
    where $\tau(t)$ is the torsion of the error curve, which is given in Eq.~\eqref{eq:curvature_torsion}.
    \item If desired, the holonomic trajectory on the Bloch sphere $\bm{h}(t)$ can be obtained from
    \begin{equation}
        \begin{aligned}
            \theta(t)=\arccos(\dot z), \quad             \phi(t)=\int^t_0 \frac{\dot{x}\ddot{y}-\dot y\ddot x}{\dot z(1-\dot z^2)}dt'.
        \end{aligned}
    \end{equation} 
    \item The geometric phase is given by
    \begin{equation}
        \beta_g(T)=-\frac{1}{2}\int_0^T\frac{\dot x\ddot y-\dot y\ddot x}{\dot z(1+\dot z)}dt.
    \end{equation}
\end{enumerate}
We refer to the resulting gates as doubly geometric (DoG) gates.

Before we demonstrate this technique with explicit examples, we point out that the above procedure reveals that there is a unique holonomic trajectory associated with each smooth closed error curve. In contrast, there are infinitely many cyclic (non-holonomic) trajectories on the Bloch sphere that can be associated with the same closed error curve. A distinct geometric phase can be defined for each of these non-holonomic trajectories. In the absence of the parallel transport constraint, it is not possible to uniquely relate the curvature and torsion of the error curve to the three fields of the control Hamiltonian $H_c(t)$. This is evident from Eq.~\eqref{eq:curvature_torsion}, where the torsion depends only on the difference $\dot{\Phi}(t)-\Delta(t)$. Two Hamiltonians that have distinct $\Phi(t)$ and $\Delta(t)$ but the same $\Omega(t)$ and $\dot{\Phi}(t)-\Delta(t)$ will generate the same error curve but different Bloch sphere trajectories. The infinite family of cyclic evolutions associated with a particular error curve are related to each other via unitary transformations that commute with the error term $\delta_z\sigma_z$.

\begin{figure}
    \centering
    \includegraphics[width=0.5\textwidth]{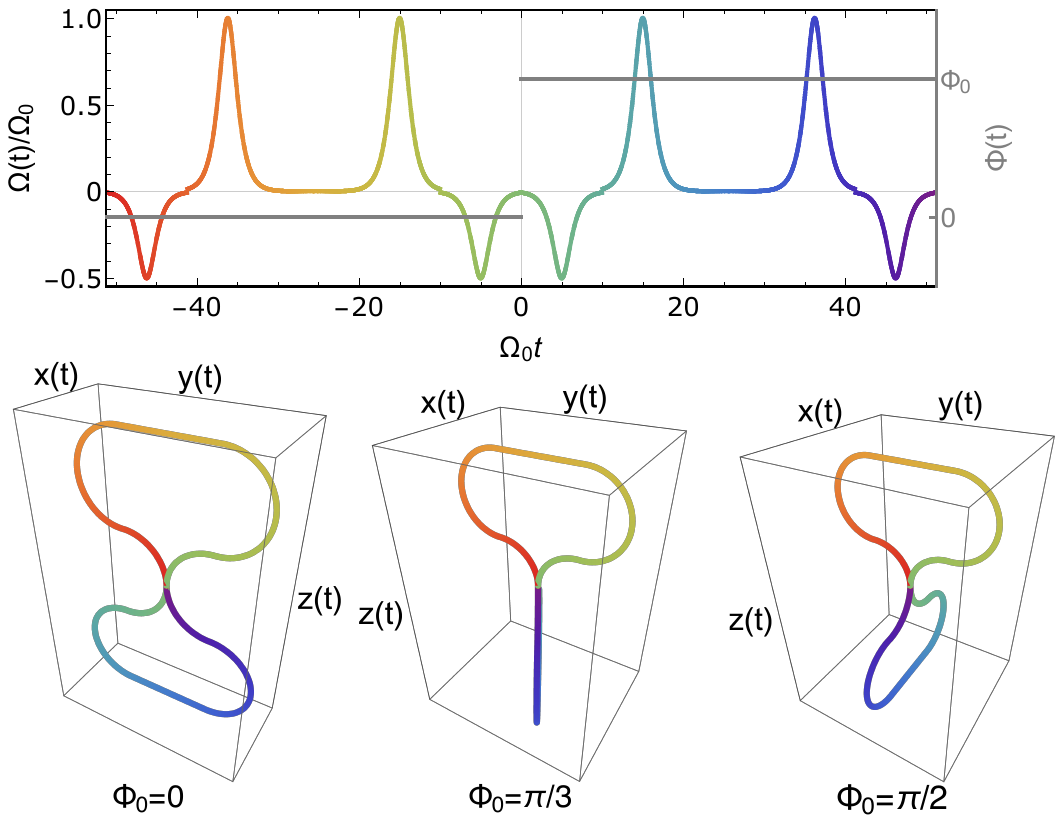}
    \caption{The two-field control (top) and error curves (bottom) of the extended orange-slice DoG gate for three different gate angles $\Phi_0$. Each DoG gate consists of two pole-to-pole evolutions along different geodesic lines on the Bloch sphere. Each pole-to-pole evolution is generated by four sech pulses that collectively implement a $\pi$ pulse. 
    Each closed $\bm{r}(t)$  consists of two closed planar lobes that correspond to each pole-to-pole evolution. The colors indicate which part of the curve corresponds to which part of the pulse.}
    \label{fig:fig5_2D_orange_sllice_DCHG}
\end{figure}

It is also important to note that the above procedure, as stated, generates arbitrary rotations about the $z$ axis. This is easily generalized to arbitrary rotation axes by 
designing the error curve such that $\dot{\bm{r}}(T)=\dot{\bm{r}}(0)\ne\hat z$. Other choices of the initial tangent vector correspond to initial states for the holonomic evolution that differ from the logical basis states, i.e., $\bm{h}(t)$ starts away from the poles of the Bloch sphere. Note that even though such a curve might be equivalent to a $z$-basis curve in terms of its curvature and torsion, the parallel transport conditions will be different, resulting in different control fields and Bloch sphere trajectories. In addition to designing a new curve with different boundary conditions, one can also generate other types of gates by starting from a $z$-basis curve and redefining the start/end point to lie elsewhere on the curve. This will again result in $\bm{h}(t)$ starting away from the Bloch sphere poles. Note, however, that if one wants a control field that starts with zero amplitude, then this may restrict which points along an error curve can serve as the start/end point.

We now construct a class of DoG gates by modifying the orange-slice model. In particular, we generalize it from single-pulse  to multi-pulse driving along each geodesic. The pulses along each geodesic vary in amplitude and sign but combine to generate a net $\pi$ rotation that drives the system from one pole of the Bloch sphere to the other. One can also design the pulses in each half of the evolution such that the error curve forms a closed planar curve. The two sets of pulses (one set for each geodesic) together generate a closed holonomic trajectory $\bm{h}(t)$ and a closed error curve $\bm{r}(t)$. The extended orange-slice DoG gate always has a closed error curve, in contrast to the standard orange-slice model. These driving fields and their corresponding error curves are shown in Fig.~\ref{fig:fig5_2D_orange_sllice_DCHG}. The DoG method affords tremendous flexibility in terms of the pulse shapes that can be used since we can choose any smooth, closed error curve $\bm{r}(t)$. In particular, we can ensure that the control waveforms are  experimentally friendly, which is crucial for practical implementations. With this in mind, we focus on designing DoG gates based on sech or sech-like pulses $\Omega(t)$, which, due to their analytical properties, have been proposed for quantum gates~\cite{PhysRevB.74.205415,EconomouPRL2007,EconomouPRB2012} and used in experiments with quantum dots~\cite{Greilich2009} and superconducting qubits~\cite{KuPRA2017}. More details about the control fields and error curves in Fig.~\ref{fig:fig5_2D_orange_sllice_DCHG} can be found in Appendix~\ref{appendix:orange_slice-DCHG}.

\begin{figure}
    \centering
    \includegraphics[width=0.5\textwidth]{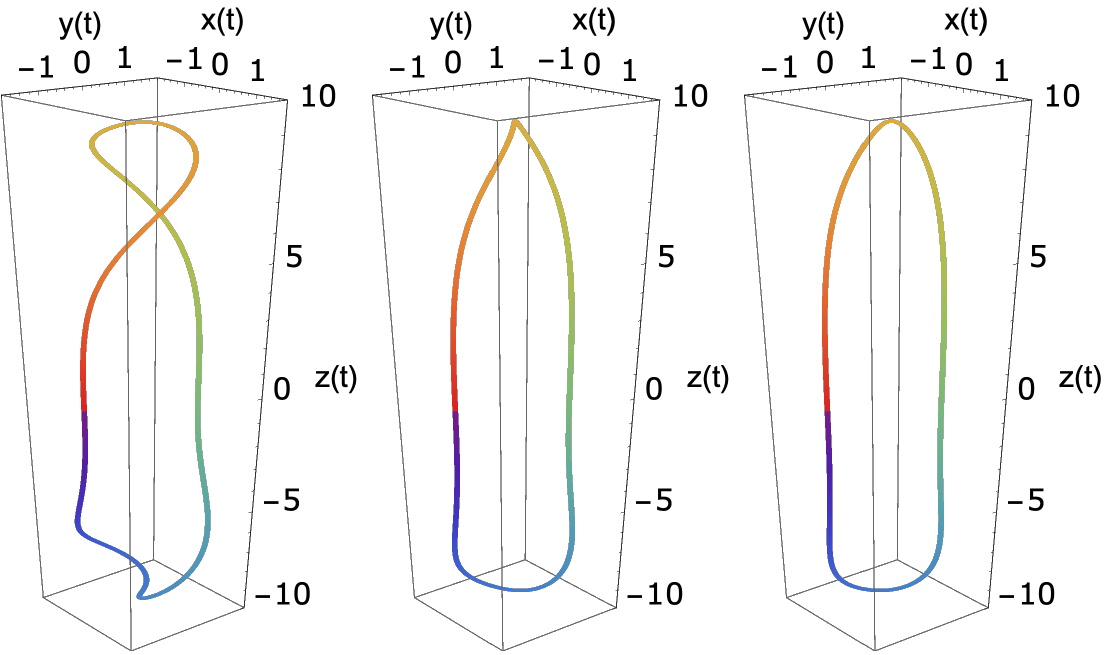}
    \caption{Twisted error curves $\bm{r}_\xi(t)$ for three values of the twist parameter $\xi$: $\pi/1000$ (left), $\pi/3000$ (middle), and $\pi/5000$ (right). As $\xi$ is reduced to zero, the curve flattens out and lies in a vertical plane. Note that $\dot{\bm{r}}_\xi(T)=\dot{\bm{r}}_\xi(0)=\hat{z}$ (corresponding to the point where the red and purple segments meet) for all values of $\xi$.}
    \label{fig:fig6_3D_DCHG_ErrorCurve}
\end{figure}

\begin{figure}
    \centering
    \includegraphics[width=0.5\textwidth]{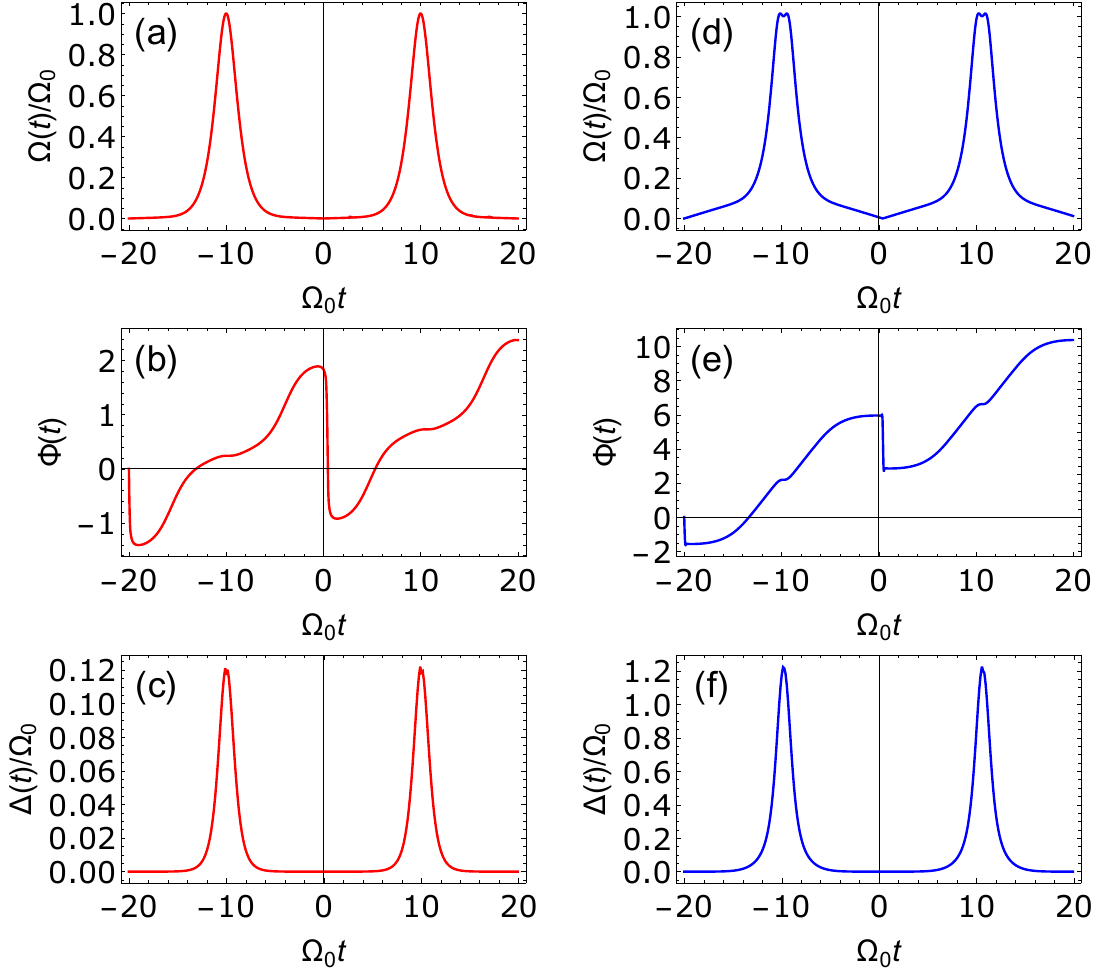}
    \caption{3D DoG control fields $\Omega(t)$, $\Phi(t)$, $\Delta(t)$ generated from a twisted error curve $\bm{r}_\xi(t)$ with  $\xi=\pi/20000$ (left) and  $\xi=\pi/2000$ (right). Time is shifted such that the zero points on the abscissas are centered in all plots. $\Omega(t)$ and $\Delta(t)$ both have near-sech envelopes.  }
    \label{fig:fig7_3D_DCHG_field}
\end{figure}

Like the original orange-slice model, the above class of DoG gates is based on restricting the field $\Phi(t)$ to be piece-wise constant, which corresponds to error curves that have zero torsion everywhere except at a single point. Here, we instead consider an entirely different class of DoG gates in which the torsion is allowed to vary continuously throughout the curve rather than only in the vicinity of a single point. In this case, the error curves are truly three-dimensional, and not locally two-dimensional like in the case of the extended orange-slice model. We thus refer to the former as 3D DoG gates, and the latter 2D DoG gates. As an explicit example of a family of 3D DoG gates, we again start from a planar curve with sech curvature. This time, however, we continuously twist the curve about the $z$ axis so that it fills all three dimensions. The twisting causes the curvature to deviate from a sech function, and it creates nonzero torsion. All three control fields are then nonzero in this case. The precise construction of this family of three-dimensional error curves is as follows. We start from an untwisted planar curve $\bm{r}_0(t)=y(t)\hat y+z(t)\hat z$ (here we choose it to lie in the $yz$ plane for concreteness). We then obtain a twisted version of the curve from
\begin{equation}
    \widetilde{\bm{r}}_\xi(t)= -(y-\pi/2)\sin(\xi z^3)\hat x+(y-\pi/2)\cos(\xi z^3)\hat y+z\hat z,
\end{equation}
where we write $y=y(t)$ and $z=z(t)$ for brevity, and the $\pi/2$ shift in $y$ is included so that the twist is about the line parallel to $\hat z$ that bisects $\bm{r}_0(t)$. The ``twist constant" $\xi$ controls the amount of twisting. Because the tantrix of the twisted curve is not normalized, $\|\dot{\widetilde{\bm{r}}}_\xi(t)\|\ne1$, the curve must be reparameterized after the twist to restore $t$ as the arc-length. We denote by $\bm{r}_\xi(t)$ the curve we obtain after this reparameterization. Details about this step can be found in Appendix~\ref{appendix:DCHG_calculate}. The curvature of $\bm{r}_0(t)$ is equal to two sequential sech functions, each given by $\Omega(t)=\Omega_0\hbox{sech}(\Omega_0t)$ with $\Omega_0t\in[-10,10]$. Larger twist values $\xi$ cause the curvature of $\bm{r}_\xi(t)$ to deviate more strongly from the sech function form it assumes in the untwisted ($\xi=0$) case. The twisted $\bm{r}_\xi(t)$ are  three-dimensional closed curves. Examples are shown in Fig.~\ref{fig:fig6_3D_DCHG_ErrorCurve} for three different values of $\xi$.

By smoothly changing the twist parameter $\xi$, we can accordingly smoothly deform both $\bm{r}_\xi(t)$ and $\bm{h}(t)$ to construct different DoG gates. We use the DoG construction procedure described above to calculate the three control fields that result from a twisted error curve $\bm{r}_\xi(t)$. The detailed algebraic steps are given in Appendix~\ref{appendix:DCHG_calculate}.  
In Fig.~\ref{fig:fig7_3D_DCHG_field}, we plot the  control fields for two twist parameters that give the DoG gates $U(\xi=\pi/20000)= \text{diag}\{e^{-i1.15\pi},e^{i1.15\pi}\}$ and $U(\xi=\pi/2000)= \text{diag}\{e^{-i0.41\pi},e^{i0.41\pi}\}$, respectively. The error curves, tantrix curves, and holonomic trajectories on the Bloch sphere used to obtain these control fields are shown in Figs.~\ref{fig:figS2_error_tantrix_Bloch_xi_xmall} and \ref{fig:figS2_error_tantrix_Bloch_xi_large} of Appendix~\ref{appendix:DCHG_support-figs}. Any geometric phase can be achieved by choosing the twist parameter $\xi$ appropriately, so that this class of curves yields a universal set of single-qubit DoG gates. We note that the geometric phase in the DoG gate is linear in $\xi$ (see Fig.~\ref{fig:figS1_figLienarMap}); therefore one can easily determine what twist value is needed given a target geometric phase.

\begin{figure}
    \centering
    \includegraphics[width=0.5\textwidth]{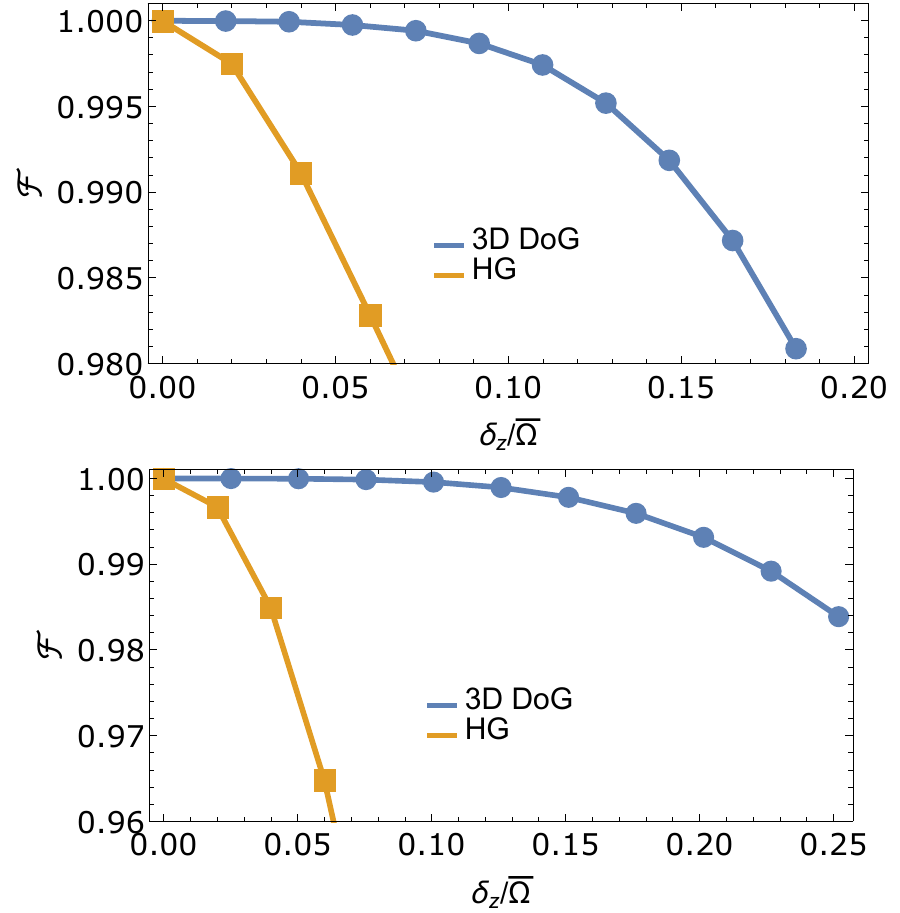}
    \caption{Gate fidelities of two 3D DoG gates constructed from twisted error curves $\bm{r}_\xi(t)$ versus detuning error rate (the ratio of the detuning error $\delta_z$ to the time-averaged driving strength $\bar{\Omega}$). Results for the standard orange-slice model-based holonomic gate (HG) are shown for comparison. The top (bottom) plot corresponds to the target gate $U(\xi=\pi/2000)$ ($U(\xi=\pi/20000)$).}
    \label{fig:fig8_DCHG_fidelity_detun_Error}
\end{figure}

The DoG construction guarantees that the gates are more robust to detuning errors compared to more standard holonomic gate designs. We calculate the 3D DoG gate fidelity $\mathcal{F}=\frac{1}{6}\text{Tr}(U_r^{\dagger}U_r)+\frac{1}{6}|\text{Tr}(U^{\dagger}_iU_r)|^2$ \cite{PEDERSEN200747} ($U_i$ and $U_{r}$ represent ideal and noisy gates, respectively) of $U(\xi=\pi/2000)$ and $U(\xi=\pi/20000)$ in the presence of a small detuning error $\delta_z$. For comparison, we also compute the fidelities of the same gates, but this time designed using the original orange-slice model. As Fig.~\ref{fig:fig8_DCHG_fidelity_detun_Error} shows,  the designed 3D DoG gates provide substantial robustness to detuning errors compared to the standard holonomic gates. In the case of the DoG gates, the fidelities for the two target gates shown in the figure are almost the same due to the fact that the error curves are nearly identical in these two cases. In contrast, the fidelity of the standard orange-slice holonomic gate depends more on the target gate; this is because the openness of the error curve depends on which gate is being performed in this case.

\begin{figure}
    \centering
    \includegraphics[width=0.5\textwidth]{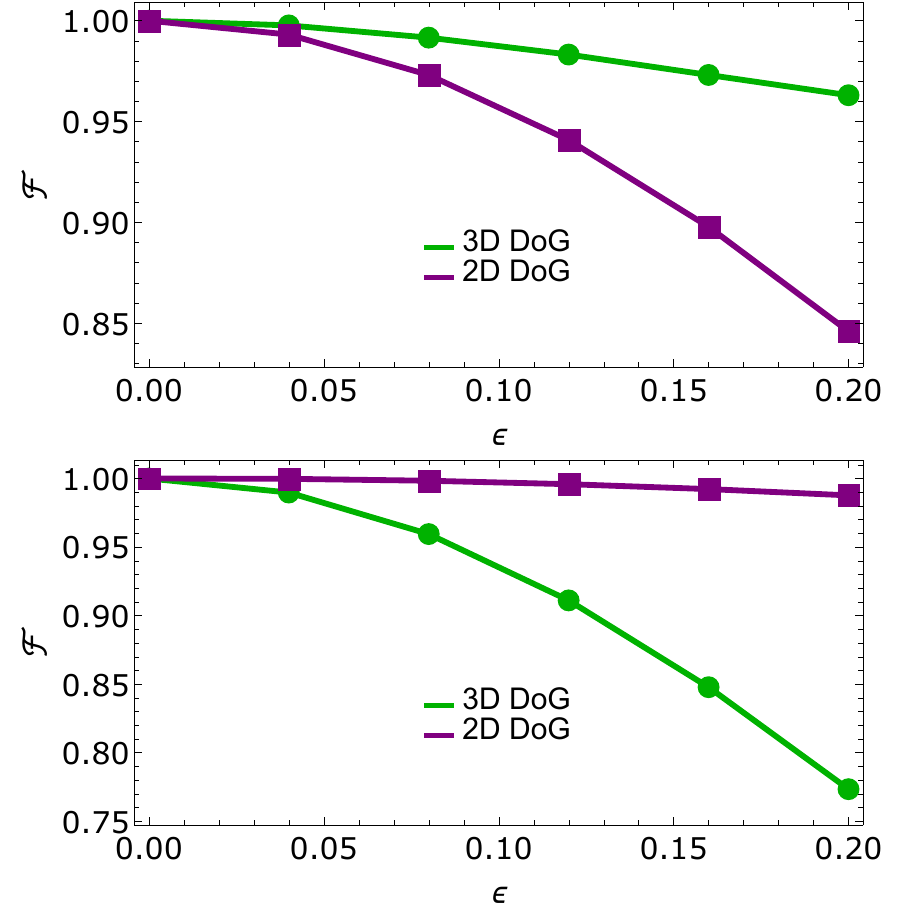}
    \caption{Comparing the performance of 3D and 2D DoG gates. Both panels show gate fidelity versus pulse amplitude error rate ($\epsilon$ in $\Omega'(t) =\Omega(t)(1+\epsilon)$). The 3D DoG gates are obtained from twisted 3D error curves $\bm{r}_\xi(t)$. The top (bottom) panel shows results for the target gate $U(\xi=\pi/2000)$ ($U(\xi=\pi/20000)$). 
    }
    \label{fig:fig9_DCHG_fidelity_Omega_Error}
\end{figure}

The 3D and 2D DoG gates have a similar tolerance to detuning errors since they both correspond to closed error curves. Their tolerance to errors in the pulse amplitude  $\Omega(t)$, however, is different. We compare the gate fidelity of 3D and 2D DoG gates in the presence of pulse amplitude errors in Fig.~\ref{fig:fig9_DCHG_fidelity_Omega_Error}. We see that which DoG gate family works better depends on the target gate, which is determined by $\xi$ or $\Phi_0$. 
For a given target gate, one should then choose the DoG gate family that gives better performance for that particular gate.
Note that the 2D DoG gates inherit the same pulse amplitude error robustness from the  simple orange-slice holonomic gate (see Appendix~\ref{appendix:HG_pulse_error}).  

One advantage that the 3D DoG gate family has over the 2D  DoG gates is that the former does not require pulses that change sign in order to form a closed error curve. This is important for systems in which the sign of the control field is fixed, which for example is the case in singlet-triplet spin qubits~\cite{Wang.NatCom.2012,Kestner.PRL.2013}.

\subsection{\textcolor{black}{Quantum speed limits and experimental implementations}}

\textcolor{black}{
The time it takes to perform a quantum operation is fundamentally bounded from below by a quantum speed limit~\cite{Margolus.PhysicaD.1998,Deffner.PRL.2013,Deffner.JPA.2017}, which is usually determined by the number and structure of the control terms in the Hamiltonian and by the maximum pulse strengths or bandwidths. DoG gates are not an exception. Since the length of the error curve is equal to the gate time, designing the fastest DoG gate relies on designing an error curve with the least total length~\cite{ZengPRA2018} while respecting curvature constraints that correspond to experimental pulse-shaping limitations. In the case of a fully-controllable system such as the three-axis control Hamiltonian given in Eq.~\eqref{eq:3_field_hamiltonian}, a simple way to reduce the gate time is to rescale the curve. 
When an error curve is scaled down by some factor, the curvature and torsion, and hence the control field amplitudes, increase by the same factor, such that the pulse areas, and correspondingly the evolution operator/quantum gate, remain invariant. Thus, the speed limit for each DoG gate can be approached by rescaling the curve until the pulse constraints are saturated. After this rescaling, different curves will deviate from the quantum speed limit by different amounts. In the case of the 2D and 3D DoG gates, the 3D DoG gate is faster for a given pulse amplitude bound. This can be readily explained by the fixed-sign pulses in the 3D model, while the 2D model requires more pulses (eight pulses in Fig.~\ref{fig:fig5_2D_orange_sllice_DCHG}) to realize dynamical correction.  
A natural question arises: For a particular pulse amplitude bound, is there a fastest DoG gate (beyond the two classes of DoG gates presented here)? Because the geometric error curve method is completely general (any pulse that cancels transverse noise corresponds to a closed error curve), this approach is well suited to answer this question. This could potentially be addressed by combining our analytical techniques with numerical optimization~\cite{Stefanatos.EPL.2020,Glaser.TEPJ.2015} (notice that a similar case has been studied in Ref.~\cite{ZengPRA2018}). We leave this to future work.
}

\textcolor{black}{
Since rescaling the error curve amounts to rescaling the driving field amplitudes and the gate time in opposite directions, we choose the maximum pulse amplitude $\Omega_0$ as the basic unit for all dimensionful quantities in this work. The gate fidelities shown in Figs.~\ref{fig:fig8_DCHG_fidelity_detun_Error} and Fig.~\ref{fig:fig9_DCHG_fidelity_Omega_Error} can be interpreted in the context of various experimental platforms, as long as the static noise is small ($\delta_z \ll \Omega_0$). The values of pulse amplitudes and noise levels for several qubit platforms are listed in Table~\ref{tab:exp_data}, based on which  we also listed the evaluated 3D DoG gate time $T=40/\Omega_0=2\pi/\bar{\Omega}$ (considering $\Omega_0 T=40$ in Fig.~\ref{fig:fig7_3D_DCHG_field}). In all cases shown in the table, $T$ is well below the coherence time of the given platform.
We use an NV center spin qubit in diamond to illustrate how it fits the DoG model. Consider an NV center spin qubit that is driven by 40 MHz microwave pulses~\cite{vanderSar_Nature12}. The onsite $^{14}$N nuclear spin constantly dephases the NV center at $\sim2$ MHz through the hyperfine interaction~\cite{vanderSar_Nature12}. This corresponds to $\delta_z/\bar{\Omega}\sim0.05$, in which case the DoG gate achieves $>0.999$ fidelity (see Fig.\ref{fig:fig8_DCHG_fidelity_detun_Error}). The 3D DoG gate time $T=240$ ns is much shorter than the qubit coherence time~\cite{vanderSar_Nature12}. We can also consider the DoG gate performance for an optically driven NV spin. In Ref.~\cite{Zhou2016}, an optical dephasing time of $T_\phi=18$ ns was measured, corresponding to an rms noise value of $\delta_z/2\pi\sim\sqrt{2}/T_{\phi}=12.5$ MHz. Gate times are limited by pulse rise times, which were found to be lower-bounded by 1.2 ns in Ref.~\cite{BrianZhou.PRL.2017}, implying a maximum average pulse strength of $\bar\Omega/2\pi=0.5/(2.4~\mathrm{ns})\approx208$ MHz for the standard orange-slice model. On the other hand, imposing the same rise time constraint on the sech pulses shown in Fig.~\ref{fig:fig7_3D_DCHG_field} for the 3D DoG gates yields $\bar\Omega/2\pi\approx106$ MHz. Thus, for an optically controlled NV spin, we estimate that $\delta_z/\bar\Omega\approx0.06$ is the lowest possible effective noise strength for the standard orangle-slice holonomic gate, while $\delta_z/\bar\Omega\approx0.12$ for the 3D DoG gate example shown in Fig.~\ref{fig:fig7_3D_DCHG_field}. From Fig.~\ref{fig:fig8_DCHG_fidelity_detun_Error}, we see that the fidelity of the 3D DoG gate is still substantially higher than that of the standard holonomic gate despite the increase in gate time and effective noise strength (both are larger by a factor of $\sim 2$ compared to the standard orange-slice model). This is a direct consequence of the built-in suppression of transverse noise enjoyed by the DoG gate. 
\begin{table}
    \centering
    \caption{Experimental parameters, and the 3D DoG gate time $T=2\pi/\bar{\Omega}$ (see text), in superconducting qubits (SC), trapped ions (Ions), NV centers (NV), and silicon quantum dot (QD).}
    \begin{tabular}{|c| c|c | c|}
    \hline
    Platform & $\bar{\Omega}/2\pi$ (MHz)  & $\delta_z/2\pi$ (MHz)   & $T$ ($\mu$s) \\
    \hline
       SC  & $10\sim100$~\cite{Burnett.npjQI.2019}  & $0.001\sim0.003$~\cite{Krantz.APR.2019} & $0.01\sim0.1$ \\
        Ions  & $<1$~\cite{Bruzewicz.APR.2019}  & $<0.1$~\cite{Milne.PRApp.2020} & $>1$ \\
        NV   & 10 $\sim$ 40~\cite{Rong.NatCom.2015,vanderSar_Nature12}  & $\sim2$~\cite{vanderSar_Nature12} & $0.025\sim0.1$\\
        QD   & $\sim0.3$~\cite{CHYang.NatEle.2019}  & $\sim0.02$~\cite{CHYang.NatEle.2019}& $3.3$\\
        \hline
    \end{tabular}
    \label{tab:exp_data}
\end{table}
}

\subsection{\textcolor{black}{Two-qubit DoG gates}}

\textcolor{black}{Thus far, we have focused on applying the DoG formalism to single-qubit gates. However, both holonomic gates and the concept of error curves can also be applied to multi-level or multi-qubit quantum systems. In the higher-dimensional case, noise cancellation still requires the error curves to close; however, for systems with more than two levels, the curves generally live in higher dimensions~\cite{Buterakos.PRXQ.2021}. Therefore, we can generalize the DoG formalism to any finite-dimensional system by starting from closed error curves embedded in the appropriate number of dimensions and imposing the parallel transport condition.}

\textcolor{black}{
Here, we exemplify these ideas by designing two-qubit DoG gates. The concrete example we focus on is a system in which the qubits interact through a tunable coupling described by the following Hamiltonian:
\begin{equation}
\begin{aligned}
        H_{\text{couple}}(t) =&\frac{\lambda_{\perp}(t)}{2} \sigma^{(1)}_z\Big[-\sin \Phi(t) \sigma^{(2)}_x  +\cos \Phi(t) \sigma^{(2)}_y  \Big]\\
        +&  \frac{\lambda_{z}(t)}{2}\sigma^{(1)}_z\sigma^{(2)}_z ,    
\end{aligned}
\end{equation}
where $\lambda_{z}(t)$ and $\lambda_{\perp}(t)$ represent tunable interaction amplitudes along two directions, $\Phi(t)$ is the tunable phase field of the transverse part, and $\sigma^{(j)}_{\mu}$ is the Pauli $\mu$ matrix acting on the $j$th qubit.  We assume further that the frequency of qubit 1 is fixed while that of qubit 2 is tunable. Specifically, we let qubit 2 be driven by the following three-field Hamiltonian:
\begin{equation}
\label{eq:H_ctrl_2qb}
    H_{\text{drive}}(t)=\frac{\Omega(t)}{2}\left[\cos\Phi(t)\sigma_x^{(2)}+\sin\Phi(t)\sigma_y^{(2)}\right] +\frac{\Delta(t)}{2}\sigma_z^{(2)},
\end{equation}
where the $\Phi(t)$ here is the same one appearing in $H_{\text{couple}}(t)$. Qubit 1 is left idle. The total Hamiltonian is then block diagonal as follows:
\begin{equation}
\begin{aligned}
        &H_{\text{c}}(t)= H_{\text{couple}}(t)+ H_{\text{drive}}(t)= \frac{1}{2} \times \\
         & \resizebox{\linewidth}{!}{ 
 $\begin{pmatrix}
 \Delta (t)+\lambda_{z}(t) &\tilde{ \Omega}(t)e^{-i[ \Phi (t)+\tilde{\Phi}(t)]}  & 0 & 0 \\
\tilde{ \Omega} (t)e^{i[ \Phi (t)+\tilde{\Phi}(t)]} & -\Delta (t)-\lambda_{z}(t)& 0 & 0 \\
 0 & 0 & \Delta (t)-\lambda_{z}(t)& \tilde{ \Omega}(t)e^{-i[ \Phi (t)-\tilde{\Phi}(t)]} \\
 0 & 0 &\tilde{ \Omega}(t)e^{i[\Phi (t)-\tilde{\Phi}(t)]} & \lambda_{z}(t)-\Delta (t) \\
\end{pmatrix}$
},
\end{aligned}
\end{equation}
where for brevity we have introduced the following quantities: 
\begin{equation}
\begin{aligned}
\tilde{\Omega}(t)& = \sqrt{\Omega^2(t)+\lambda^2_{\perp}(t)}, \\
\tilde{\Phi}(t) &=\text{arg} [\Omega(t)+i \lambda_{\perp}(t)].
\end{aligned}
\end{equation}
We take the error Hamiltonian to be $H_{\text{error}}=\delta_z \sigma^{(2)}_z$, which represents quasistatic fluctuations of the energy splitting of qubit 2~\cite{Krantz.APR.2019,Burnett.npjQI.2019}.
Naively, the error curve for the total two-qubit Hamiltonian, $H_{\text{c}}(t)+H_{\text{error}}$, lives in six dimensions~\cite{Buterakos.PRXQ.2021}. However, this Hamiltonian is block diagonal, and each $2\times2$ block can also be interpreted as an effective two-level system, implying that the six-dimensional curve can be factorized into two three-dimensional curves.  Notice that these two curves are not independent because the two blocks both involve the same $\tilde{\Omega}(t)$ and $\Delta(t)$. This means that their error curves have the same curvature yet different torsions:
\begin{equation}
\label{eq:dual_curve_curvatures}
\begin{aligned}
            \kappa_1(t)&  = \tilde{\Omega}(t),  \\
            \tau_1(t) & = \dot{\Phi}(t)+\dot{\tilde{\Phi}}(t)-\Delta(t)-\lambda_{z}(t),\\
            \kappa_2(t)&  = \tilde{\Omega}(t),  \\
            \tau_2(t) & = \dot{\Phi}(t)-\dot{\tilde{\Phi}}(t)-\Delta(t)+\lambda_{z}(t).\\
\end{aligned}
\end{equation}
Moreover, these two curves must have identical  total length since both lengths are set by the evolution time.
These geometric constraints can be satisfied by using a pair of geometric curves that we used to construct single-qubit 3D DoG gates, $\bm{r}_1(t)=\bm{r}_{\xi}(t)$ and $\bm{r}_2(t)=\bm{r}_{-\xi}(t)$ (see Fig.~\ref{fig:fig10_two_qubit_DoG}), with opposite twist parameters. Due to the opposite twist constant $\xi$, $\bm{r}_1(t)$ and $\bm{r}_2(t)$ have opposite torsions. Such a \textit{torsion symmetry} has the extra benefit of reducing the number of independent control fields needed to implement the gate, as will become clear shortly. We can use this pair of curves to design two-qubit DoG gates. For each curve, we identify the unique holonomic evolution and compute the corresponding diagonal control fields, yielding 
\begin{equation}
\begin{aligned}
            2\Delta(t)&= (\dot{x}_1\ddot{y}_1-\dot{y}_1\ddot{x}_1)/\dot{z}_1+(\dot{x}_2\ddot{y}_2-\dot{y}_2\ddot{x}_2)/\dot{z}_2,  \\
            2\lambda_z(t) &=(\dot{x}_1\ddot{y}_1-\dot{y}_1\ddot{x}_1)/\dot{z}_1- (\dot{x}_2\ddot{y}_2-\dot{y}_2\ddot{x}_2)/\dot{z}_2 .
\end{aligned}
\end{equation}
The phase fields are then obtained as
\begin{equation}
    \begin{aligned}
                \Phi(t)=&\frac{1}{2}\int^t_0[\tau_1(t')+\tau_2(t')]dt'+\int^t_0\Delta(t')dt', \\ \tilde{\Phi}(t)=&\frac{1}{2}\int^t_0[\tau_1(t')-\tau_2(t')]dt'+\int^t_0\lambda_z(t')dt'.
    \end{aligned}
\end{equation}
Due to the torsion symmetry of such dual curves, two of the fields are identically zero:
\begin{equation}
    \Delta(t)=0, \quad \Phi(t)= 0.
\end{equation}
Consequently, the remaining nontrivial control fields $\lambda_z(t)$, $\lambda_{\perp}(t)$, and $\Omega(t)$ are sufficient to realize DoG control in this system. In Fig.~\ref{fig:fig10_two_qubit_DoG}, we plot the control fields synthesized from $\xi=\pi/2000$ dual curves.
The control fields give the two-qubit DoG gate
\begin{equation}
    U_c(T) = \begin{pmatrix}
        e^{i\beta_{g,\xi}(T)} &  0 & 0 & 0 \\
         0 &  e^{-i\beta_{g,\xi}(T)} & 0 & 0 \\ 
          0 &  0&  e^{i\beta_{g,-\xi}(T)}& 0 \\  
          0 & 0 & 0 & e^{-i\beta_{g,-\xi}(T)} \\
    \end{pmatrix},
\end{equation}
where $\beta_{g,\xi}(T)$ and $\beta_{g,-\xi}(T)$ are the geometric phases of the two subsystems. Another consequence of the torsion  symmetry is that these geometric phases are related by a minus sign: $\beta_{g,\xi}(T)=-\beta_{g,-\xi}(T)$. Different values of $\xi$ give different two-qubit gates (see Fig.~\ref{fig:figS1_figLienarMap}). $\xi=0.08\pi/1000$ corresponds to  $\beta_{g,\xi}(T)=-1.25\pi$, and the  DoG gate is $\text{diag}\{e^{3 i \pi/ 4},e^{-3 i \pi/ 4},e^{-3 i \pi/ 4},e^{3 i \pi/ 4}\}$, which is equivalent to the controlled-Z gate~\cite{CZ.equiv}. Notice that $\beta_{g,\xi}(T)=-1.25\pi$ up to an integer multiplies of $2\pi$ corresponds to different control fields yet same CZ equivalent gate. 
\begin{figure}
    \centering
    \includegraphics[width=0.5\textwidth]{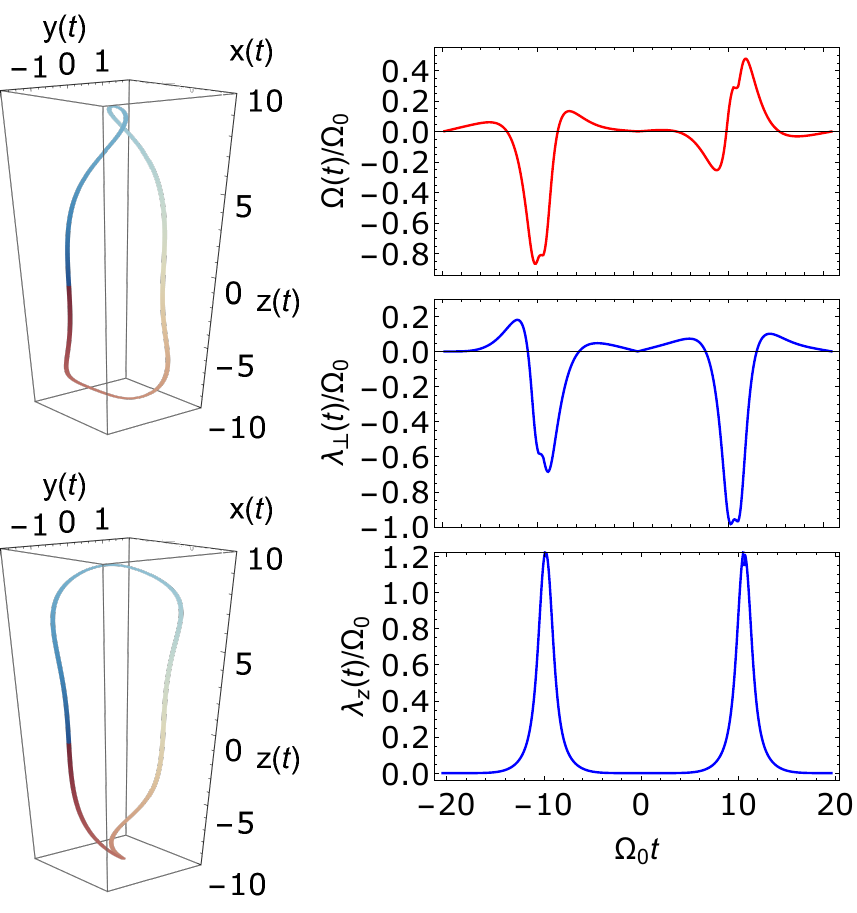}
    \caption{Two-qubit DoG gate by tuning a tunable  coupler and by driving qubit 2. (Left) A pair of dual curves $r_1(t)=\bm{r}_{\xi}(t)$ and $r_2(t)=\bm{r}_{-\xi}(t)$  with $\xi=\pi/2000$ that have same curve length, curvature and opposite torsion. (Right) Control fields that yield the two-qubit DoG gate: $\lambda_{z}(t)$ and $\lambda_{\perp}(t)$ in inter-qubit coupling [blue] and $\Omega(t)$ in qubit 2 driving [red]. }
    \label{fig:fig10_two_qubit_DoG}
\end{figure}
}

\textcolor{black}{It is interesting to point out that other designs of dual curves without the torsion symmetry, say those used in \cite{Buterakos.PRXQ.2021}, can require more independent control fields to implement~\cite{donovan.curves.comment}. In Appendix~\ref{appendix:two_qubit_molmer_sorrensen} we also show an alternative approach to constructing two-qubit DoG gates that does not employ dual curves.}
~

\section{Conclusions}\label{sec:conslusion}
In conclusion, we showed how to combine two types of geometry to design quantum gate operations that are simultaneously robust to multiple types of errors. In particular, we presented a systematic procedure for designing qubit evolutions that exhibit both closed holonomy loops and closed error curves---three-dimensional curves that quantify the effect of transverse noise. In addition, we provided a general recipe for determining the control fields that generate such evolutions, which we refer to as Doubly Geometric (DoG) gates. These gates are robust against both control errors and transverse noise errors, thus extending the error-correcting capabilities of holonomic gate designs. We demonstrated our formalism by constructing \textcolor{black}{several types of single-qubit and two-qubit} DoG gates. Our approach provides a new perspective on achieving robust quantum control in the presence of multiple sources of error. \textcolor{black}{In future work, it would be interesting to combine our DoG formalism, which provides a global view of the space of noise-cancelling pulses, with numerical approaches to quantum control~\cite{Glaser.TEPJ.2015}, and with more insights of various types of noises~\cite{Frey.PRApp.2020,chalermpusitarak2020framebased}. Combining these methods could yield globally time-optimal pulses that cancel noise while respecting experimental constraints.}

\begin{acknowledgments}
W.D. thanks  Vlad Shkolnikov and Bikun Li  for useful discussions. This work was supported by the Army Research Office (W911NF-17-0287). E.B. also acknowledges support by the U.S. Office of Naval Research (N00014-17-1-2971). S.E.E. acknowledges support by NSF (DMR-1737921).
\end{acknowledgments}

\appendix

\setcounter{figure}{0} 
\makeatletter 
\renewcommand{\thefigure}{S\@arabic\c@figure}
\makeatother


\section{Definition of dynamical phase}
\label{appendix:def_DP}
Even though the dynamical phase is often expressed as the integral of $\braket{\psi(t)|H(t)|\psi(t)}$, one must be careful in interpreting this as the integral of the expectation value of the Hamiltonian. Generally, the expectation value of the Hamiltonian is invariant under frame transformations: $\braket{\psi(t)|H(t)|\psi(t)}=\braket{\psi(t)|R^{\dagger}(t)R(t)H(t)R^{\dagger}(t)R(t)|\psi(t)}$ for any unitary transformation $R(t)$. The dynamical phase can only be written in terms of an expectation value in the Schr\"odinger picture, which is the only frame considered in Aharonov and Anandan's work~\cite{PhysRevLett.58.1593}. Here, we show this by starting from the definition of the geometric phase in the projected space. We show that under a basis transformation, the integrand in the dynamical phase should transform according to $\braket{\psi(t)|H(t)|\psi(t)}$ $\rightarrow$ $ \braket{\psi(t)|R(t)R^{\dagger}(t)H(t)R(t)R^{\dagger}(t)|\psi(t)}$ $ -i\bra{\psi(t)}R(t)R^{\dagger}(t)\dot{R}(t) R^{\dagger}(t)\ket{\psi(t)}$. No matter  which frame one works in, the dynamical phase should be defined in terms of the Hamiltonian that generates the cylic evolution, which means that in a rotating frame, one should use the effective Hamiltonian $H_{\text{eff}}(t)=R^{\dagger}(t)H(t)R(t)-iR^{\dagger}(t)\dot{R}(t)$. This in turn means that the dynamical phase is not invariant under frame transformations.

We begin by considering the Hamiltonian in the Schr\"odinger picture: 
\begin{equation}
      H(t)=   \frac{\Omega(t)}{2} 
\left(\begin{array}{cc}
 0 & e^{-i \Phi(t)} \\
 e^{i \Phi(t)} & 0 \\
\end{array} \right) + \frac{\Delta(t)}{2}
\left(\begin{array}{cc}
 1 & 0 \\
0 & -1 \\
\end{array} \right).
\end{equation}
Suppose that this Hamiltonian generates the cyclic evolution $\ket{\psi_S(t)}=e^{i \alpha_S (t)} \left(
\begin{array}{c}
 \cos \frac{\theta_S (t)}{2} \\
 e^{i \phi_S (t)} \sin \frac{\theta_S (t)}{2} \\
\end{array}
\right) = e^{i\alpha_S(t)}\ket{\tilde{\psi}_S(t)}$, where $\ket{\tilde{\psi}_S(t)}$ is the projected state that defines the geometric phase $i\bra{\tilde{\psi}_S(t)}\partial_t\ket{\tilde{\psi}_S(t)}$ and which is cyclic (starts/ends at the north pole of the Bloch sphere). We can subtract the geometric phase  from the total phase to obtain (the time derivative of) the dynamical phase:
\begin{equation}
        \dot{\alpha}_S(t)-i\bra{\tilde{\psi}_S(t)}\partial_t\ket{\tilde{\psi}_S(t)}
        = \dot{\alpha}_S(t) +\sin^2\frac{\theta_S(t)}{2}\dot{\phi}_S(t).
\end{equation}
It is readily shown that this is equivalent to $\braket{\psi(t)|H(t)|\psi(t)}$, the expectation value in the Schr\"odinger picture.

Now we define the basis transformation $R(t)=e^{-i\int^t_0\frac{\Delta(t')}{2}\sigma_z dt'}$. It is immediately clear that (i) the transformed evolution is also cyclic, making the geometric phase well-defined in the new frame, and (ii) the state in the rotating frame (labelled by $I$) is:
$\ket{\psi_I(t)}=e^{i \alpha_I (t)} \left(
\begin{array}{c}
 \cos \frac{\theta_I (t)}{2} \\
 e^{i \phi_I (t)} \sin \frac{\theta_I (t)}{2} \\
\end{array}
\right) = e^{i\alpha_I(t)}\ket{\tilde{\psi}_I(t)}$. One can check that the parameters in the two frames are related through:
\begin{equation}
    \theta_I(t)=\theta_S(t),  \dot{\alpha}_I(t)= \dot{\alpha}_S(t)+\frac{1}{2}\Delta(t),  \dot{\phi}
    _I(t) =\dot{\phi}_S(t)-\Delta(t).  
\end{equation}
Accordingly, we can calculate the dynamical phase in the rotating frame:
\begin{equation}
    \begin{aligned}
        &\dot{\alpha}_I(t)-i\bra{\tilde{\psi}_I(t)}\partial_t\ket{\tilde{\psi}_I(t)} \\
        = &\dot{\alpha}_I(t) +\sin^2\frac{\theta_I(t)}{2}\dot{\phi}_I(t) \\
        = & \dot{\alpha}_S(t) +\sin^2\frac{\theta_S(t)}{2}\dot{\phi}_S(t) +\frac{\Delta(t)}{2} (1-2\sin^2\frac{\theta_S(t)}{2}).
    \end{aligned}
\end{equation}
Clearly, the dynamical phase is not invariant under the frame change. Moreover, one can check that it is equal to the integral of $\bra{\psi_I(t)}H_{\text{eff}}(t)\ket{\psi_I(t)}$.

\section{Robustness of geometric phases}
\label{appendix:robustness}
In this appendix, we examine the robustness of the geometric phase to perturbations in the Hamiltonian. 

Following the simple model from~\cite{Sjoqvist.IJQC.2015}, we consider a two-level system with Hamiltonian $H=E_n\ket{n}\bra{n}+E_m\ket{m}\bra{m}$, $E_n>E_m$ and state $\ket{\psi(t)}=a e^{-i E_n t}\ket{n}+b e^{-iE_m t }\ket{m}$ with $|a|^2+|b|^2=1$.  At time $T=\frac{2\pi}{E_n-E_m}$, the evolution is cyclic:  $\ket{\psi(T)}=a e^{-iE_n T}\ket{n}+b e^{-iE_m T}\ket{m}=e^{-i\frac{ 2\pi E_n }{E_n-E_m}} (a\ket{n}+b\ket{m})$. The  total phase is $\beta_{\text{total}}=-2\pi\frac{E_n}{E_n-E_m}$, and the dynamical phase is $\beta_d=-\int^T_0 \bra{\psi(t)}H(t)\ket{\psi(t)} =-\frac{2\pi}{E_n-E_m}(|a|^2E_n+|b|^2E_m) $. The geometric phase is $\beta_g=\beta_{\text{total}}-\beta_d=-2\pi |b|^2$. Clearly, the geometric phase depends on the initial state. 
Now consider a perturbation in the system energy such that both energies are shifted to $E'_n \neq E_n$, $E'_m \neq E_m$, where $\delta E'=  E'_n-E'_m\neq\delta E=E_n-E_m$. We can define a dimensionless quantity $\epsilon = (\delta E- \delta E')/\delta E $ that characterizes the strength of the perturbation. 
We will now show that if $\epsilon \ll 1 $, i.e., the perturbation is small compared to the original energy splitting, then the error in the geometric phase vanishes to first order in $\epsilon$, while this is not the case for the total phase and the dynamical phase. 

The state at time $T$ with the perturbation included is $\ket{\psi'(T)}=a e^{-iE'_n T}\ket{n}+b e^{-iE'_m T}\ket{m}=e^{-i\frac{ 2\pi E'_n }{E_n-E_m}} (a\ket{n}+be^{i \frac{2\pi(E'_n-E'_m)}{E_n-E_m}}\ket{m})$. The total and dynamical phases become $\beta'_{\text{total}}=\text{Arg}\big[e^{-i2\pi \frac{E'_n}{E_n-E_m}} (|a|^2+|b|^2e^{i2\pi(1-\epsilon)})\big]$ and  $\beta'_d=-\frac{2\pi}{E_n-E_m} (|a|^2E'_n+|b|^2E'_m)$. 
The geometric phase is then
\begin{equation}
	\begin{aligned}
		\beta'_g=& \beta'_{\text{total}}- \beta'_d \\
		=& \frac{2\pi}{E_n-E_m}\bigg(-E'_n+|a|^2E'_n+|b|^2E'_m \bigg)\\
		 &+  \arctan \bigg(\frac{|b|^2\text{sin}((1-\epsilon)2\pi)}{|b|^2 \text{cos}((1-\epsilon)2\pi)+|a|^2 } \bigg).
	\end{aligned}
\end{equation}
Expanding this to first order in $\epsilon$, we find
\begin{equation}
	\begin{aligned}
		\beta'_g&=-2\pi|b|^2(1-\epsilon)
		+ \textcolor{black}{{\arctan} \bigg(\frac{|b|^2\text{sin}((1-\epsilon)2\pi)}{|b|^2 \text{cos}((1-\epsilon)2\pi)+|a|^2 }} \bigg) \\
		& = -2\pi|b|^2(1-\epsilon) - 2\pi|b|^2 \epsilon+\mathcal{O}(\epsilon^3) \\
		& = -2\pi|b|^2 +\mathcal{O}(\epsilon^3).
	\end{aligned}
\end{equation}
We see that the two $\mathcal{O}(\epsilon)$ terms cancel out, and so the geometric phase is first-order robust against such energy perturbations. 

The above analysis is also applicable to the case of Rabi driving. Parameterizing the state as $\ket{\psi(0)}=\cos\theta\ket{e}+\sin\theta e^{i\phi}\ket{g}$ and considering the case of a real Rabi frequency $\Omega(t)$, the evolution operator is $U(t)=\exp(-i\int^t_0\Omega(t')dt'\sigma_x)$. The cycle time $T$ is determined from the condition $\int^T_0\Omega(t)dt=\pi$. The geometric phase in this case is given by $\beta_g=\pi(1-\sin2\theta\cos\phi)$. If we assume the noise is parallel to the evolution path, $\int^T_0\Omega(t)dt=\pi+\epsilon_x$, where $\epsilon_x\ll 1$, then following a similar algebraic procedure reveals that $\beta'_g=\beta_g+\mathcal{O}(\epsilon_x^3)$. However, if the noise is perpendicular to the travel path, say $\Omega(t)\sigma_x \rightarrow \Omega(t)\big((1-\epsilon_x)\sigma_x+\epsilon_z\sigma_z\big)$, then one finds that $\beta'_g$ contains terms of order $\mathcal{O}(\epsilon_z)$.

\section{Driving fields that generate holonomic gates}
\label{appendix:holonomy_derive}
Assume the  state $\ket{\psi(t)}=e^{i\alpha(t)}\left(\begin{array}{c}
    \cos \frac{\theta(t)}{2}  \\
    \sin\frac{\theta(t)}{2}e^{i\phi(t)} 
\end{array}\right)$ evolves under the Hamiltonian $H_c(t)=\frac{\Omega(t)}{2}
\left(\begin{array}{cc}
 0 & e^{-i \Phi(t)} \\
 e^{i \Phi(t)} & 0 \\
\end{array} \right) + \frac{\Delta(t)}{2}
\left(\begin{array}{cc}
 1 & 0 \\
0 & -1 \\
\end{array} \right)$ in the lab frame. The three-field control Hamiltonian includes the Rabi amplitude $\Omega(t)$, the phase field $\Phi(t)$ and the detuning field $\Delta(t)$, which must all be chosen in such a way as to make the evolution holonomic. From the  expression for state $\ket{\psi(t)}$, we can write the  evolution operator as
\begin{equation}
    U(t)=\left(
\begin{array}{cc}
 e^{i \alpha (t)} \cos \frac{\theta (t)}{2} & -e^{-i (\alpha (t)+\phi (t))} \sin \frac{\theta (t)}{2} \\
 e^{i (\alpha (t)+\phi (t))} \sin \frac{\theta (t)}{2} & e^{-i \alpha (t)} \cos \frac{\theta (t)}{2}\\
\end{array}
\right),
\end{equation}
in which case the Schr\"odinger equation yields the following pair of equations:
\begin{align}
    -\Delta(t) +\dot{\phi}(t)-\cos\theta(t)\big( 2\dot{\alpha}(t)+\dot{\phi}(t)\big) &=0,\nonumber\\
    -e^{i\Phi(t)}\Omega(t)+e^{i\phi(t)}\big[i\dot{\theta}(t)-\sin\theta(t)(2\dot{\alpha}(t)+\dot{\phi}(t))\big]&=0.
\end{align}
Combining these with the parallel transport condition, $\dot{\alpha}(t)=-\frac{1}{2}(1-\cos\theta)\dot{\phi}(t)$, we obtain
\begin{align}
    &\Delta(t)=\sin^2\theta(t)\dot{\phi}(t),\nonumber\\
    &\Omega(t)\sin[\Phi(t)-\phi(t)] = \dot{\theta}(t), \\
    &\Omega(t)\cos{[\Phi(t)-\phi(t)]}=-\cos\theta(t)\sin\theta(t)\dot{\phi}(t).\nonumber
\end{align}
These three equations can be re-arranged to yield Eq.~\eqref{eq:field_path_map_TDM} from the main text.

\section{Holonomic gates in the presence of pulse errors}
\label{appendix:HG_pulse_error}
Despite the robust nature of the geometric phase, there is generally no absolute advantage of holonomic gates over non-holonomic ones. Even in the case of noise that is parallel to the evolution path, it is still possible that holonomic gates exhibit worse gate fidelity compared to non-holonomic ones. 

Here we compare the fidelity of the orange-slice holonomic gate with that of a straightforward non-holonomic gate in the presence of pulse errors. We focus on three types of errors: (i) parallel noise, (ii) perpendicular noise, and (iii) $\Omega$ noise. In the first two types of noise, the ideal orange-slice evolution trajectory connecting the two poles is preserved but ``appended" by a perturbation parallel (or perpendicular) to the ideal geodesic line. The $\Omega$ noise will break the ideal evolution path since the second geodesic line will be spoiled if there exists a pulse area error along the first geodesic line. It is worth pointing out that due to different trajectories for the holonomic and non-holonomic gates, what constitutes ``parallel” noise for one gate is not necessarily ``parallel” for another gate.

We begin by writing explicitly the Hamiltonians for the holonomic and non-holonomic gates, respectively:
\begin{equation}
\begin{aligned}
     H_{HG}(t)&= \begin{cases}
     \frac{\Omega}{2} \sigma_x \quad  \quad  \quad  \quad  \quad  \quad  \quad  \quad \quad  \quad 0\leq t\leq\frac{\pi}{\Omega} \\
     \frac{\Omega}{2} (\cos \Phi\sigma_x+\sin \Phi\sigma_y) \quad  \quad \frac{\pi}{\Omega}<t\leq\frac{2\pi}{\Omega}
    \end{cases}, \\
    H_{NHG}(t)&= \frac{\Omega}{2}\sigma_z \quad \quad 0\leq t\leq \frac{2\Phi}{\Omega},
\end{aligned}
\end{equation}
where $\Omega$ and $\Phi$ are constants, i.e., both gates are implemented using square pulses.

We first examine the case of ``parallel" noise, where the perturbed Hamiltonian can be written as:
\begin{equation}
    \begin{aligned}
        H^{\epsilon,\parallel}_{HG}(t)&=
        \begin{cases}
         \frac{\Omega}{2} \sigma_x \quad  \quad  \quad  \quad  \quad  \quad  \quad \quad  \quad  \quad \quad  \quad 0\leq t\leq\frac{\pi}{\Omega} \\
     \frac{\Omega(1+\epsilon)}{2} (\cos \Phi\sigma_x+\sin \Phi\sigma_y) \quad  \quad \frac{\pi}{\Omega}<t\leq\frac{2\pi}{\Omega}
        \end{cases}\\
    H^{\epsilon,\parallel}_{NHG}(t)&=\frac{\Omega(1+\epsilon)}{2}\sigma_z \quad \quad 0\leq t\leq \frac{2\Phi}{\Omega},
    \end{aligned}
\end{equation}
where $\epsilon$ represents a small ($\epsilon \ll1$) deviation in the pulse area caused by noise.
Note that these Hamiltonians are meant to serve as effective toy models in which the perturbation has the intended consequence on the evolution, e.g., in $H^{\epsilon,\parallel}_{HG}(t)$ the noise does not necessarily happen only on the second geodesic line, but could arise along the first line instead (but in $\Omega$ and $\Phi$) such that the ideal trajectory is preserved until the state returns to north pole. The gate fidelities \cite{PEDERSEN200747} for the holonomic and non-holonomic gates are
\begin{equation}
\begin{aligned}
    \mathcal{F}^{\parallel}_{HG} = \frac{1}{3}(2+\cos\pi\epsilon),  \\
    \mathcal{F}^{\parallel}_{NHG} = \frac{1}{3}(2+\cos2\Phi\epsilon), \\
\end{aligned}
\end{equation}
 and so we find that for parallel noise, the non-holonomic gate fidelity depends on the gate angle $\Phi$, while the fidelity of the holonomic gate does not. The relative performance thus depends on $\Phi$.

Next we consider ``perpendicular" noise, where the perturbed Hamiltonians are now
\begin{equation}
    \begin{aligned}
        H^{\epsilon,\perp}_{HG}(t)&=
        \begin{cases}
         \frac{\Omega}{2} \sigma_x \quad  \quad  \quad  \quad  \quad  \quad  \quad \quad  \quad  \quad \quad  \quad 0\leq t\leq\frac{\pi}{\Omega} \\
     \frac{\Omega}{2} (\cos \Phi\sigma_x+\sin \Phi\sigma_y) \quad  \quad\quad \quad  \frac{\pi}{\Omega}<t\leq\frac{2\pi}{\Omega} \\
     \frac{\epsilon\Omega}{\pi}\sigma_z \quad \quad  \quad \quad \quad \quad \quad \quad \quad  \frac{2\pi}{\Omega}< t \leq \frac{2\pi}{\Omega}+\frac{\pi}{\Omega}
        \end{cases}\\
    H^{\epsilon,\perp}_{NHG}(t)&=\begin{cases}
    \frac{\Omega}{2}\sigma_z \quad \quad \quad \quad \quad \quad\quad \quad  \quad \quad \quad0\leq t\leq \frac{2\Phi}{\Omega} \\
    \frac{\epsilon\Omega}{\pi} \sigma_x \quad \quad\quad \quad\quad \quad\quad \quad \frac{2\Phi}{\Omega}<t\leq \frac{2\Phi}{\Omega}+\frac{\pi}{\Omega}.
    \end{cases}
    \end{aligned}
\end{equation}
Again this toy model simply appends an extra perturbation that is transverse to the ideal evolution. The corresponding fidelities are
\begin{equation}
\begin{aligned}
    \mathcal{F}^{\perp}_{HG} = \frac{1}{3}(2+\cos2\epsilon) , \\
    \mathcal{F}^{\perp}_{NHG} = \frac{1}{3}(2+\cos2\epsilon), \\
\end{aligned}
\end{equation}
and they obviously have equal performance.

Finally in the case of $\Omega$ noise, the corresponding Hamiltonians are 
\begin{equation}
\begin{aligned}
     H^{\Omega}_{HG}(t)&= \begin{cases}
     \frac{\Omega(1+\epsilon)}{2} \sigma_x \quad  \quad \quad \quad  \quad  \quad  \quad  \quad \quad  \quad 0\leq t\leq\frac{\pi}{\Omega} \\
     \frac{\Omega(1+\epsilon)}{2} (\cos \Phi\sigma_x+\sin \Phi\sigma_y) \quad   \quad \frac{\pi}{\Omega}<t\leq\frac{2\pi}{\Omega}
    \end{cases} \\
    H^{\Omega}_{NHG}(t)&= \frac{\Omega(1+\epsilon)}{2}\sigma_z \quad  \quad  \quad \quad \quad \quad \quad \quad \quad 0\leq t\leq \frac{2\Phi}{\Omega}.
\end{aligned}
\end{equation}
The fidelity is readily calculated:
\begin{equation}
\begin{aligned}
    \mathcal{F}^{\Omega}_{HG} &\approx \frac{1}{3} \left(-\pi ^2 \epsilon ^2 \cos \Phi -\pi ^2 \epsilon ^2+3\right) , \\
    \mathcal{F}^{\Omega}_{NHG}& \approx 1-\frac{2}{3}\Phi^2\epsilon^2 ,
\end{aligned}
\end{equation}
where we have kept terms up to $\mathcal{O}(\epsilon^2)$.
We see that the holonomic gate outperforms the non-holonomic one when $\Phi > 0.595\pi$.

It is interesting to point out that for non-abelian quantum holonomy, where an ancillary level is required \cite{Sjoqvist,Xu.PRL.2012}, non-unitary errors occur when the cyclicity is broken \cite{ZhengShibiao.PRA.2016,Ribeiro.PRX.2017}, which could reduce the gate fidelity below the non-holonomic one.

\section{Geometric formalism for dynamically corrected gates}
\label{appendix:geometric_formalism}
The first-order term in the Magnus expansion of the interaction-picture evolution operator $U_I(t)$ and its derivatives are
\begin{align}
    A_1(t)=&\bm{r}(t)\cdot \bm{\sigma}=\int^t_0H_I(t')dt'= \nonumber \\
    =& x(t)\sigma_x+y(t)\sigma_y+z(t)\sigma(z), \\
    \dot{A}_1(t)=& \dot{\bm{r}}(t)\cdot\bm{\sigma}=H_I(t),  \\
    \ddot{A}_1(t)=&\ddot{\bm{r}}(t)\cdot\bm{\sigma}=iU^{\dagger}_c(t)[H_c(t),\sigma_z]U_c(t), \\
    \dddot{A}_1(t)=&\dddot{\bm{r}}(t)\cdot\bm{\sigma}=-U^{\dagger}_c(t)H_c(t) [H_c(t),\sigma_z]U_c(t)\\
    &\quad \quad \quad +iU^{\dagger}_c(t)[\dot{H}_c(t),\sigma_z]U_c(t)\\
    &\quad \quad \quad +U^{\dagger}_c(t)[H_c(t),\sigma_z]H_c(t)U_c(t).
\end{align}
The explicit form of the error curve and its derivative (the tantrix) are given in Eqs.~\ref{eq:error_curve} and \ref{eq:tantrix} of the main text. To obtain the curvature and torsion from the error curve, we use  the dimension-scaled Frobenius norm of matrices defined as  $\|A \|_F=\sqrt{\sum^n_{i,j}|A_{ij}|^2}/\sqrt{n}$.
Using that  $\|AU \|_F=\|A \|_F$ for arbitrary unitary $U$, we have
\begin{equation}
\begin{aligned}
   & \|\ddot{A}_1(t) \|_F = \| [H_0(t),\sigma_z] \|_F \\
    &=\Omega(t) \Big\|\left(
    \begin{array}{cc}
    0     & e^{-i\Phi(t)} \\
    e^{i\Phi(t)} &  0
    \end{array}
    \right)\Big\|_F=\Omega(t).\end{aligned}
\end{equation}
So, $\Omega(t)$ is the curvature $\kappa(t)=\|\ddot{\bm{r}}(t)\|$ of error curve $\bm{r}(t)$.

Next, using the result
 \begin{widetext}
\begin{equation}
\begin{aligned}
        \dot{A}_1(t)&\ddot{A}_1(t)\dddot{A_1}(t)\\
        =&iU^{\dagger}_c(t)\left(
    \begin{array}{cc}
        -\Omega(t)\big[\Delta(t)\Omega(t)-\Omega(t)\dot{\Phi}(t)+i\dot{\Omega}(t)\big] & -e^{-i\Phi(t)}\Omega^3(t) \\
       e^{-i\Phi(t)}\Omega^3(t)  & -\Omega(t)\big[\Delta(t)\Omega(t)-\Omega(t)\dot{\Phi}(t)-i\dot{\Omega}(t)\big]
    \end{array}
    \right)U_c(t),\\
\|[\dot{A}_1(t)&,\ddot{A}_1(t)]\|_F  = \|\sigma_z[H_c(t),\sigma_z] -[H_c(t),\sigma_z]\sigma_z\|_F=2\Omega(t),  
\end{aligned}
\end{equation}
\end{widetext}
we obtain 
\begin{equation}
    -2i \frac{\text{Tr}\big\{\dot{A}_1(t)\ddot{A}_1(t)\dddot{A_1}(t)\big\}}{\big\|[\dot{A}_1(t),\ddot{A}_1(t)]\big\|^2_F} = \dot{\Phi}(t)-\Delta(t).
\end{equation}

Geometrically, it is also equivalent to
\begin{equation}
\begin{aligned}
    -2i& \frac{\text{Tr}\big\{\dot{A}_1(t)\ddot{A}_1(t)\dddot{A_1}(t)\big\}}{\big\|[\dot{A}_1(t),\ddot{A}_1(t)]\big\|^2_F} \\
    = & -i 2i \frac{\text{Tr}\big\{ [(\dot{\bm{r}}\times \ddot{\bm{r}})\cdot \bm{\sigma}] \cdot (\dddot{\bm{r}}\cdot \bm{\sigma}) \big\}}{\big\|2i(\dot{\bm{r}}\times \ddot{\bm{r}})\cdot\bm{\sigma} \big\|^2_F} \\
    = &\frac{(\dot{\bm{r}}\times \ddot{\bm{r}})\cdot\dddot{\bm{r}} }{\big\| \dot{\bm{r}}\times \ddot{\bm{r}}  \big\|^2}
    =\tau(t),
\end{aligned}
\end{equation}
which is the torsion of the error curve.

\section{DoG gates from an extended orange-slice model}
\label{appendix:orange_slice-DCHG}
\begin{figure}
    \centering
    \includegraphics[width=0.5\textwidth]{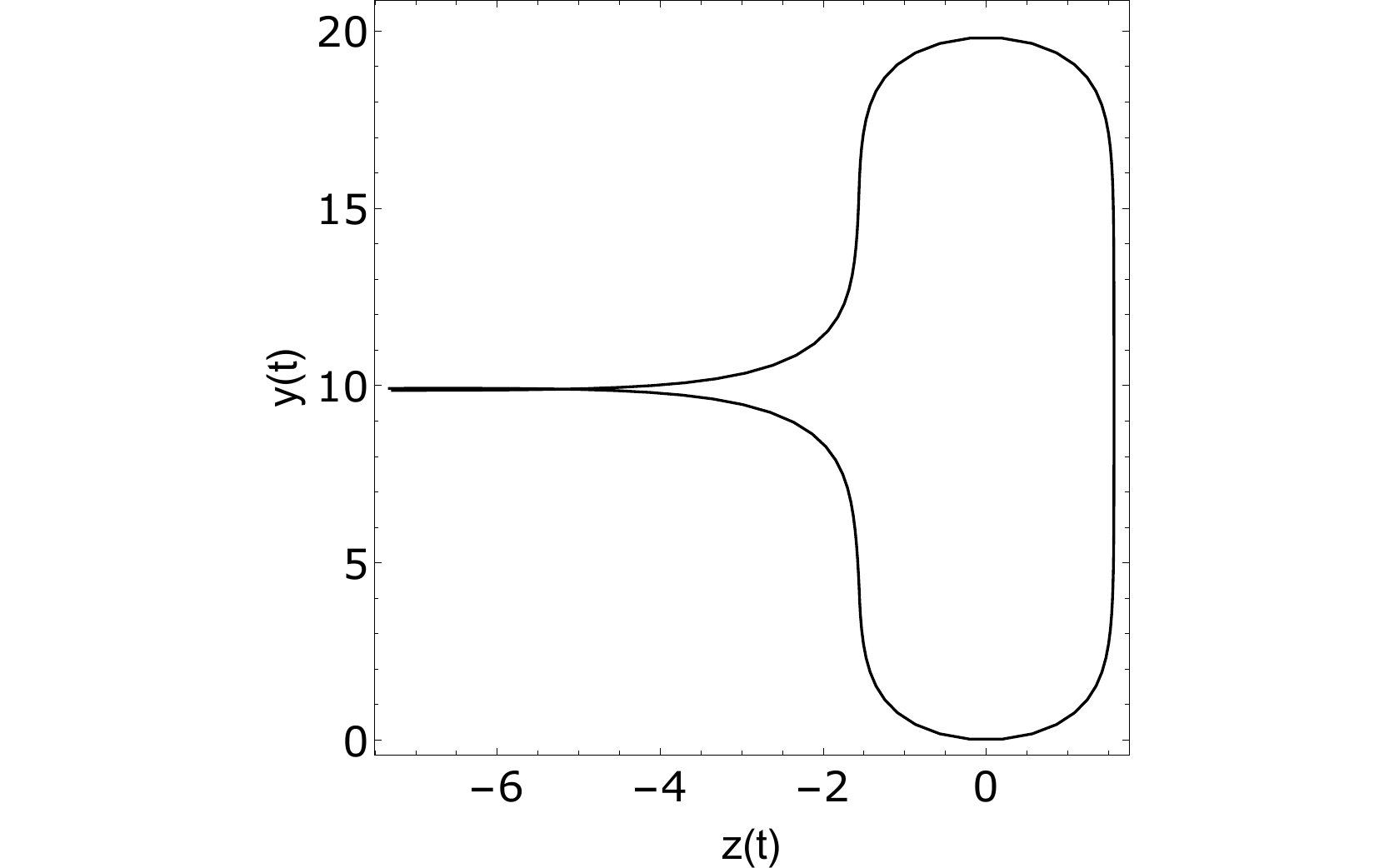}
    \caption{The error curve generated by the control field $\Omega_1(t)$ described in Eq.~\eqref{eq:quadSech}.  This corresponds to one of the planar lobes of the 3D error curves plotted in Fig.~\ref{fig:fig5_2D_orange_sllice_DCHG}.}
    \label{fig:figS3_DCHG_planar_Lobe}
\end{figure}

In this appendix, we provide additional details about the DoG gate shown in Fig.~\ref{fig:fig5_2D_orange_sllice_DCHG}, which is based on a modified orange-slice model. In this example, the driving field is constructed from 8 hyperbolic secant pulses, as shown in the figure. This 8-pulse sequence is obtained by applying the following 4-pulse sequence twice in succession:
\begin{equation}
    \label{eq:quadSech}
   \Omega_1(t)= \begin{cases}
          \begin{array}{cc}
\begin{array}{cc}
 -\frac{\Omega_0}{2} \text{sech}(u+20.6) & -25.6\leq u\leq -15.6 \\
 \Omega_0\text{sech}(u+10.6) & -15.6\leq u\leq 0 \\
 \Omega_0\text{sech}(10.6\, -u) & 0\leq u\leq 15.6 \\
 -\frac{\Omega_0}{2} \text{sech}(20.6\, -u) & 15.6\leq u\leq 25.6 \\
\end{array}
 \\
\end{array}
    \end{cases}
\end{equation}
Here, $u=\Omega_0t$. The first 4-pulse sequence drives the system along a geodesic from the north pole to the south pole of the Bloch sphere, while the second sequence takes the system along a second geodesic back to the north pole. Which geodesic is traversed during the return trip is controlled by the value of $\Phi(t)=\Phi_0$. The error curve generated by the 4-pulse sequence (one planar lobe of the whole 3D closed error curve in \textcolor{black}{Fig.~\ref{fig:fig5_2D_orange_sllice_DCHG}}) is shown in Fig.~\ref{fig:figS3_DCHG_planar_Lobe}. 

\section{DoG gates from twisted three-dimensional error curves}
\label{appendix:DCHG_calculate}
Here we discuss details of the DoG gate construction based on twisted three-dimensional error curves $\bm{r}_\xi(t)$. An important aspect of this procedure is that after one constructs a closed error curve $\widetilde{\bm{r}}_\xi$ by twisting $\bm{r}_0$, one needs to restore the arc-length parameterization to obtain the control fields that implement the DoG gate.

Suppose we construct $\bm{r}_{\xi\neq0}(t)$  by twisting $\bm{r}_0(t)$. It is important to note that the arc-length parameterization $t$ of $\bm{r}_0(t)$ is not the same as that of the twisted curve $\bm{r}_{\xi\neq0}$. This is evident considering the arc-length on $\bm{r}_0(t)$ is not invariant after the non-rigid body twist. If $\widetilde{\bm{r}}_\xi(w)$ is parameterized by $w$, then the arc-length parameterization is obtained from $t(w')=\int^{w'}_{w'_{\text{min}}}\sqrt{\big(\frac{dx(w)}{dw}\big)^2+\big(\frac{dy(w)}{dw}\big)^2+\big(\frac{dz(w)}{w}\big)^2}dw$, where $w'_{\text{min}}\leq w'\leq w'_{\text{max}}$, and $x,y,z$ are the components of $\widetilde{\bm{r}}_\xi(w)$. After performing the integration, we can invert the result to obtain $\bm{r}_\xi(t)=\widetilde{\bm{r}}_\xi(w(t))$. Hereafter, we will use the $w$ parameterization ($w=t$ for $\bm{r}_0(t)$) for calculations; $w$ is the arc-length only for the untwisted curve. 

In terms of the $w$ parameterization, we can write the tantrix as
\begin{equation}
    \begin{aligned}
        \dot{\bm{r}}(w)& = d{\bm{r}}(w)/dt =\mathcal{N}\big( d\bm{r}(w)/dw \big) =  \mathcal{N}\big( \bm{r}'(w) \big)\\
        &= [-\sin\eta(w) \cos\zeta(w),   -\sin\eta(w)\sin\zeta(w),\cos\eta(w)],
    \end{aligned}
\end{equation}
where $\mathcal{N}$ is the normalization operator.  
We can determine $\eta(w)$ and $\zeta(w)$ from the length of $\bm{r}'(w)$.

We can derive the Bloch sphere parameters using
\begin{equation}
    \begin{aligned}
        \dot{\theta}(t)&=\frac{d\theta}{dw}\Big/\frac{dt}{dw}=\frac{d\eta(w)}{dw}\Big/\frac{dt}{dw}, \\
        \dot{\phi}(t)&=\frac{d\phi}{dw}\Big/\frac{dt}{dw}=\frac{\zeta'(w)}{\cos\eta(w)}\Big/\frac{dt}{dw},
    \end{aligned}
\end{equation}
where $f(t)=f(t(w'))$ for simplicity. The Bloch sphere coordinates are then
\begin{equation}
    \begin{aligned}
        \theta(t)& = \eta [w(t)], \\
        \phi(t) &= \int^{w(t)}_0 \frac{\zeta'(w')}{\cos\eta(w')}dw',
    \end{aligned}
\end{equation} 
where $w(t)$ is the $w$ value that maps from $t$.

The three control fields are:
\begin{equation}
    \begin{aligned}
        \Omega(t) &=\sqrt{\Big(\frac{d\theta}{dw}\Big/\frac{dt}{dw}\Big)^2+\sin^2\theta(w)\cos^2\theta(w)\Big(\frac{d\phi}{dw}\Big/\frac{dt}{dw}\Big)^2 } \\
        &= \sqrt{{\eta}'(w)^2+\sin^2\eta(w){\zeta}'(w)^2}\Big/  \frac{dt}{dw}  ,  \\
        \Phi(t)
        &=  \arg \bigg(-\big[\eta'(w) \sin \textcolor{black}{\phi(w)}+{\zeta}'(w) \sin \eta(w) \cos \textcolor{black}{\phi(w)} \big]\Big/ \frac{dt}{dw} \\
             &\quad +i\big[\eta'(w) \cos \textcolor{black}{\phi(w)}-{\zeta}'(w) \sin \eta(w) \sin \textcolor{black}{\phi(w)} \big]\Big/ \frac{dt}{dw} \bigg) , \\
        \Delta(t) 
        &= \sin^2\eta(w) \frac{{\zeta}'(w)}{\cos\eta(w)}\Big/ \frac{dt}{dw},
    \end{aligned}
\end{equation}
where   $w=w(t)$ ($t=t(w)$) and $\frac{dt}{dw}=\sqrt{\big(\frac{dx(w)}{dw}\big)^2+\big(\frac{dy(w)}{dw}\big)^2+\big(\frac{dz(w)}{w}\big)^2}$. Note that to calculate $\Phi(t)$ we need $\phi(w)$, which is an integral involving $\eta$ and $\zeta$; this integral cannot be obtained analytically in general. The same is true of the integral the yields $t(w)$. These integrals must be computed numerically in general.

\section{3D DoG gate}
\label{appendix:DCHG_support-figs}
\begin{figure}
    \centering
    \includegraphics[width=0.5\textwidth]{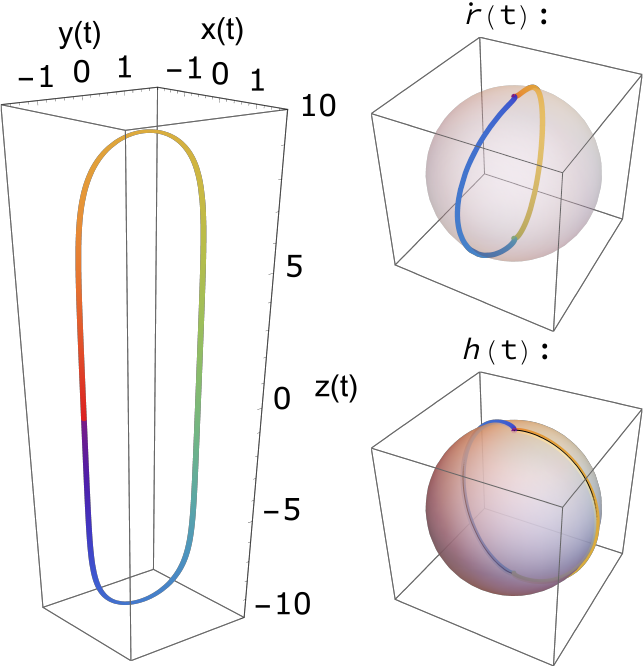}
    \caption{Error curve $\bm{r}_\xi(t)$ with $\xi=\pi/20000$, and the corresponding tantrix $\dot{\bm{r}}_\xi(t)$ and the evolution trajectory on the Bloch sphere $\bm{h}(t)$.}
    \label{fig:figS2_error_tantrix_Bloch_xi_xmall}
\end{figure}

\begin{figure}
    \centering
    \includegraphics[width=0.5\textwidth]{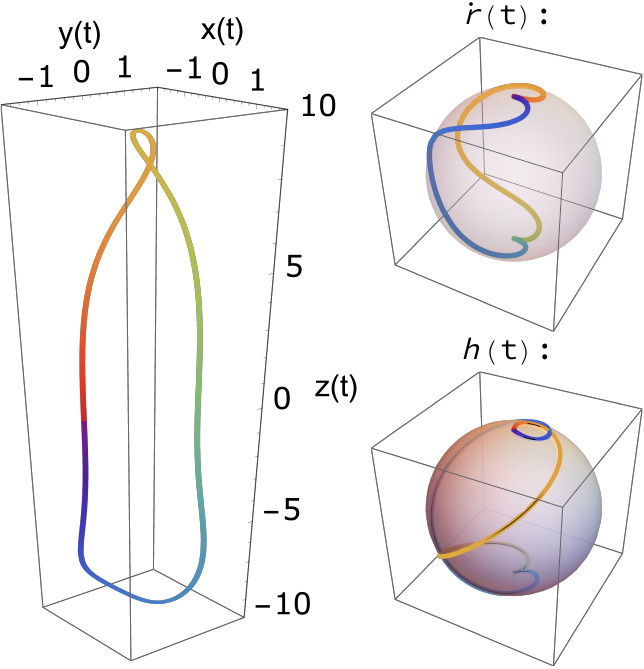}
    \caption{Error curve $\bm{r}_\xi(t)$ with $\xi=\pi/2000)$, and the corresponding tantrix $\dot{\bm{r}}_\xi(t)$ and the evolution trajectory on the Bloch sphere $\bm{h}(t)$.}
    \label{fig:figS2_error_tantrix_Bloch_xi_large}
\end{figure}

In the main text, we show the error curves (Fig.~\ref{fig:fig6_3D_DCHG_ErrorCurve}) and control fields (Fig.~\ref{fig:fig7_3D_DCHG_field}) that generate 3D DoG gates with different error curve twist values $\xi=\pi/2000$ and $\xi=\pi/20000$. In Fig.~\ref{fig:figS2_error_tantrix_Bloch_xi_xmall} and Fig.~\ref{fig:figS2_error_tantrix_Bloch_xi_large} we plot the tantrix $\dot{\bm{r}}(t)$ and holonomic Bloch sphere trajectory $\bm{h}(t)$ corresponding to these two $\xi$ values. 

\begin{figure}
    \centering
    \includegraphics[width=0.5\textwidth]{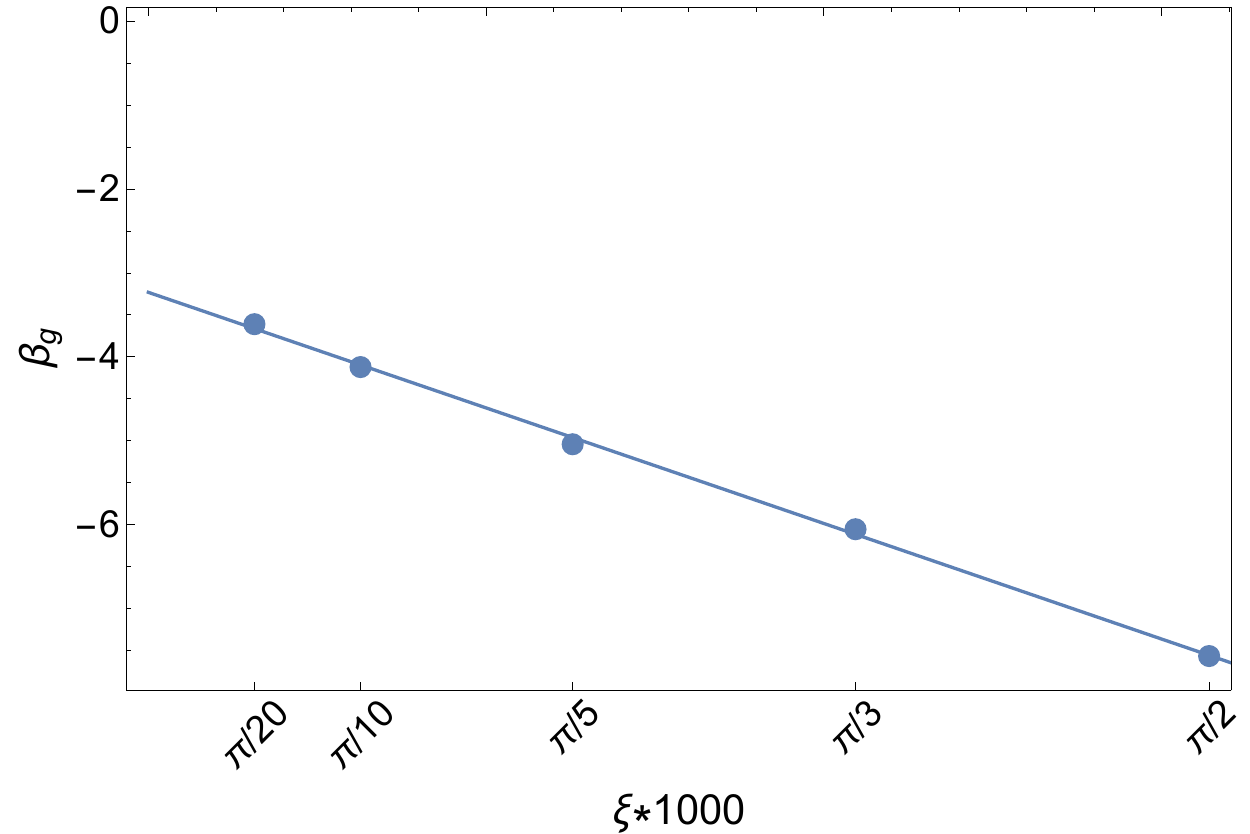}
    \caption{The geometric phase $\beta_g$ of a 3D DoG gate versus the scaled twist parameter $1000\xi$ of the error curve.   }
    \label{fig:figS1_figLienarMap}
\end{figure}

In Fig.~\ref{fig:figS1_figLienarMap} we show the relationship between the twist parameter $\xi$ and the geometric phase encoded in the DoG gate. It is apparent from the figure that different $\xi$ values yield different geometric phases.

\section{Two-qubit DoG gates using a single 3D curve}
\label{appendix:two_qubit_molmer_sorrensen}
\textcolor{black}{
Here we discuss how the single-qubit DoG formalism straightforwardly can be extended to construct two-qubit DoG gates. First, following the procedure in \cite{LiKZ.PRR.2020}, we focus on a diagonal entangling gate by enforcing the projected bases to be  
\begin{equation}
\begin{aligned}
     \ket{\nu_1(t)}&=\ket{00}, \\
     \ket{\nu_2(t)}&= \cos \frac{\theta(t)}{2} \ket{01}+ \sin\frac{\theta(t)}{2}e^{i\phi(t)} \ket{10}, \\
     \ket{\nu_3(t)}&=\sin\frac{\theta(t)}{2}e^{-i\phi(t)} \ket{01}-\cos \frac{\theta(t)}{2} \ket{10}, \\
     \ket{\nu_4(t)}&= \ket{11}.
\end{aligned}
\end{equation}
Note that $\ket{\nu_2(t)}$ and $\ket{\nu_3(t)}$ span a two-dimensional subsystem that is described by the Bloch sphere geometry. Cyclic evolution dictates $\ket{\nu_2(T)}=\ket{\nu_2(0)}$ and $\ket{\nu_3(T)}=\ket{\nu_3(0)}$.
The two-qubit holonomic gate is
\begin{equation}
    U_c(T) = \begin{pmatrix}
        1 &  0 & 0 & 0 \\
         0 &  e^{-i \beta_g(T)} & 0 & 0 \\ 
          0 &  0&  e^{i \beta_g(T)}& 0 \\  
          0 & 0 & 0 & 1 \\
    \end{pmatrix},
\end{equation}
which is a trivial generalization of Eq.~\eqref{eq:holonomic_gate_two_base}. 
The Hamiltonian that gives such a holonomic gate is
\begin{equation}
\begin{aligned}
       H_c(t) =& -\frac{1}{2}\big[\dot{\theta}\sin\phi + \frac{\dot{\phi}}{2}\sin 2\theta\cos \phi\big] R^x \\
        & -\frac{1}{2}\big[\dot{\theta}\cos\phi - \frac{\dot{\phi}}{2}\sin 2\theta\sin \phi\big] R^y +\frac{1}{2} \dot{\phi} \sin ^2 \theta R^z,
\end{aligned}
\end{equation}
where we make the time-dependence of the Bloch parameters implicit ($\theta:=\theta(t)$, $\phi:=\phi(t)$), and $R^x=(\sigma^{(1)}_x\sigma^{(2)}_x+\sigma^{(1)}_y\sigma^{(2)}_y)/2$, $R^y=(\sigma^{(1)}_x\sigma^{(2)}_y-\sigma^{(1)}_y\sigma^{(2)}_x)/2$, $R^z=(\sigma^{(1)}_z-\sigma^{(2)}_z)/2$~\cite{LiKZ.PRR.2020}. 
The driving fields can be achieved in the S{\o}rensen-M{\o}lmer setting~\cite{Sorrensen.Molmer.PRL.1999,Sorrensen.Molmer.PRA.2000}. In the Lamb-Dicke limit, the Hamiltonian translates into 
\begin{equation}
    \tilde{H}_c(t)= \Omega_{\text{eff}}(t) \ket{01}\bra{10}+\text{h.c.},
\end{equation} 
where $\Omega_{\text{eff}}(t)\sim \Omega^*_1(t)/\Omega_2(t)$~\cite{LiKZ.PRR.2020} is an effective Hamiltonian which can be real or complex. 
Obviously, such a Hamiltonian can  implement the 2D DoG gate in the two relevant subspaces, yielding a two-qubit DoG gate in the full space. 
The static noise here is taken to be $\delta H_z = \delta_z(\ket{01}\bra{01}-\ket{10}\bra{10})$, which corresponds physically to vibration-inducing energy shifts between the intermediate states $\ket{01;n}$ and $\ket{10;n}$ (where $n$ is the quantum number for the relevant vibrational mode of the trap in the S{\o}rensen-M{\o}lmer setting ~\cite{Sorrensen.Molmer.PRL.1999}). 
}

\clearpage

\newpage


\begin{thebibliography}{112}%
\makeatletter
\providecommand \@ifxundefined [1]{%
 \@ifx{#1\undefined}
}%
\providecommand \@ifnum [1]{%
 \ifnum #1\expandafter \@firstoftwo
 \else \expandafter \@secondoftwo
 \fi
}%
\providecommand \@ifx [1]{%
 \ifx #1\expandafter \@firstoftwo
 \else \expandafter \@secondoftwo
 \fi
}%
\providecommand \natexlab [1]{#1}%
\providecommand \enquote  [1]{``#1''}%
\providecommand \bibnamefont  [1]{#1}%
\providecommand \bibfnamefont [1]{#1}%
\providecommand \citenamefont [1]{#1}%
\providecommand \href@noop [0]{\@secondoftwo}%
\providecommand \href [0]{\begingroup \@sanitize@url \@href}%
\providecommand \@href[1]{\@@startlink{#1}\@@href}%
\providecommand \@@href[1]{\endgroup#1\@@endlink}%
\providecommand \@sanitize@url [0]{\catcode `\\12\catcode `\$12\catcode
  `\&12\catcode `\#12\catcode `\^12\catcode `\_12\catcode `\%12\relax}%
\providecommand \@@startlink[1]{}%
\providecommand \@@endlink[0]{}%
\providecommand \url  [0]{\begingroup\@sanitize@url \@url }%
\providecommand \@url [1]{\endgroup\@href {#1}{\urlprefix }}%
\providecommand \urlprefix  [0]{URL }%
\providecommand \Eprint [0]{\href }%
\providecommand \doibase [0]{http://dx.doi.org/}%
\providecommand \selectlanguage [0]{\@gobble}%
\providecommand \bibinfo  [0]{\@secondoftwo}%
\providecommand \bibfield  [0]{\@secondoftwo}%
\providecommand \translation [1]{[#1]}%
\providecommand \BibitemOpen [0]{}%
\providecommand \bibitemStop [0]{}%
\providecommand \bibitemNoStop [0]{.\EOS\space}%
\providecommand \EOS [0]{\spacefactor3000\relax}%
\providecommand \BibitemShut  [1]{\csname bibitem#1\endcsname}%
\let\auto@bib@innerbib\@empty
\bibitem [{\citenamefont {Glaser}\ \emph {et~al.}(2015)\citenamefont {Glaser},
  \citenamefont {Boscain}, \citenamefont {Calarco}, \citenamefont {Koch},
  \citenamefont {K{\"o}ckenberger}, \citenamefont {Kosloff}, \citenamefont
  {Kuprov}, \citenamefont {Luy}, \citenamefont {Schirmer}, \citenamefont
  {Schulte-Herbr{\"u}ggen} \emph {et~al.}}]{Glaser.TEPJ.2015}%
  \BibitemOpen
  \bibfield  {author} {\bibinfo {author} {\bibfnamefont {Steffen~J}\
  \bibnamefont {Glaser}}, \bibinfo {author} {\bibfnamefont {Ugo}\ \bibnamefont
  {Boscain}}, \bibinfo {author} {\bibfnamefont {Tommaso}\ \bibnamefont
  {Calarco}}, \bibinfo {author} {\bibfnamefont {Christiane~P}\ \bibnamefont
  {Koch}}, \bibinfo {author} {\bibfnamefont {Walter}\ \bibnamefont
  {K{\"o}ckenberger}}, \bibinfo {author} {\bibfnamefont {Ronnie}\ \bibnamefont
  {Kosloff}}, \bibinfo {author} {\bibfnamefont {Ilya}\ \bibnamefont {Kuprov}},
  \bibinfo {author} {\bibfnamefont {Burkhard}\ \bibnamefont {Luy}}, \bibinfo
  {author} {\bibfnamefont {Sophie}\ \bibnamefont {Schirmer}}, \bibinfo {author}
  {\bibfnamefont {Thomas}\ \bibnamefont {Schulte-Herbr{\"u}ggen}},  \emph
  {et~al.},\ }\bibfield  {title} {\enquote {\bibinfo {title} {Training
  schr{\"o}dinger’s cat: quantum optimal control},}\ }\href
  {https://link.springer.com/article/10.1140/epjd/e2015-60464-1} {\bibfield
  {journal} {\bibinfo  {journal} {The European Physical Journal D}\ }\textbf
  {\bibinfo {volume} {69}},\ \bibinfo {pages} {1--24} (\bibinfo {year}
  {2015})}\BibitemShut {NoStop}%
\bibitem [{\citenamefont {Stefanatos}\ and\ \citenamefont
  {Paspalakis}(2020{\natexlab{a}})}]{Stefanatos.EPL.2020}%
  \BibitemOpen
  \bibfield  {author} {\bibinfo {author} {\bibfnamefont {D.}~\bibnamefont
  {Stefanatos}}\ and\ \bibinfo {author} {\bibfnamefont {E.}~\bibnamefont
  {Paspalakis}},\ }\bibfield  {title} {\enquote {\bibinfo {title} {A shortcut
  tour of quantum control methods for modern quantum technologies},}\ }\href
  {\doibase 10.1209/0295-5075/132/60001} {\bibfield  {journal} {\bibinfo
  {journal} {{EPL} (Europhysics Letters)}\ }\textbf {\bibinfo {volume} {132}},\
  \bibinfo {pages} {60001} (\bibinfo {year} {2020}{\natexlab{a}})}\BibitemShut
  {NoStop}%
\bibitem [{\citenamefont {Vandersypen}\ and\ \citenamefont
  {Chuang}(2005)}]{Vandersypen.RMP.2005}%
  \BibitemOpen
  \bibfield  {author} {\bibinfo {author} {\bibfnamefont {L.~M.~K.}\
  \bibnamefont {Vandersypen}}\ and\ \bibinfo {author} {\bibfnamefont {I.~L.}\
  \bibnamefont {Chuang}},\ }\bibfield  {title} {\enquote {\bibinfo {title} {Nmr
  techniques for quantum control and computation},}\ }\href {\doibase
  10.1103/RevModPhys.76.1037} {\bibfield  {journal} {\bibinfo  {journal} {Rev.
  Mod. Phys.}\ }\textbf {\bibinfo {volume} {76}},\ \bibinfo {pages}
  {1037--1069} (\bibinfo {year} {2005})}\BibitemShut {NoStop}%
\bibitem [{\citenamefont {Johansson}\ \emph {et~al.}(2012)\citenamefont
  {Johansson}, \citenamefont {Nation},\ and\ \citenamefont
  {Nori}}]{johansson2012qutip}%
  \BibitemOpen
  \bibfield  {author} {\bibinfo {author} {\bibfnamefont {J~Robert}\
  \bibnamefont {Johansson}}, \bibinfo {author} {\bibfnamefont {Paul~D}\
  \bibnamefont {Nation}}, \ and\ \bibinfo {author} {\bibfnamefont {Franco}\
  \bibnamefont {Nori}},\ }\bibfield  {title} {\enquote {\bibinfo {title}
  {Qutip: An open-source python framework for the dynamics of open quantum
  systems},}\ }\href
  {https://www.sciencedirect.com/science/article/abs/pii/S0010465512000835}
  {\bibfield  {journal} {\bibinfo  {journal} {Computer Physics Communications}\
  }\textbf {\bibinfo {volume} {183}},\ \bibinfo {pages} {1760--1772} (\bibinfo
  {year} {2012})}\BibitemShut {NoStop}%
\bibitem [{\citenamefont {Li}\ and\ \citenamefont
  {Khaneja}(2006)}]{Jr-Shin.Li.PRA.2006}%
  \BibitemOpen
  \bibfield  {author} {\bibinfo {author} {\bibfnamefont {Jr-Shin}\ \bibnamefont
  {Li}}\ and\ \bibinfo {author} {\bibfnamefont {Navin}\ \bibnamefont
  {Khaneja}},\ }\bibfield  {title} {\enquote {\bibinfo {title} {Control of
  inhomogeneous quantum ensembles},}\ }\href {\doibase
  10.1103/PhysRevA.73.030302} {\bibfield  {journal} {\bibinfo  {journal} {Phys.
  Rev. A}\ }\textbf {\bibinfo {volume} {73}},\ \bibinfo {pages} {030302}
  (\bibinfo {year} {2006})}\BibitemShut {NoStop}%
\bibitem [{\citenamefont {Ruschhaupt}\ \emph {et~al.}(2012)\citenamefont
  {Ruschhaupt}, \citenamefont {Chen}, \citenamefont {Alonso},\ and\
  \citenamefont {Muga}}]{Ruschhaupt.NJP.2012}%
  \BibitemOpen
  \bibfield  {author} {\bibinfo {author} {\bibfnamefont {A}~\bibnamefont
  {Ruschhaupt}}, \bibinfo {author} {\bibfnamefont {Xi}~\bibnamefont {Chen}},
  \bibinfo {author} {\bibfnamefont {D}~\bibnamefont {Alonso}}, \ and\ \bibinfo
  {author} {\bibfnamefont {J~G}\ \bibnamefont {Muga}},\ }\bibfield  {title}
  {\enquote {\bibinfo {title} {Optimally robust shortcuts to population
  inversion in two-level quantum systems},}\ }\href {\doibase
  10.1088/1367-2630/14/9/093040} {\bibfield  {journal} {\bibinfo  {journal}
  {New Journal of Physics}\ }\textbf {\bibinfo {volume} {14}},\ \bibinfo
  {pages} {093040} (\bibinfo {year} {2012})}\BibitemShut {NoStop}%
\bibitem [{\citenamefont {Daems}\ \emph {et~al.}(2013)\citenamefont {Daems},
  \citenamefont {Ruschhaupt}, \citenamefont {Sugny},\ and\ \citenamefont
  {Gu\'erin}}]{Daems.PRL.2013}%
  \BibitemOpen
  \bibfield  {author} {\bibinfo {author} {\bibfnamefont {D.}~\bibnamefont
  {Daems}}, \bibinfo {author} {\bibfnamefont {A.}~\bibnamefont {Ruschhaupt}},
  \bibinfo {author} {\bibfnamefont {D.}~\bibnamefont {Sugny}}, \ and\ \bibinfo
  {author} {\bibfnamefont {S.}~\bibnamefont {Gu\'erin}},\ }\bibfield  {title}
  {\enquote {\bibinfo {title} {Robust quantum control by a single-shot shaped
  pulse},}\ }\href {\doibase 10.1103/PhysRevLett.111.050404} {\bibfield
  {journal} {\bibinfo  {journal} {Phys. Rev. Lett.}\ }\textbf {\bibinfo
  {volume} {111}},\ \bibinfo {pages} {050404} (\bibinfo {year}
  {2013})}\BibitemShut {NoStop}%
\bibitem [{\citenamefont {Dridi}\ \emph
  {et~al.}(2020{\natexlab{a}})\citenamefont {Dridi}, \citenamefont {Liu},\ and\
  \citenamefont {Gu\'erin}}]{Dridi.PRL.2020}%
  \BibitemOpen
  \bibfield  {author} {\bibinfo {author} {\bibfnamefont {Ghassen}\ \bibnamefont
  {Dridi}}, \bibinfo {author} {\bibfnamefont {Kaipeng}\ \bibnamefont {Liu}}, \
  and\ \bibinfo {author} {\bibfnamefont {St\'ephane}\ \bibnamefont
  {Gu\'erin}},\ }\bibfield  {title} {\enquote {\bibinfo {title} {Optimal robust
  quantum control by inverse geometric optimization},}\ }\href {\doibase
  10.1103/PhysRevLett.125.250403} {\bibfield  {journal} {\bibinfo  {journal}
  {Phys. Rev. Lett.}\ }\textbf {\bibinfo {volume} {125}},\ \bibinfo {pages}
  {250403} (\bibinfo {year} {2020}{\natexlab{a}})}\BibitemShut {NoStop}%
\bibitem [{\citenamefont {Dridi}\ \emph
  {et~al.}(2020{\natexlab{b}})\citenamefont {Dridi}, \citenamefont {Mejatty},
  \citenamefont {Glaser},\ and\ \citenamefont {Sugny}}]{Dridi.PRA.2020}%
  \BibitemOpen
  \bibfield  {author} {\bibinfo {author} {\bibfnamefont {G.}~\bibnamefont
  {Dridi}}, \bibinfo {author} {\bibfnamefont {M.}~\bibnamefont {Mejatty}},
  \bibinfo {author} {\bibfnamefont {S.~J.}\ \bibnamefont {Glaser}}, \ and\
  \bibinfo {author} {\bibfnamefont {D.}~\bibnamefont {Sugny}},\ }\bibfield
  {title} {\enquote {\bibinfo {title} {Robust control of a not gate by
  composite pulses},}\ }\href {\doibase 10.1103/PhysRevA.101.012321} {\bibfield
   {journal} {\bibinfo  {journal} {Phys. Rev. A}\ }\textbf {\bibinfo {volume}
  {101}},\ \bibinfo {pages} {012321} (\bibinfo {year}
  {2020}{\natexlab{b}})}\BibitemShut {NoStop}%
\bibitem [{\citenamefont {Tian}\ \emph {et~al.}(2020)\citenamefont {Tian},
  \citenamefont {Liu}, \citenamefont {Liu}, \citenamefont {Yang}, \citenamefont
  {Betzholz}, \citenamefont {Said}, \citenamefont {Jelezko},\ and\
  \citenamefont {Cai}}]{Tian.PRA.2020}%
  \BibitemOpen
  \bibfield  {author} {\bibinfo {author} {\bibfnamefont {Jiazhao}\ \bibnamefont
  {Tian}}, \bibinfo {author} {\bibfnamefont {Haibin}\ \bibnamefont {Liu}},
  \bibinfo {author} {\bibfnamefont {Yu}~\bibnamefont {Liu}}, \bibinfo {author}
  {\bibfnamefont {Pengcheng}\ \bibnamefont {Yang}}, \bibinfo {author}
  {\bibfnamefont {Ralf}\ \bibnamefont {Betzholz}}, \bibinfo {author}
  {\bibfnamefont {Ressa~S.}\ \bibnamefont {Said}}, \bibinfo {author}
  {\bibfnamefont {Fedor}\ \bibnamefont {Jelezko}}, \ and\ \bibinfo {author}
  {\bibfnamefont {Jianming}\ \bibnamefont {Cai}},\ }\bibfield  {title}
  {\enquote {\bibinfo {title} {Quantum optimal control using phase-modulated
  driving fields},}\ }\href {\doibase 10.1103/PhysRevA.102.043707} {\bibfield
  {journal} {\bibinfo  {journal} {Phys. Rev. A}\ }\textbf {\bibinfo {volume}
  {102}},\ \bibinfo {pages} {043707} (\bibinfo {year} {2020})}\BibitemShut
  {NoStop}%
\bibitem [{\citenamefont {Lapert}\ \emph {et~al.}(2010)\citenamefont {Lapert},
  \citenamefont {Zhang}, \citenamefont {Braun}, \citenamefont {Glaser},\ and\
  \citenamefont {Sugny}}]{Lapert.PRL.2010}%
  \BibitemOpen
  \bibfield  {author} {\bibinfo {author} {\bibfnamefont {M.}~\bibnamefont
  {Lapert}}, \bibinfo {author} {\bibfnamefont {Y.}~\bibnamefont {Zhang}},
  \bibinfo {author} {\bibfnamefont {M.}~\bibnamefont {Braun}}, \bibinfo
  {author} {\bibfnamefont {S.~J.}\ \bibnamefont {Glaser}}, \ and\ \bibinfo
  {author} {\bibfnamefont {D.}~\bibnamefont {Sugny}},\ }\bibfield  {title}
  {\enquote {\bibinfo {title} {Singular extremals for the time-optimal control
  of dissipative spin $\frac{1}{2}$ particles},}\ }\href {\doibase
  10.1103/PhysRevLett.104.083001} {\bibfield  {journal} {\bibinfo  {journal}
  {Phys. Rev. Lett.}\ }\textbf {\bibinfo {volume} {104}},\ \bibinfo {pages}
  {083001} (\bibinfo {year} {2010})}\BibitemShut {NoStop}%
\bibitem [{\citenamefont {Yuan}\ \emph {et~al.}(2012)\citenamefont {Yuan},
  \citenamefont {Koch}, \citenamefont {Salamon},\ and\ \citenamefont
  {Tannor}}]{YuanHaidong.PRA.2012}%
  \BibitemOpen
  \bibfield  {author} {\bibinfo {author} {\bibfnamefont {Haidong}\ \bibnamefont
  {Yuan}}, \bibinfo {author} {\bibfnamefont {Christiane~P.}\ \bibnamefont
  {Koch}}, \bibinfo {author} {\bibfnamefont {Peter}\ \bibnamefont {Salamon}}, \
  and\ \bibinfo {author} {\bibfnamefont {David~J.}\ \bibnamefont {Tannor}},\
  }\bibfield  {title} {\enquote {\bibinfo {title} {Controllability on
  relaxation-free subspaces: On the relationship between adiabatic population
  transfer and optimal control},}\ }\href {\doibase 10.1103/PhysRevA.85.033417}
  {\bibfield  {journal} {\bibinfo  {journal} {Phys. Rev. A}\ }\textbf {\bibinfo
  {volume} {85}},\ \bibinfo {pages} {033417} (\bibinfo {year}
  {2012})}\BibitemShut {NoStop}%
\bibitem [{\citenamefont {Lapert}\ \emph {et~al.}(2012)\citenamefont {Lapert},
  \citenamefont {Zhang}, \citenamefont {Janich}, \citenamefont {Glaser},\ and\
  \citenamefont {Sugny}}]{Lampert.SciRep.2012}%
  \BibitemOpen
  \bibfield  {author} {\bibinfo {author} {\bibfnamefont {Marc}\ \bibnamefont
  {Lapert}}, \bibinfo {author} {\bibfnamefont {Yun}\ \bibnamefont {Zhang}},
  \bibinfo {author} {\bibfnamefont {Martin~A}\ \bibnamefont {Janich}}, \bibinfo
  {author} {\bibfnamefont {Steffen~J}\ \bibnamefont {Glaser}}, \ and\ \bibinfo
  {author} {\bibfnamefont {Dominique}\ \bibnamefont {Sugny}},\ }\bibfield
  {title} {\enquote {\bibinfo {title} {Exploring the physical limits of
  saturation contrast in magnetic resonance imaging},}\ }\href
  {https://www.nature.com/articles/srep00589} {\bibfield  {journal} {\bibinfo
  {journal} {Scientific Reports}\ }\textbf {\bibinfo {volume} {2}},\ \bibinfo
  {pages} {1--5} (\bibinfo {year} {2012})}\BibitemShut {NoStop}%
\bibitem [{\citenamefont {Doria}\ \emph {et~al.}(2011)\citenamefont {Doria},
  \citenamefont {Calarco},\ and\ \citenamefont {Montangero}}]{Doria.PRL.2011}%
  \BibitemOpen
  \bibfield  {author} {\bibinfo {author} {\bibfnamefont {Patrick}\ \bibnamefont
  {Doria}}, \bibinfo {author} {\bibfnamefont {Tommaso}\ \bibnamefont
  {Calarco}}, \ and\ \bibinfo {author} {\bibfnamefont {Simone}\ \bibnamefont
  {Montangero}},\ }\bibfield  {title} {\enquote {\bibinfo {title} {Optimal
  control technique for many-body quantum dynamics},}\ }\href {\doibase
  10.1103/PhysRevLett.106.190501} {\bibfield  {journal} {\bibinfo  {journal}
  {Phys. Rev. Lett.}\ }\textbf {\bibinfo {volume} {106}},\ \bibinfo {pages}
  {190501} (\bibinfo {year} {2011})}\BibitemShut {NoStop}%
\bibitem [{\citenamefont {Sarandy}\ and\ \citenamefont
  {Lidar}(2005)}]{Sarandy.PRL.2005}%
  \BibitemOpen
  \bibfield  {author} {\bibinfo {author} {\bibfnamefont {M.~S.}\ \bibnamefont
  {Sarandy}}\ and\ \bibinfo {author} {\bibfnamefont {D.~A.}\ \bibnamefont
  {Lidar}},\ }\bibfield  {title} {\enquote {\bibinfo {title} {Adiabatic quantum
  computation in open systems},}\ }\href {\doibase
  10.1103/PhysRevLett.95.250503} {\bibfield  {journal} {\bibinfo  {journal}
  {Phys. Rev. Lett.}\ }\textbf {\bibinfo {volume} {95}},\ \bibinfo {pages}
  {250503} (\bibinfo {year} {2005})}\BibitemShut {NoStop}%
\bibitem [{\citenamefont {Goerz}\ \emph {et~al.}(2014)\citenamefont {Goerz},
  \citenamefont {Reich},\ and\ \citenamefont {Koch}}]{Goerz.NJP.2014}%
  \BibitemOpen
  \bibfield  {author} {\bibinfo {author} {\bibfnamefont {Michael~H}\
  \bibnamefont {Goerz}}, \bibinfo {author} {\bibfnamefont {Daniel~M}\
  \bibnamefont {Reich}}, \ and\ \bibinfo {author} {\bibfnamefont
  {Christiane~P}\ \bibnamefont {Koch}},\ }\bibfield  {title} {\enquote
  {\bibinfo {title} {Optimal control theory for a unitary operation under
  dissipative evolution},}\ }\href
  {https://iopscience.iop.org/article/10.1088/1367-2630/16/5/055012} {\bibfield
   {journal} {\bibinfo  {journal} {New Journal of Physics}\ }\textbf {\bibinfo
  {volume} {16}},\ \bibinfo {pages} {055012} (\bibinfo {year}
  {2014})}\BibitemShut {NoStop}%
\bibitem [{\citenamefont {Marx}\ \emph {et~al.}(2010)\citenamefont {Marx},
  \citenamefont {Fahmy}, \citenamefont {Kauffman}, \citenamefont {Lomonaco},
  \citenamefont {Sp\"orl}, \citenamefont {Pomplun}, \citenamefont
  {Schulte-Herbr\"uggen}, \citenamefont {Myers},\ and\ \citenamefont
  {Glaser}}]{Marx.PRA.2010}%
  \BibitemOpen
  \bibfield  {author} {\bibinfo {author} {\bibfnamefont {Raimund}\ \bibnamefont
  {Marx}}, \bibinfo {author} {\bibfnamefont {Amr}\ \bibnamefont {Fahmy}},
  \bibinfo {author} {\bibfnamefont {Louis}\ \bibnamefont {Kauffman}}, \bibinfo
  {author} {\bibfnamefont {Samuel}\ \bibnamefont {Lomonaco}}, \bibinfo {author}
  {\bibfnamefont {Andreas}\ \bibnamefont {Sp\"orl}}, \bibinfo {author}
  {\bibfnamefont {Nikolas}\ \bibnamefont {Pomplun}}, \bibinfo {author}
  {\bibfnamefont {Thomas}\ \bibnamefont {Schulte-Herbr\"uggen}}, \bibinfo
  {author} {\bibfnamefont {John~M.}\ \bibnamefont {Myers}}, \ and\ \bibinfo
  {author} {\bibfnamefont {Steffen~J.}\ \bibnamefont {Glaser}},\ }\bibfield
  {title} {\enquote {\bibinfo {title} {Nuclear-magnetic-resonance quantum
  calculations of the jones polynomial},}\ }\href {\doibase
  10.1103/PhysRevA.81.032319} {\bibfield  {journal} {\bibinfo  {journal} {Phys.
  Rev. A}\ }\textbf {\bibinfo {volume} {81}},\ \bibinfo {pages} {032319}
  (\bibinfo {year} {2010})}\BibitemShut {NoStop}%
\bibitem [{\citenamefont {Schulte-Herbr{\"u}ggen}\ \emph
  {et~al.}(2012)\citenamefont {Schulte-Herbr{\"u}ggen}, \citenamefont {Marx},
  \citenamefont {Fahmy}, \citenamefont {Kauffman}, \citenamefont {Lomonaco},
  \citenamefont {Khaneja},\ and\ \citenamefont {Glaser}}]{schulte2012control}%
  \BibitemOpen
  \bibfield  {author} {\bibinfo {author} {\bibfnamefont {Thomas}\ \bibnamefont
  {Schulte-Herbr{\"u}ggen}}, \bibinfo {author} {\bibfnamefont {Raimund}\
  \bibnamefont {Marx}}, \bibinfo {author} {\bibfnamefont {Amr}\ \bibnamefont
  {Fahmy}}, \bibinfo {author} {\bibfnamefont {Louis}\ \bibnamefont {Kauffman}},
  \bibinfo {author} {\bibfnamefont {Samuel}\ \bibnamefont {Lomonaco}}, \bibinfo
  {author} {\bibfnamefont {Navin}\ \bibnamefont {Khaneja}}, \ and\ \bibinfo
  {author} {\bibfnamefont {Steffen~J}\ \bibnamefont {Glaser}},\ }\bibfield
  {title} {\enquote {\bibinfo {title} {Control aspects of quantum computing
  using pure and mixed states},}\ }\href
  {https://royalsocietypublishing.org/doi/10.1098/rsta.2011.0513} {\bibfield
  {journal} {\bibinfo  {journal} {Philosophical Transactions of the Royal
  Society A: Mathematical, Physical and Engineering Sciences}\ }\textbf
  {\bibinfo {volume} {370}},\ \bibinfo {pages} {4651--4670} (\bibinfo {year}
  {2012})}\BibitemShut {NoStop}%
\bibitem [{\citenamefont {Martinis}\ and\ \citenamefont
  {Geller}(2014)}]{Martinis_and_Geller.PRA.2014}%
  \BibitemOpen
  \bibfield  {author} {\bibinfo {author} {\bibfnamefont {John~M.}\ \bibnamefont
  {Martinis}}\ and\ \bibinfo {author} {\bibfnamefont {Michael~R.}\ \bibnamefont
  {Geller}},\ }\bibfield  {title} {\enquote {\bibinfo {title} {Fast adiabatic
  qubit gates using only ${\ensuremath{\sigma}}_{z}$ control},}\ }\href
  {\doibase 10.1103/PhysRevA.90.022307} {\bibfield  {journal} {\bibinfo
  {journal} {Phys. Rev. A}\ }\textbf {\bibinfo {volume} {90}},\ \bibinfo
  {pages} {022307} (\bibinfo {year} {2014})}\BibitemShut {NoStop}%
\bibitem [{\citenamefont {Stefanatos}\ and\ \citenamefont
  {Paspalakis}(2019)}]{Stefanatos.PRA.2019}%
  \BibitemOpen
  \bibfield  {author} {\bibinfo {author} {\bibfnamefont {Dionisis}\
  \bibnamefont {Stefanatos}}\ and\ \bibinfo {author} {\bibfnamefont {Emmanuel}\
  \bibnamefont {Paspalakis}},\ }\bibfield  {title} {\enquote {\bibinfo {title}
  {Resonant shortcuts for adiabatic rapid passage with only $z$-field
  control},}\ }\href {\doibase 10.1103/PhysRevA.100.012111} {\bibfield
  {journal} {\bibinfo  {journal} {Phys. Rev. A}\ }\textbf {\bibinfo {volume}
  {100}},\ \bibinfo {pages} {012111} (\bibinfo {year} {2019})}\BibitemShut
  {NoStop}%
\bibitem [{\citenamefont {Stefanatos}\ and\ \citenamefont
  {Paspalakis}(2020{\natexlab{b}})}]{Stefanatos.JPAMT.2020}%
  \BibitemOpen
  \bibfield  {author} {\bibinfo {author} {\bibfnamefont {Dionisis}\
  \bibnamefont {Stefanatos}}\ and\ \bibinfo {author} {\bibfnamefont {Emmanuel}\
  \bibnamefont {Paspalakis}},\ }\bibfield  {title} {\enquote {\bibinfo {title}
  {Speeding up adiabatic passage with an optimal modified roland--cerf
  protocol},}\ }\href
  {https://iopscience.iop.org/article/10.1088/1751-8121/ab7423} {\bibfield
  {journal} {\bibinfo  {journal} {Journal of Physics A: Mathematical and
  Theoretical}\ }\textbf {\bibinfo {volume} {53}},\ \bibinfo {pages} {115304}
  (\bibinfo {year} {2020}{\natexlab{b}})}\BibitemShut {NoStop}%
\bibitem [{\citenamefont {Khaneja}\ \emph {et~al.}(2005)\citenamefont
  {Khaneja}, \citenamefont {Reiss}, \citenamefont {Kehlet}, \citenamefont
  {Schulte-Herbr{\"u}ggen},\ and\ \citenamefont {Glaser}}]{khaneja.JMR.2005}%
  \BibitemOpen
  \bibfield  {author} {\bibinfo {author} {\bibfnamefont {Navin}\ \bibnamefont
  {Khaneja}}, \bibinfo {author} {\bibfnamefont {Timo}\ \bibnamefont {Reiss}},
  \bibinfo {author} {\bibfnamefont {Cindie}\ \bibnamefont {Kehlet}}, \bibinfo
  {author} {\bibfnamefont {Thomas}\ \bibnamefont {Schulte-Herbr{\"u}ggen}}, \
  and\ \bibinfo {author} {\bibfnamefont {Steffen~J}\ \bibnamefont {Glaser}},\
  }\bibfield  {title} {\enquote {\bibinfo {title} {Optimal control of coupled
  spin dynamics: design of nmr pulse sequences by gradient ascent
  algorithms},}\ }\href {https://pubmed.ncbi.nlm.nih.gov/15649756/} {\bibfield
  {journal} {\bibinfo  {journal} {Journal of magnetic resonance}\ }\textbf
  {\bibinfo {volume} {172}},\ \bibinfo {pages} {296--305} (\bibinfo {year}
  {2005})}\BibitemShut {NoStop}%
\bibitem [{\citenamefont {Machnes}\ \emph {et~al.}(2011)\citenamefont
  {Machnes}, \citenamefont {Sander}, \citenamefont {Glaser}, \citenamefont
  {de~Fouqui\`eres}, \citenamefont {Gruslys}, \citenamefont {Schirmer},\ and\
  \citenamefont {Schulte-Herbr\"uggen}}]{Machnes.PRA.2011}%
  \BibitemOpen
  \bibfield  {author} {\bibinfo {author} {\bibfnamefont {S.}~\bibnamefont
  {Machnes}}, \bibinfo {author} {\bibfnamefont {U.}~\bibnamefont {Sander}},
  \bibinfo {author} {\bibfnamefont {S.~J.}\ \bibnamefont {Glaser}}, \bibinfo
  {author} {\bibfnamefont {P.}~\bibnamefont {de~Fouqui\`eres}}, \bibinfo
  {author} {\bibfnamefont {A.}~\bibnamefont {Gruslys}}, \bibinfo {author}
  {\bibfnamefont {S.}~\bibnamefont {Schirmer}}, \ and\ \bibinfo {author}
  {\bibfnamefont {T.}~\bibnamefont {Schulte-Herbr\"uggen}},\ }\bibfield
  {title} {\enquote {\bibinfo {title} {Comparing, optimizing, and benchmarking
  quantum-control algorithms in a unifying programming framework},}\ }\href
  {\doibase 10.1103/PhysRevA.84.022305} {\bibfield  {journal} {\bibinfo
  {journal} {Phys. Rev. A}\ }\textbf {\bibinfo {volume} {84}},\ \bibinfo
  {pages} {022305} (\bibinfo {year} {2011})}\BibitemShut {NoStop}%
\bibitem [{\citenamefont {Reich}\ \emph {et~al.}(2012)\citenamefont {Reich},
  \citenamefont {Ndong},\ and\ \citenamefont {Koch}}]{Reich.JCP.2012}%
  \BibitemOpen
  \bibfield  {author} {\bibinfo {author} {\bibfnamefont {Daniel~M}\
  \bibnamefont {Reich}}, \bibinfo {author} {\bibfnamefont {Mamadou}\
  \bibnamefont {Ndong}}, \ and\ \bibinfo {author} {\bibfnamefont
  {Christiane~P}\ \bibnamefont {Koch}},\ }\bibfield  {title} {\enquote
  {\bibinfo {title} {Monotonically convergent optimization in quantum control
  using krotov's method},}\ }\href
  {https://aip.scitation.org/doi/abs/10.1063/1.3691827} {\bibfield  {journal}
  {\bibinfo  {journal} {The Journal of chemical physics}\ }\textbf {\bibinfo
  {volume} {136}},\ \bibinfo {pages} {104103} (\bibinfo {year}
  {2012})}\BibitemShut {NoStop}%
\bibitem [{\citenamefont {Palao}\ and\ \citenamefont
  {Kosloff}(2002)}]{Palao.PRL.2002}%
  \BibitemOpen
  \bibfield  {author} {\bibinfo {author} {\bibfnamefont {Jos\'e~P.}\
  \bibnamefont {Palao}}\ and\ \bibinfo {author} {\bibfnamefont {Ronnie}\
  \bibnamefont {Kosloff}},\ }\bibfield  {title} {\enquote {\bibinfo {title}
  {Quantum computing by an optimal control algorithm for unitary
  transformations},}\ }\href {\doibase 10.1103/PhysRevLett.89.188301}
  {\bibfield  {journal} {\bibinfo  {journal} {Phys. Rev. Lett.}\ }\textbf
  {\bibinfo {volume} {89}},\ \bibinfo {pages} {188301} (\bibinfo {year}
  {2002})}\BibitemShut {NoStop}%
\bibitem [{\citenamefont {Palao}\ and\ \citenamefont
  {Kosloff}(2003)}]{Palao.PRA.2003}%
  \BibitemOpen
  \bibfield  {author} {\bibinfo {author} {\bibfnamefont {Jos\'e~P.}\
  \bibnamefont {Palao}}\ and\ \bibinfo {author} {\bibfnamefont {Ronnie}\
  \bibnamefont {Kosloff}},\ }\bibfield  {title} {\enquote {\bibinfo {title}
  {Optimal control theory for unitary transformations},}\ }\href {\doibase
  10.1103/PhysRevA.68.062308} {\bibfield  {journal} {\bibinfo  {journal} {Phys.
  Rev. A}\ }\textbf {\bibinfo {volume} {68}},\ \bibinfo {pages} {062308}
  (\bibinfo {year} {2003})}\BibitemShut {NoStop}%
\bibitem [{\citenamefont {Tesch}\ and\ \citenamefont
  {de~Vivie-Riedle}(2002)}]{Tesch.PRL.2002}%
  \BibitemOpen
  \bibfield  {author} {\bibinfo {author} {\bibfnamefont {Carmen~M.}\
  \bibnamefont {Tesch}}\ and\ \bibinfo {author} {\bibfnamefont {Regina}\
  \bibnamefont {de~Vivie-Riedle}},\ }\bibfield  {title} {\enquote {\bibinfo
  {title} {Quantum computation with vibrationally excited molecules},}\ }\href
  {\doibase 10.1103/PhysRevLett.89.157901} {\bibfield  {journal} {\bibinfo
  {journal} {Phys. Rev. Lett.}\ }\textbf {\bibinfo {volume} {89}},\ \bibinfo
  {pages} {157901} (\bibinfo {year} {2002})}\BibitemShut {NoStop}%
\bibitem [{\citenamefont {Zanardi}\ and\ \citenamefont
  {Rasetti}(1999)}]{ZANARDI199994}%
  \BibitemOpen
  \bibfield  {author} {\bibinfo {author} {\bibnamefont {Zanardi}}\ and\
  \bibinfo {author} {\bibnamefont {Rasetti}},\ }\bibfield  {title} {\enquote
  {\bibinfo {title} {Holonomic quantum computation},}\ }\href {\doibase
  https://doi.org/10.1016/S0375-9601(99)00803-8} {\bibfield  {journal}
  {\bibinfo  {journal} {Physics Letters A}\ }\textbf {\bibinfo {volume}
  {264}},\ \bibinfo {pages} {94 -- 99} (\bibinfo {year} {1999})}\BibitemShut
  {NoStop}%
\bibitem [{\citenamefont {Anandan}(1988)}]{ANANDAN1988171}%
  \BibitemOpen
  \bibfield  {author} {\bibinfo {author} {\bibfnamefont {J.}~\bibnamefont
  {Anandan}},\ }\bibfield  {title} {\enquote {\bibinfo {title} {Non-adiabatic
  non-abelian geometric phase},}\ }\href {\doibase
  https://doi.org/10.1016/0375-9601(88)91010-9} {\bibfield  {journal} {\bibinfo
   {journal} {Physics Letters A}\ }\textbf {\bibinfo {volume} {133}},\ \bibinfo
  {pages} {171 -- 175} (\bibinfo {year} {1988})}\BibitemShut {NoStop}%
\bibitem [{\citenamefont {Wilczek}\ and\ \citenamefont
  {Zee}(1984)}]{Wilczek.Zee}%
  \BibitemOpen
  \bibfield  {author} {\bibinfo {author} {\bibfnamefont {Frank}\ \bibnamefont
  {Wilczek}}\ and\ \bibinfo {author} {\bibfnamefont {A.}~\bibnamefont {Zee}},\
  }\bibfield  {title} {\enquote {\bibinfo {title} {Appearance of gauge
  structure in simple dynamical systems},}\ }\href {\doibase
  10.1103/PhysRevLett.52.2111} {\bibfield  {journal} {\bibinfo  {journal}
  {Phys. Rev. Lett.}\ }\textbf {\bibinfo {volume} {52}},\ \bibinfo {pages}
  {2111--2114} (\bibinfo {year} {1984})}\BibitemShut {NoStop}%
\bibitem [{\citenamefont {Berry}(2009)}]{Berry_2009}%
  \BibitemOpen
  \bibfield  {author} {\bibinfo {author} {\bibfnamefont {M~V}\ \bibnamefont
  {Berry}},\ }\bibfield  {title} {\enquote {\bibinfo {title} {Transitionless
  quantum driving},}\ }\href {\doibase 10.1088/1751-8113/42/36/365303}
  {\bibfield  {journal} {\bibinfo  {journal} {Journal of Physics A:
  Mathematical and Theoretical}\ }\textbf {\bibinfo {volume} {42}},\ \bibinfo
  {pages} {365303} (\bibinfo {year} {2009})}\BibitemShut {NoStop}%
\bibitem [{\citenamefont {Solinas}\ \emph {et~al.}(2004)\citenamefont
  {Solinas}, \citenamefont {Zanardi},\ and\ \citenamefont
  {Zangh\`{\i}}}]{SolinasPRA2004}%
  \BibitemOpen
  \bibfield  {author} {\bibinfo {author} {\bibfnamefont {Paolo}\ \bibnamefont
  {Solinas}}, \bibinfo {author} {\bibfnamefont {Paolo}\ \bibnamefont
  {Zanardi}}, \ and\ \bibinfo {author} {\bibfnamefont {Nino}\ \bibnamefont
  {Zangh\`{\i}}},\ }\bibfield  {title} {\enquote {\bibinfo {title} {Robustness
  of non-abelian holonomic quantum gates against parametric noise},}\ }\href
  {\doibase 10.1103/PhysRevA.70.042316} {\bibfield  {journal} {\bibinfo
  {journal} {Phys. Rev. A}\ }\textbf {\bibinfo {volume} {70}},\ \bibinfo
  {pages} {042316} (\bibinfo {year} {2004})}\BibitemShut {NoStop}%
\bibitem [{\citenamefont {Sjöqvist}(2015)}]{Sjoqvist.IJQC.2015}%
  \BibitemOpen
  \bibfield  {author} {\bibinfo {author} {\bibfnamefont {Erik}\ \bibnamefont
  {Sjöqvist}},\ }\bibfield  {title} {\enquote {\bibinfo {title} {Geometric
  phases in quantum information},}\ }\href {\doibase
  https://doi.org/10.1002/qua.24941} {\bibfield  {journal} {\bibinfo  {journal}
  {International Journal of Quantum Chemistry}\ }\textbf {\bibinfo {volume}
  {115}},\ \bibinfo {pages} {1311--1326} (\bibinfo {year} {2015})}\BibitemShut
  {NoStop}%
\bibitem [{\citenamefont {Xu}\ \emph {et~al.}(2012)\citenamefont {Xu},
  \citenamefont {Zhang}, \citenamefont {Tong}, \citenamefont {Sj\"oqvist},\
  and\ \citenamefont {Kwek}}]{Xu.PRL.2012}%
  \BibitemOpen
  \bibfield  {author} {\bibinfo {author} {\bibfnamefont {G.~F.}\ \bibnamefont
  {Xu}}, \bibinfo {author} {\bibfnamefont {J.}~\bibnamefont {Zhang}}, \bibinfo
  {author} {\bibfnamefont {D.~M.}\ \bibnamefont {Tong}}, \bibinfo {author}
  {\bibfnamefont {Erik}\ \bibnamefont {Sj\"oqvist}}, \ and\ \bibinfo {author}
  {\bibfnamefont {L.~C.}\ \bibnamefont {Kwek}},\ }\bibfield  {title} {\enquote
  {\bibinfo {title} {Nonadiabatic holonomic quantum computation in
  decoherence-free subspaces},}\ }\href {\doibase
  10.1103/PhysRevLett.109.170501} {\bibfield  {journal} {\bibinfo  {journal}
  {Phys. Rev. Lett.}\ }\textbf {\bibinfo {volume} {109}},\ \bibinfo {pages}
  {170501} (\bibinfo {year} {2012})}\BibitemShut {NoStop}%
\bibitem [{\citenamefont {Güngördü}\ \emph {et~al.}(2014)\citenamefont
  {Güngördü}, \citenamefont {Wan},\ and\ \citenamefont
  {Nakahara}}]{Utkan.JPSJ.2014}%
  \BibitemOpen
  \bibfield  {author} {\bibinfo {author} {\bibfnamefont {Utkan}\ \bibnamefont
  {Güngördü}}, \bibinfo {author} {\bibfnamefont {Yidun}\ \bibnamefont
  {Wan}}, \ and\ \bibinfo {author} {\bibfnamefont {Mikio}\ \bibnamefont
  {Nakahara}},\ }\bibfield  {title} {\enquote {\bibinfo {title} {Non-adiabatic
  universal holonomic quantum gates based on abelian holonomies},}\ }\href
  {\doibase 10.7566/JPSJ.83.034001} {\bibfield  {journal} {\bibinfo  {journal}
  {Journal of the Physical Society of Japan}\ }\textbf {\bibinfo {volume}
  {83}},\ \bibinfo {pages} {034001} (\bibinfo {year} {2014})}\BibitemShut
  {NoStop}%
\bibitem [{\citenamefont {Berry}(1984)}]{Berry.Geometric.Phase}%
  \BibitemOpen
  \bibfield  {author} {\bibinfo {author} {\bibfnamefont {Michael~Victor}\
  \bibnamefont {Berry}},\ }\bibfield  {title} {\enquote {\bibinfo {title}
  {Quantal phase factors accompanying adiabatic changes},}\ }\href {\doibase
  10.1098/rspa.1984.0023} {\bibfield  {journal} {\bibinfo  {journal}
  {Proceedings of the Royal Society of London. A. Mathematical and Physical
  Sciences}\ }\textbf {\bibinfo {volume} {392}},\ \bibinfo {pages} {45--57}
  (\bibinfo {year} {1984})}\BibitemShut {NoStop}%
\bibitem [{\citenamefont {De~Chiara}\ and\ \citenamefont
  {Palma}(2003)}]{DeChiara.PRL.2003}%
  \BibitemOpen
  \bibfield  {author} {\bibinfo {author} {\bibfnamefont {Gabriele}\
  \bibnamefont {De~Chiara}}\ and\ \bibinfo {author} {\bibfnamefont
  {G.~Massimo}\ \bibnamefont {Palma}},\ }\bibfield  {title} {\enquote {\bibinfo
  {title} {Berry phase for a spin $1/2$ particle in a classical fluctuating
  field},}\ }\href {\doibase 10.1103/PhysRevLett.91.090404} {\bibfield
  {journal} {\bibinfo  {journal} {Phys. Rev. Lett.}\ }\textbf {\bibinfo
  {volume} {91}},\ \bibinfo {pages} {090404} (\bibinfo {year}
  {2003})}\BibitemShut {NoStop}%
\bibitem [{\citenamefont {Yale}\ \emph {et~al.}(2016)\citenamefont {Yale},
  \citenamefont {Heremans}, \citenamefont {Zhou}, \citenamefont {Auer},
  \citenamefont {Burkard},\ and\ \citenamefont {Awschalom}}]{Yale2016}%
  \BibitemOpen
  \bibfield  {author} {\bibinfo {author} {\bibfnamefont {Christopher~G}\
  \bibnamefont {Yale}}, \bibinfo {author} {\bibfnamefont {F~Joseph}\
  \bibnamefont {Heremans}}, \bibinfo {author} {\bibfnamefont {Brian~B}\
  \bibnamefont {Zhou}}, \bibinfo {author} {\bibfnamefont {Adrian}\ \bibnamefont
  {Auer}}, \bibinfo {author} {\bibfnamefont {Guido}\ \bibnamefont {Burkard}}, \
  and\ \bibinfo {author} {\bibfnamefont {David~D}\ \bibnamefont {Awschalom}},\
  }\bibfield  {title} {\enquote {\bibinfo {title} {Optical manipulation of the
  berry phase in a solid-state spin qubit},}\ }\href
  {https://www.nature.com/articles/nphoton.2015.278} {\bibfield  {journal}
  {\bibinfo  {journal} {Nature photonics}\ }\textbf {\bibinfo {volume} {10}},\
  \bibinfo {pages} {184--189} (\bibinfo {year} {2016})}\BibitemShut {NoStop}%
\bibitem [{\citenamefont {Aharonov}\ and\ \citenamefont
  {Anandan}(1987)}]{PhysRevLett.58.1593}%
  \BibitemOpen
  \bibfield  {author} {\bibinfo {author} {\bibfnamefont {Y.}~\bibnamefont
  {Aharonov}}\ and\ \bibinfo {author} {\bibfnamefont {J.}~\bibnamefont
  {Anandan}},\ }\bibfield  {title} {\enquote {\bibinfo {title} {Phase change
  during a cyclic quantum evolution},}\ }\href {\doibase
  10.1103/PhysRevLett.58.1593} {\bibfield  {journal} {\bibinfo  {journal}
  {Phys. Rev. Lett.}\ }\textbf {\bibinfo {volume} {58}},\ \bibinfo {pages}
  {1593--1596} (\bibinfo {year} {1987})}\BibitemShut {NoStop}%
\bibitem [{\citenamefont {Sjöqvist}\ \emph {et~al.}(2012)\citenamefont
  {Sjöqvist}, \citenamefont {Tong}, \citenamefont {Andersson}, \citenamefont
  {Hessmo}, \citenamefont {Johansson},\ and\ \citenamefont
  {Singh}}]{Sj_qvist_2012}%
  \BibitemOpen
  \bibfield  {author} {\bibinfo {author} {\bibfnamefont {Erik}\ \bibnamefont
  {Sjöqvist}}, \bibinfo {author} {\bibfnamefont {D~M}\ \bibnamefont {Tong}},
  \bibinfo {author} {\bibfnamefont {L~Mauritz}\ \bibnamefont {Andersson}},
  \bibinfo {author} {\bibfnamefont {Björn}\ \bibnamefont {Hessmo}}, \bibinfo
  {author} {\bibfnamefont {Markus}\ \bibnamefont {Johansson}}, \ and\ \bibinfo
  {author} {\bibfnamefont {Kuldip}\ \bibnamefont {Singh}},\ }\bibfield  {title}
  {\enquote {\bibinfo {title} {Non-adiabatic holonomic quantum computation},}\
  }\href {\doibase 10.1088/1367-2630/14/10/103035} {\bibfield  {journal}
  {\bibinfo  {journal} {New Journal of Physics}\ }\textbf {\bibinfo {volume}
  {14}},\ \bibinfo {pages} {103035} (\bibinfo {year} {2012})}\BibitemShut
  {NoStop}%
\bibitem [{\citenamefont {Sjöqvist}(2016)}]{Sjoqvist}%
  \BibitemOpen
  \bibfield  {author} {\bibinfo {author} {\bibfnamefont {Erik}\ \bibnamefont
  {Sjöqvist}},\ }\bibfield  {title} {\enquote {\bibinfo {title} {Nonadiabatic
  holonomic single-qubit gates in off-resonant lambda systems},}\ }\href
  {\doibase https://doi.org/10.1016/j.physleta.2015.10.006} {\bibfield
  {journal} {\bibinfo  {journal} {Physics Letters A}\ }\textbf {\bibinfo
  {volume} {380}},\ \bibinfo {pages} {65 -- 67} (\bibinfo {year}
  {2016})}\BibitemShut {NoStop}%
\bibitem [{\citenamefont {Hong}\ \emph {et~al.}(2018)\citenamefont {Hong},
  \citenamefont {Liu}, \citenamefont {Cai}, \citenamefont {Zhang},
  \citenamefont {Hu}, \citenamefont {Wang},\ and\ \citenamefont
  {Xue}}]{XueZhengyuan.PRA.2018}%
  \BibitemOpen
  \bibfield  {author} {\bibinfo {author} {\bibfnamefont {Zhuo-Ping}\
  \bibnamefont {Hong}}, \bibinfo {author} {\bibfnamefont {Bao-Jie}\
  \bibnamefont {Liu}}, \bibinfo {author} {\bibfnamefont {Jia-Qi}\ \bibnamefont
  {Cai}}, \bibinfo {author} {\bibfnamefont {Xin-Ding}\ \bibnamefont {Zhang}},
  \bibinfo {author} {\bibfnamefont {Yong}\ \bibnamefont {Hu}}, \bibinfo
  {author} {\bibfnamefont {Z.~D.}\ \bibnamefont {Wang}}, \ and\ \bibinfo
  {author} {\bibfnamefont {Zheng-Yuan}\ \bibnamefont {Xue}},\ }\bibfield
  {title} {\enquote {\bibinfo {title} {Implementing universal nonadiabatic
  holonomic quantum gates with transmons},}\ }\href {\doibase
  10.1103/PhysRevA.97.022332} {\bibfield  {journal} {\bibinfo  {journal} {Phys.
  Rev. A}\ }\textbf {\bibinfo {volume} {97}},\ \bibinfo {pages} {022332}
  (\bibinfo {year} {2018})}\BibitemShut {NoStop}%
\bibitem [{\citenamefont {Ribeiro}\ and\ \citenamefont
  {Clerk}(2019)}]{Ribeiro.PRA.2019}%
  \BibitemOpen
  \bibfield  {author} {\bibinfo {author} {\bibfnamefont {Hugo}\ \bibnamefont
  {Ribeiro}}\ and\ \bibinfo {author} {\bibfnamefont {Aashish~A.}\ \bibnamefont
  {Clerk}},\ }\bibfield  {title} {\enquote {\bibinfo {title} {Accelerated
  adiabatic quantum gates: Optimizing speed versus robustness},}\ }\href
  {\doibase 10.1103/PhysRevA.100.032323} {\bibfield  {journal} {\bibinfo
  {journal} {Phys. Rev. A}\ }\textbf {\bibinfo {volume} {100}},\ \bibinfo
  {pages} {032323} (\bibinfo {year} {2019})}\BibitemShut {NoStop}%
\bibitem [{\citenamefont {Liu}\ \emph {et~al.}(2019)\citenamefont {Liu},
  \citenamefont {Song}, \citenamefont {Xue}, \citenamefont {Wang},\ and\
  \citenamefont {Yung}}]{LiuBoajie.PRL.2019}%
  \BibitemOpen
  \bibfield  {author} {\bibinfo {author} {\bibfnamefont {Bao-Jie}\ \bibnamefont
  {Liu}}, \bibinfo {author} {\bibfnamefont {Xue-Ke}\ \bibnamefont {Song}},
  \bibinfo {author} {\bibfnamefont {Zheng-Yuan}\ \bibnamefont {Xue}}, \bibinfo
  {author} {\bibfnamefont {Xin}\ \bibnamefont {Wang}}, \ and\ \bibinfo {author}
  {\bibfnamefont {Man-Hong}\ \bibnamefont {Yung}},\ }\bibfield  {title}
  {\enquote {\bibinfo {title} {Plug-and-play approach to nonadiabatic geometric
  quantum gates},}\ }\href {\doibase 10.1103/PhysRevLett.123.100501} {\bibfield
   {journal} {\bibinfo  {journal} {Phys. Rev. Lett.}\ }\textbf {\bibinfo
  {volume} {123}},\ \bibinfo {pages} {100501} (\bibinfo {year}
  {2019})}\BibitemShut {NoStop}%
\bibitem [{\citenamefont {Ying}\ \emph {et~al.}(2020)\citenamefont {Ying},
  \citenamefont {Gentile}, \citenamefont {Baltan\'as}, \citenamefont
  {Frustaglia}, \citenamefont {Ortix},\ and\ \citenamefont
  {Cuoco}}]{YingZuJian.PRR.2020}%
  \BibitemOpen
  \bibfield  {author} {\bibinfo {author} {\bibfnamefont {Zu-Jian}\ \bibnamefont
  {Ying}}, \bibinfo {author} {\bibfnamefont {Paola}\ \bibnamefont {Gentile}},
  \bibinfo {author} {\bibfnamefont {Jos\'e~Pablo}\ \bibnamefont {Baltan\'as}},
  \bibinfo {author} {\bibfnamefont {Diego}\ \bibnamefont {Frustaglia}},
  \bibinfo {author} {\bibfnamefont {Carmine}\ \bibnamefont {Ortix}}, \ and\
  \bibinfo {author} {\bibfnamefont {Mario}\ \bibnamefont {Cuoco}},\ }\bibfield
  {title} {\enquote {\bibinfo {title} {Geometric driving of two-level quantum
  systems},}\ }\href {\doibase 10.1103/PhysRevResearch.2.023167} {\bibfield
  {journal} {\bibinfo  {journal} {Phys. Rev. Research}\ }\textbf {\bibinfo
  {volume} {2}},\ \bibinfo {pages} {023167} (\bibinfo {year}
  {2020})}\BibitemShut {NoStop}%
\bibitem [{\citenamefont {Shkolnikov}\ \emph {et~al.}(2020)\citenamefont
  {Shkolnikov}, \citenamefont {Mauch},\ and\ \citenamefont
  {Burkard}}]{Shkolnikov.PRB.2020}%
  \BibitemOpen
  \bibfield  {author} {\bibinfo {author} {\bibfnamefont {V.~O.}\ \bibnamefont
  {Shkolnikov}}, \bibinfo {author} {\bibfnamefont {Roman}\ \bibnamefont
  {Mauch}}, \ and\ \bibinfo {author} {\bibfnamefont {Guido}\ \bibnamefont
  {Burkard}},\ }\bibfield  {title} {\enquote {\bibinfo {title} {All-microwave
  holonomic control of an electron-nuclear two-qubit register in diamond},}\
  }\href {\doibase 10.1103/PhysRevB.101.155306} {\bibfield  {journal} {\bibinfo
   {journal} {Phys. Rev. B}\ }\textbf {\bibinfo {volume} {101}},\ \bibinfo
  {pages} {155306} (\bibinfo {year} {2020})}\BibitemShut {NoStop}%
\bibitem [{\citenamefont {Li}\ and\ \citenamefont
  {Xue}(2020)}]{li2020dynamically}%
  \BibitemOpen
  \bibfield  {author} {\bibinfo {author} {\bibfnamefont {Sai}\ \bibnamefont
  {Li}}\ and\ \bibinfo {author} {\bibfnamefont {Zheng-Yuan}\ \bibnamefont
  {Xue}},\ }\href@noop {} {\enquote {\bibinfo {title} {Dynamically corrected
  nonadiabatic holonomic quantum gates},}\ } (\bibinfo {year} {2020}),\ \Eprint
  {http://arxiv.org/abs/2012.09034} {arXiv:2012.09034 [quant-ph]} \BibitemShut
  {NoStop}%
\bibitem [{\citenamefont {Ji}\ \emph {et~al.}(2021)\citenamefont {Ji},
  \citenamefont {Ding}, \citenamefont {Chen},\ and\ \citenamefont
  {Xue}}]{Ji2021noncyclic}%
  \BibitemOpen
  \bibfield  {author} {\bibinfo {author} {\bibfnamefont {Li-Na}\ \bibnamefont
  {Ji}}, \bibinfo {author} {\bibfnamefont {Cheng-Yun}\ \bibnamefont {Ding}},
  \bibinfo {author} {\bibfnamefont {Tao}\ \bibnamefont {Chen}}, \ and\ \bibinfo
  {author} {\bibfnamefont {Zheng-Yuan}\ \bibnamefont {Xue}},\ }\href@noop {}
  {\enquote {\bibinfo {title} {Noncyclic and nonadiabatic geometric quantum
  gates with smooth paths},}\ } (\bibinfo {year} {2021}),\ \Eprint
  {http://arxiv.org/abs/2102.00893} {arXiv:2102.00893 [quant-ph]} \BibitemShut
  {NoStop}%
\bibitem [{\citenamefont {Zhao}\ \emph {et~al.}(2021)\citenamefont {Zhao},
  \citenamefont {Wu},\ and\ \citenamefont {Tong}}]{Zhao.holonomicDD.PRL.2021}%
  \BibitemOpen
  \bibfield  {author} {\bibinfo {author} {\bibfnamefont {P.~Z.}\ \bibnamefont
  {Zhao}}, \bibinfo {author} {\bibfnamefont {X.}~\bibnamefont {Wu}}, \ and\
  \bibinfo {author} {\bibfnamefont {D.~M.}\ \bibnamefont {Tong}},\ }\bibfield
  {title} {\enquote {\bibinfo {title} {Dynamical-decoupling-protected
  nonadiabatic holonomic quantum computation},}\ }\href {\doibase
  10.1103/PhysRevA.103.012205} {\bibfield  {journal} {\bibinfo  {journal}
  {Phys. Rev. A}\ }\textbf {\bibinfo {volume} {103}},\ \bibinfo {pages}
  {012205} (\bibinfo {year} {2021})}\BibitemShut {NoStop}%
\bibitem [{\citenamefont {Yan}\ \emph {et~al.}(2019)\citenamefont {Yan},
  \citenamefont {Liu}, \citenamefont {Xu}, \citenamefont {Song}, \citenamefont
  {Liu}, \citenamefont {Zhang}, \citenamefont {Deng}, \citenamefont {Yan},
  \citenamefont {Rong}, \citenamefont {Huang}, \citenamefont {Yung},
  \citenamefont {Chen},\ and\ \citenamefont {Yu}}]{YanTongxing.PRL.2019}%
  \BibitemOpen
  \bibfield  {author} {\bibinfo {author} {\bibfnamefont {Tongxing}\
  \bibnamefont {Yan}}, \bibinfo {author} {\bibfnamefont {Bao-Jie}\ \bibnamefont
  {Liu}}, \bibinfo {author} {\bibfnamefont {Kai}\ \bibnamefont {Xu}}, \bibinfo
  {author} {\bibfnamefont {Chao}\ \bibnamefont {Song}}, \bibinfo {author}
  {\bibfnamefont {Song}\ \bibnamefont {Liu}}, \bibinfo {author} {\bibfnamefont
  {Zhensheng}\ \bibnamefont {Zhang}}, \bibinfo {author} {\bibfnamefont {Hui}\
  \bibnamefont {Deng}}, \bibinfo {author} {\bibfnamefont {Zhiguang}\
  \bibnamefont {Yan}}, \bibinfo {author} {\bibfnamefont {Hao}\ \bibnamefont
  {Rong}}, \bibinfo {author} {\bibfnamefont {Keqiang}\ \bibnamefont {Huang}},
  \bibinfo {author} {\bibfnamefont {Man-Hong}\ \bibnamefont {Yung}}, \bibinfo
  {author} {\bibfnamefont {Yuanzhen}\ \bibnamefont {Chen}}, \ and\ \bibinfo
  {author} {\bibfnamefont {Dapeng}\ \bibnamefont {Yu}},\ }\bibfield  {title}
  {\enquote {\bibinfo {title} {Experimental realization of nonadiabatic
  shortcut to non-abelian geometric gates},}\ }\href {\doibase
  10.1103/PhysRevLett.122.080501} {\bibfield  {journal} {\bibinfo  {journal}
  {Phys. Rev. Lett.}\ }\textbf {\bibinfo {volume} {122}},\ \bibinfo {pages}
  {080501} (\bibinfo {year} {2019})}\BibitemShut {NoStop}%
\bibitem [{\citenamefont {Xu}\ \emph {et~al.}(2020)\citenamefont {Xu},
  \citenamefont {Hua}, \citenamefont {Chen}, \citenamefont {Pan}, \citenamefont
  {Li}, \citenamefont {Han}, \citenamefont {Cai}, \citenamefont {Ma},
  \citenamefont {Wang}, \citenamefont {Song}, \citenamefont {Xue},\ and\
  \citenamefont {Sun}}]{SunLuYan.PRL.2020}%
  \BibitemOpen
  \bibfield  {author} {\bibinfo {author} {\bibfnamefont {Y.}~\bibnamefont
  {Xu}}, \bibinfo {author} {\bibfnamefont {Z.}~\bibnamefont {Hua}}, \bibinfo
  {author} {\bibfnamefont {Tao}\ \bibnamefont {Chen}}, \bibinfo {author}
  {\bibfnamefont {X.}~\bibnamefont {Pan}}, \bibinfo {author} {\bibfnamefont
  {X.}~\bibnamefont {Li}}, \bibinfo {author} {\bibfnamefont {J.}~\bibnamefont
  {Han}}, \bibinfo {author} {\bibfnamefont {W.}~\bibnamefont {Cai}}, \bibinfo
  {author} {\bibfnamefont {Y.}~\bibnamefont {Ma}}, \bibinfo {author}
  {\bibfnamefont {H.}~\bibnamefont {Wang}}, \bibinfo {author} {\bibfnamefont
  {Y.~P.}\ \bibnamefont {Song}}, \bibinfo {author} {\bibfnamefont {Zheng-Yuan}\
  \bibnamefont {Xue}}, \ and\ \bibinfo {author} {\bibfnamefont
  {L.}~\bibnamefont {Sun}},\ }\bibfield  {title} {\enquote {\bibinfo {title}
  {Experimental implementation of universal nonadiabatic geometric quantum
  gates in a superconducting circuit},}\ }\href {\doibase
  10.1103/PhysRevLett.124.230503} {\bibfield  {journal} {\bibinfo  {journal}
  {Phys. Rev. Lett.}\ }\textbf {\bibinfo {volume} {124}},\ \bibinfo {pages}
  {230503} (\bibinfo {year} {2020})}\BibitemShut {NoStop}%
\bibitem [{\citenamefont {Duan}\ \emph {et~al.}(2001)\citenamefont {Duan},
  \citenamefont {Cirac},\ and\ \citenamefont {Zoller}}]{Duan.Science.2001}%
  \BibitemOpen
  \bibfield  {author} {\bibinfo {author} {\bibfnamefont {L.-M.}\ \bibnamefont
  {Duan}}, \bibinfo {author} {\bibfnamefont {J.~I.}\ \bibnamefont {Cirac}}, \
  and\ \bibinfo {author} {\bibfnamefont {P.}~\bibnamefont {Zoller}},\
  }\bibfield  {title} {\enquote {\bibinfo {title} {Geometric manipulation of
  trapped ions for quantum computation},}\ }\href {\doibase
  10.1126/science.1058835} {\bibfield  {journal} {\bibinfo  {journal}
  {Science}\ }\textbf {\bibinfo {volume} {292}},\ \bibinfo {pages} {1695--1697}
  (\bibinfo {year} {2001})}\BibitemShut {NoStop}%
\bibitem [{\citenamefont {Ai}\ \emph {et~al.}(2020)\citenamefont {Ai},
  \citenamefont {Li}, \citenamefont {Hou}, \citenamefont {He}, \citenamefont
  {Qian}, \citenamefont {Xue}, \citenamefont {Cui}, \citenamefont {Huang},
  \citenamefont {Li},\ and\ \citenamefont {Guo}}]{AiMingzhong.PRApp.2020}%
  \BibitemOpen
  \bibfield  {author} {\bibinfo {author} {\bibfnamefont {Ming-Zhong}\
  \bibnamefont {Ai}}, \bibinfo {author} {\bibfnamefont {Sai}\ \bibnamefont
  {Li}}, \bibinfo {author} {\bibfnamefont {Zhibo}\ \bibnamefont {Hou}},
  \bibinfo {author} {\bibfnamefont {Ran}\ \bibnamefont {He}}, \bibinfo {author}
  {\bibfnamefont {Zhong-Hua}\ \bibnamefont {Qian}}, \bibinfo {author}
  {\bibfnamefont {Zheng-Yuan}\ \bibnamefont {Xue}}, \bibinfo {author}
  {\bibfnamefont {Jin-Ming}\ \bibnamefont {Cui}}, \bibinfo {author}
  {\bibfnamefont {Yun-Feng}\ \bibnamefont {Huang}}, \bibinfo {author}
  {\bibfnamefont {Chuan-Feng}\ \bibnamefont {Li}}, \ and\ \bibinfo {author}
  {\bibfnamefont {Guang-Can}\ \bibnamefont {Guo}},\ }\bibfield  {title}
  {\enquote {\bibinfo {title} {Experimental realization of nonadiabatic
  holonomic single-qubit quantum gates with optimal control in a trapped
  ion},}\ }\href {\doibase 10.1103/PhysRevApplied.14.054062} {\bibfield
  {journal} {\bibinfo  {journal} {Phys. Rev. Applied}\ }\textbf {\bibinfo
  {volume} {14}},\ \bibinfo {pages} {054062} (\bibinfo {year}
  {2020})}\BibitemShut {NoStop}%
\bibitem [{\citenamefont {Zhou}\ \emph {et~al.}(2017)\citenamefont {Zhou},
  \citenamefont {Jerger}, \citenamefont {Shkolnikov}, \citenamefont {Heremans},
  \citenamefont {Burkard},\ and\ \citenamefont
  {Awschalom}}]{BrianZhou.PRL.2017}%
  \BibitemOpen
  \bibfield  {author} {\bibinfo {author} {\bibfnamefont {Brian~B.}\
  \bibnamefont {Zhou}}, \bibinfo {author} {\bibfnamefont {Paul~C.}\
  \bibnamefont {Jerger}}, \bibinfo {author} {\bibfnamefont {V.~O.}\
  \bibnamefont {Shkolnikov}}, \bibinfo {author} {\bibfnamefont {F.~Joseph}\
  \bibnamefont {Heremans}}, \bibinfo {author} {\bibfnamefont {Guido}\
  \bibnamefont {Burkard}}, \ and\ \bibinfo {author} {\bibfnamefont {David~D.}\
  \bibnamefont {Awschalom}},\ }\bibfield  {title} {\enquote {\bibinfo {title}
  {Holonomic quantum control by coherent optical excitation in diamond},}\
  }\href {\doibase 10.1103/PhysRevLett.119.140503} {\bibfield  {journal}
  {\bibinfo  {journal} {Phys. Rev. Lett.}\ }\textbf {\bibinfo {volume} {119}},\
  \bibinfo {pages} {140503} (\bibinfo {year} {2017})}\BibitemShut {NoStop}%
\bibitem [{\citenamefont {Zhou}\ \emph {et~al.}(2016)\citenamefont {Zhou},
  \citenamefont {Baksic}, \citenamefont {Ribeiro}, \citenamefont {Yale},
  \citenamefont {Heremans}, \citenamefont {Jerger}, \citenamefont {Auer},
  \citenamefont {Burkard}, \citenamefont {Clerk},\ and\ \citenamefont
  {Awschalom}}]{Zhou2016}%
  \BibitemOpen
  \bibfield  {author} {\bibinfo {author} {\bibfnamefont {Brian~B.}\
  \bibnamefont {Zhou}}, \bibinfo {author} {\bibfnamefont {Alexandre}\
  \bibnamefont {Baksic}}, \bibinfo {author} {\bibfnamefont {Hugo}\ \bibnamefont
  {Ribeiro}}, \bibinfo {author} {\bibfnamefont {Christopher}\ \bibnamefont
  {Yale}}, \bibinfo {author} {\bibfnamefont {Joseph}\ \bibnamefont {Heremans}},
  \bibinfo {author} {\bibfnamefont {Paul}\ \bibnamefont {Jerger}}, \bibinfo
  {author} {\bibfnamefont {Adrian}\ \bibnamefont {Auer}}, \bibinfo {author}
  {\bibfnamefont {Guido}\ \bibnamefont {Burkard}}, \bibinfo {author}
  {\bibfnamefont {Aashish~A.}\ \bibnamefont {Clerk}}, \ and\ \bibinfo {author}
  {\bibfnamefont {David~D.}\ \bibnamefont {Awschalom}},\ }\bibfield  {title}
  {\enquote {\bibinfo {title} {Accelerated quantum control using superadiabatic
  dynamics in a solid-state lambda system},}\ }\href
  {https://doi.org/10.1038/nphys3967} {\bibfield  {journal} {\bibinfo
  {journal} {Nature Physics}\ }\textbf {\bibinfo {volume} {13}},\ \bibinfo
  {pages} {330 EP --} (\bibinfo {year} {2016})}\BibitemShut {NoStop}%
\bibitem [{\citenamefont {Sekiguchi}\ \emph {et~al.}(2017)\citenamefont
  {Sekiguchi}, \citenamefont {Niikura}, \citenamefont {Kuroiwa}, \citenamefont
  {Kano},\ and\ \citenamefont {Kosaka}}]{Sekiguchi.2017}%
  \BibitemOpen
  \bibfield  {author} {\bibinfo {author} {\bibfnamefont {Yuhei}\ \bibnamefont
  {Sekiguchi}}, \bibinfo {author} {\bibfnamefont {Naeko}\ \bibnamefont
  {Niikura}}, \bibinfo {author} {\bibfnamefont {Ryota}\ \bibnamefont
  {Kuroiwa}}, \bibinfo {author} {\bibfnamefont {Hiroki}\ \bibnamefont {Kano}},
  \ and\ \bibinfo {author} {\bibfnamefont {Hideo}\ \bibnamefont {Kosaka}},\
  }\bibfield  {title} {\enquote {\bibinfo {title} {Optical holonomic single
  quantum gates with a geometric spin under a zero field},}\ }\href {\doibase
  10.1038/nphoton.2017.40} {\bibfield  {journal} {\bibinfo  {journal} {Nature
  Photonics}\ }\textbf {\bibinfo {volume} {11}},\ \bibinfo {pages} {309--314}
  (\bibinfo {year} {2017})}\BibitemShut {NoStop}%
\bibitem [{\citenamefont {Goelman}\ \emph {et~al.}(1989)\citenamefont
  {Goelman}, \citenamefont {Vega},\ and\ \citenamefont {Zax}}]{Goelman_JMR89}%
  \BibitemOpen
  \bibfield  {author} {\bibinfo {author} {\bibfnamefont {G.}~\bibnamefont
  {Goelman}}, \bibinfo {author} {\bibfnamefont {S.}~\bibnamefont {Vega}}, \
  and\ \bibinfo {author} {\bibfnamefont {D.~B.}\ \bibnamefont {Zax}},\
  }\bibfield  {title} {\enquote {\bibinfo {title} {Squared amplitude-modulated
  composite pulses},}\ }\href
  {https://www.sciencedirect.com/science/article/pii/0022236489900772}
  {\bibfield  {journal} {\bibinfo  {journal} {J.\ Magn.\ Reson.}\ }\textbf
  {\bibinfo {volume} {81}},\ \bibinfo {pages} {423} (\bibinfo {year}
  {1989})}\BibitemShut {NoStop}%
\bibitem [{\citenamefont {Viola}\ and\ \citenamefont
  {Lloyd}(1998)}]{Viola.PRL.1998}%
  \BibitemOpen
  \bibfield  {author} {\bibinfo {author} {\bibfnamefont {Lorenza}\ \bibnamefont
  {Viola}}\ and\ \bibinfo {author} {\bibfnamefont {Seth}\ \bibnamefont
  {Lloyd}},\ }\bibfield  {title} {\enquote {\bibinfo {title} {Dynamical
  suppression of decoherence in two-state quantum systems},}\ }\href {\doibase
  10.1103/PhysRevA.58.2733} {\bibfield  {journal} {\bibinfo  {journal} {Phys.
  Rev. A}\ }\textbf {\bibinfo {volume} {58}},\ \bibinfo {pages} {2733--2744}
  (\bibinfo {year} {1998})}\BibitemShut {NoStop}%
\bibitem [{\citenamefont {Biercuk}\ \emph {et~al.}(2009)\citenamefont
  {Biercuk}, \citenamefont {Uys}, \citenamefont {VanDevender}, \citenamefont
  {Shiga}, \citenamefont {Itano},\ and\ \citenamefont
  {Bollinger}}]{Biercuk_Nature09}%
  \BibitemOpen
  \bibfield  {author} {\bibinfo {author} {\bibfnamefont {Michael~J.}\
  \bibnamefont {Biercuk}}, \bibinfo {author} {\bibfnamefont {Hermann}\
  \bibnamefont {Uys}}, \bibinfo {author} {\bibfnamefont {Aaron~P.}\
  \bibnamefont {VanDevender}}, \bibinfo {author} {\bibfnamefont {Nobuyasu}\
  \bibnamefont {Shiga}}, \bibinfo {author} {\bibfnamefont {Wayne~M.}\
  \bibnamefont {Itano}}, \ and\ \bibinfo {author} {\bibfnamefont {John~J.}\
  \bibnamefont {Bollinger}},\ }\bibfield  {title} {\enquote {\bibinfo {title}
  {Optimized dynamical decoupling in a model quantum memory},}\ }\href
  {https://www.nature.com/articles/nature07951} {\bibfield  {journal} {\bibinfo
   {journal} {Nature}\ }\textbf {\bibinfo {volume} {458}},\ \bibinfo {pages}
  {996} (\bibinfo {year} {2009})}\BibitemShut {NoStop}%
\bibitem [{\citenamefont {Khodjasteh}\ \emph {et~al.}(2010)\citenamefont
  {Khodjasteh}, \citenamefont {Lidar},\ and\ \citenamefont
  {Viola}}]{Khodjasteh.PRL.2010}%
  \BibitemOpen
  \bibfield  {author} {\bibinfo {author} {\bibfnamefont {Kaveh}\ \bibnamefont
  {Khodjasteh}}, \bibinfo {author} {\bibfnamefont {Daniel~A.}\ \bibnamefont
  {Lidar}}, \ and\ \bibinfo {author} {\bibfnamefont {Lorenza}\ \bibnamefont
  {Viola}},\ }\bibfield  {title} {\enquote {\bibinfo {title} {Arbitrarily
  accurate dynamical control in open quantum systems},}\ }\href {\doibase
  10.1103/PhysRevLett.104.090501} {\bibfield  {journal} {\bibinfo  {journal}
  {Phys. Rev. Lett.}\ }\textbf {\bibinfo {volume} {104}},\ \bibinfo {pages}
  {090501} (\bibinfo {year} {2010})}\BibitemShut {NoStop}%
\bibitem [{\citenamefont {Barnes}\ \emph {et~al.}(2015)\citenamefont {Barnes},
  \citenamefont {Wang},\ and\ \citenamefont {Das~Sarma}}]{Barnes.SciRep.2015}%
  \BibitemOpen
  \bibfield  {author} {\bibinfo {author} {\bibfnamefont {Edwin}\ \bibnamefont
  {Barnes}}, \bibinfo {author} {\bibfnamefont {Xin}\ \bibnamefont {Wang}}, \
  and\ \bibinfo {author} {\bibfnamefont {S.}~\bibnamefont {Das~Sarma}},\
  }\bibfield  {title} {\enquote {\bibinfo {title} {Robust quantum control using
  smooth pulses and topological winding},}\ }\href {\doibase 10.1038/srep12685}
  {\bibfield  {journal} {\bibinfo  {journal} {Scientific Reports}\ }\textbf
  {\bibinfo {volume} {5}},\ \bibinfo {pages} {12685} (\bibinfo {year}
  {2015})}\BibitemShut {NoStop}%
\bibitem [{\citenamefont {Wang}\ \emph {et~al.}(2012)\citenamefont {Wang},
  \citenamefont {Bishop}, \citenamefont {Kestner}, \citenamefont {Barnes},
  \citenamefont {Sun},\ and\ \citenamefont {Das~Sarma}}]{Wang.NatCom.2012}%
  \BibitemOpen
  \bibfield  {author} {\bibinfo {author} {\bibfnamefont {Xin}\ \bibnamefont
  {Wang}}, \bibinfo {author} {\bibfnamefont {Lev~S.}\ \bibnamefont {Bishop}},
  \bibinfo {author} {\bibfnamefont {J.~P.}\ \bibnamefont {Kestner}}, \bibinfo
  {author} {\bibfnamefont {Edwin}\ \bibnamefont {Barnes}}, \bibinfo {author}
  {\bibfnamefont {Kai}\ \bibnamefont {Sun}}, \ and\ \bibinfo {author}
  {\bibfnamefont {S.}~\bibnamefont {Das~Sarma}},\ }\bibfield  {title} {\enquote
  {\bibinfo {title} {Composite pulses for robust universal control of
  singlet--triplet qubits},}\ }\href {\doibase 10.1038/ncomms2003} {\bibfield
  {journal} {\bibinfo  {journal} {Nature Communications}\ }\textbf {\bibinfo
  {volume} {3}},\ \bibinfo {pages} {997} (\bibinfo {year} {2012})}\BibitemShut
  {NoStop}%
\bibitem [{\citenamefont {Kestner}\ \emph {et~al.}(2013)\citenamefont
  {Kestner}, \citenamefont {Wang}, \citenamefont {Bishop}, \citenamefont
  {Barnes},\ and\ \citenamefont {Das~Sarma}}]{Kestner.PRL.2013}%
  \BibitemOpen
  \bibfield  {author} {\bibinfo {author} {\bibfnamefont {J.~P.}\ \bibnamefont
  {Kestner}}, \bibinfo {author} {\bibfnamefont {Xin}\ \bibnamefont {Wang}},
  \bibinfo {author} {\bibfnamefont {Lev~S.}\ \bibnamefont {Bishop}}, \bibinfo
  {author} {\bibfnamefont {Edwin}\ \bibnamefont {Barnes}}, \ and\ \bibinfo
  {author} {\bibfnamefont {S.}~\bibnamefont {Das~Sarma}},\ }\bibfield  {title}
  {\enquote {\bibinfo {title} {Noise-resistant control for a spin qubit
  array},}\ }\href {\doibase 10.1103/PhysRevLett.110.140502} {\bibfield
  {journal} {\bibinfo  {journal} {Phys. Rev. Lett.}\ }\textbf {\bibinfo
  {volume} {110}},\ \bibinfo {pages} {140502} (\bibinfo {year}
  {2013})}\BibitemShut {NoStop}%
\bibitem [{\citenamefont {van~der Sar}\ \emph {et~al.}(2012)\citenamefont
  {van~der Sar}, \citenamefont {Wang}, \citenamefont {Blok}, \citenamefont
  {Bernien}, \citenamefont {Taminiau}, \citenamefont {Toyli}, \citenamefont
  {Lidar}, \citenamefont {Awschalom}, \citenamefont {Hanson},\ and\
  \citenamefont {Dobrovitski}}]{vanderSar_Nature12}%
  \BibitemOpen
  \bibfield  {author} {\bibinfo {author} {\bibfnamefont {T.}~\bibnamefont
  {van~der Sar}}, \bibinfo {author} {\bibfnamefont {Z.~H.}\ \bibnamefont
  {Wang}}, \bibinfo {author} {\bibfnamefont {M.~S.}\ \bibnamefont {Blok}},
  \bibinfo {author} {\bibfnamefont {H.}~\bibnamefont {Bernien}}, \bibinfo
  {author} {\bibfnamefont {T.~H.}\ \bibnamefont {Taminiau}}, \bibinfo {author}
  {\bibfnamefont {D.M.}\ \bibnamefont {Toyli}}, \bibinfo {author}
  {\bibfnamefont {D.~A.}\ \bibnamefont {Lidar}}, \bibinfo {author}
  {\bibfnamefont {D.~D.}\ \bibnamefont {Awschalom}}, \bibinfo {author}
  {\bibfnamefont {R.}~\bibnamefont {Hanson}}, \ and\ \bibinfo {author}
  {\bibfnamefont {V.~V.}\ \bibnamefont {Dobrovitski}},\ }\bibfield  {title}
  {\enquote {\bibinfo {title} {Decoherence-protected quantum gates for a hybrid
  solid-state spin register},}\ }\href
  {https://www.nature.com/articles/nature10900} {\bibfield  {journal} {\bibinfo
   {journal} {Nature}\ }\textbf {\bibinfo {volume} {484}},\ \bibinfo {pages}
  {82--86} (\bibinfo {year} {2012})}\BibitemShut {NoStop}%
\bibitem [{\citenamefont {Green}\ \emph {et~al.}(2013)\citenamefont {Green},
  \citenamefont {Sastrawan}, \citenamefont {Uys},\ and\ \citenamefont
  {Biercuk}}]{Green_NJP13}%
  \BibitemOpen
  \bibfield  {author} {\bibinfo {author} {\bibfnamefont {Todd~J}\ \bibnamefont
  {Green}}, \bibinfo {author} {\bibfnamefont {Jarrah}\ \bibnamefont
  {Sastrawan}}, \bibinfo {author} {\bibfnamefont {Hermann}\ \bibnamefont
  {Uys}}, \ and\ \bibinfo {author} {\bibfnamefont {Michael~J}\ \bibnamefont
  {Biercuk}},\ }\bibfield  {title} {\enquote {\bibinfo {title} {Arbitrary
  quantum control of qubits in the presence of universal noise},}\ }\href
  {http://stacks.iop.org/1367-2630/15/i=9/a=095004} {\bibfield  {journal}
  {\bibinfo  {journal} {New Journal of Physics}\ }\textbf {\bibinfo {volume}
  {15}},\ \bibinfo {pages} {095004} (\bibinfo {year} {2013})}\BibitemShut
  {NoStop}%
\bibitem [{\citenamefont {Merrill}\ and\ \citenamefont
  {Brown}(2014)}]{Merrill_Wiley14}%
  \BibitemOpen
  \bibfield  {author} {\bibinfo {author} {\bibfnamefont {J.~T.}\ \bibnamefont
  {Merrill}}\ and\ \bibinfo {author} {\bibfnamefont {K.~R.}\ \bibnamefont
  {Brown}},\ }\bibfield  {title} {\enquote {\bibinfo {title} {Progress in
  compensating pulse sequences for quantum computation, in {\it {quantum}
  information and computation for chemistry: {advances} in chemical
  physics}},}\ }\href {url =
  {https://onlinelibrary.wiley.com/doi/abs/10.1002/9781118742631.ch10},}
  {\bibfield  {journal} {\bibinfo  {journal} {vol. 154 (ed. S. Kais), John
  Wiley \& Sons, Inc.}\ } (\bibinfo {year} {2014})}\BibitemShut {NoStop}%
\bibitem [{\citenamefont {Calderon-Vargas}\ and\ \citenamefont
  {Kestner}(2017)}]{CalderonVargasPRL2017}%
  \BibitemOpen
  \bibfield  {author} {\bibinfo {author} {\bibfnamefont {F.~A.}\ \bibnamefont
  {Calderon-Vargas}}\ and\ \bibinfo {author} {\bibfnamefont {J.~P.}\
  \bibnamefont {Kestner}},\ }\bibfield  {title} {\enquote {\bibinfo {title}
  {Dynamically correcting a $\mathrm{CNOT}$ gate for any systematic logical
  error},}\ }\href {\doibase 10.1103/PhysRevLett.118.150502} {\bibfield
  {journal} {\bibinfo  {journal} {Phys. Rev. Lett.}\ }\textbf {\bibinfo
  {volume} {118}},\ \bibinfo {pages} {150502} (\bibinfo {year}
  {2017})}\BibitemShut {NoStop}%
\bibitem [{\citenamefont {Buterakos}\ \emph {et~al.}(2018)\citenamefont
  {Buterakos}, \citenamefont {Throckmorton},\ and\ \citenamefont
  {Das~Sarma}}]{ButerakosPRB2018}%
  \BibitemOpen
  \bibfield  {author} {\bibinfo {author} {\bibfnamefont {Donovan}\ \bibnamefont
  {Buterakos}}, \bibinfo {author} {\bibfnamefont {Robert~E.}\ \bibnamefont
  {Throckmorton}}, \ and\ \bibinfo {author} {\bibfnamefont {S.}~\bibnamefont
  {Das~Sarma}},\ }\bibfield  {title} {\enquote {\bibinfo {title} {Crosstalk
  error correction through dynamical decoupling of single-qubit gates in
  capacitively coupled singlet-triplet semiconductor spin qubits},}\ }\href
  {\doibase 10.1103/PhysRevB.97.045431} {\bibfield  {journal} {\bibinfo
  {journal} {Phys. Rev. B}\ }\textbf {\bibinfo {volume} {97}},\ \bibinfo
  {pages} {045431} (\bibinfo {year} {2018})}\BibitemShut {NoStop}%
\bibitem [{\citenamefont {G\"ung\"ord\"u}\ and\ \citenamefont
  {Kestner}(2018)}]{GungorduPRB2018}%
  \BibitemOpen
  \bibfield  {author} {\bibinfo {author} {\bibfnamefont {Utkan}\ \bibnamefont
  {G\"ung\"ord\"u}}\ and\ \bibinfo {author} {\bibfnamefont {J.~P.}\
  \bibnamefont {Kestner}},\ }\bibfield  {title} {\enquote {\bibinfo {title}
  {Pulse sequence designed for robust $c$-phase gates in simos and si/sige
  double quantum dots},}\ }\href {\doibase 10.1103/PhysRevB.98.165301}
  {\bibfield  {journal} {\bibinfo  {journal} {Phys. Rev. B}\ }\textbf {\bibinfo
  {volume} {98}},\ \bibinfo {pages} {165301} (\bibinfo {year}
  {2018})}\BibitemShut {NoStop}%
\bibitem [{\citenamefont {Zeng}\ \emph {et~al.}(2019)\citenamefont {Zeng},
  \citenamefont {Yang}, \citenamefont {Dzurak},\ and\ \citenamefont
  {Barnes}}]{Zeng.PRA.2019}%
  \BibitemOpen
  \bibfield  {author} {\bibinfo {author} {\bibfnamefont {Junkai}\ \bibnamefont
  {Zeng}}, \bibinfo {author} {\bibfnamefont {C.~H.}\ \bibnamefont {Yang}},
  \bibinfo {author} {\bibfnamefont {A.~S.}\ \bibnamefont {Dzurak}}, \ and\
  \bibinfo {author} {\bibfnamefont {Edwin}\ \bibnamefont {Barnes}},\ }\bibfield
   {title} {\enquote {\bibinfo {title} {Geometric formalism for constructing
  arbitrary single-qubit dynamically corrected gates},}\ }\href {\doibase
  10.1103/PhysRevA.99.052321} {\bibfield  {journal} {\bibinfo  {journal} {Phys.
  Rev. A}\ }\textbf {\bibinfo {volume} {99}},\ \bibinfo {pages} {052321}
  (\bibinfo {year} {2019})}\BibitemShut {NoStop}%
\bibitem [{\citenamefont {Kanaar}\ \emph {et~al.}(2021)\citenamefont {Kanaar},
  \citenamefont {Wolin}, \citenamefont {G\"ung\"ord\"u},\ and\ \citenamefont
  {Kestner}}]{Kanaar.PRB.2021}%
  \BibitemOpen
  \bibfield  {author} {\bibinfo {author} {\bibfnamefont {David~W.}\
  \bibnamefont {Kanaar}}, \bibinfo {author} {\bibfnamefont {Sidney}\
  \bibnamefont {Wolin}}, \bibinfo {author} {\bibfnamefont {Utkan}\ \bibnamefont
  {G\"ung\"ord\"u}}, \ and\ \bibinfo {author} {\bibfnamefont {J.~P.}\
  \bibnamefont {Kestner}},\ }\bibfield  {title} {\enquote {\bibinfo {title}
  {Single-tone pulse sequences and robust two-tone shaped pulses for three
  silicon spin qubits with always-on exchange},}\ }\href {\doibase
  10.1103/PhysRevB.103.235314} {\bibfield  {journal} {\bibinfo  {journal}
  {Phys. Rev. B}\ }\textbf {\bibinfo {volume} {103}},\ \bibinfo {pages}
  {235314} (\bibinfo {year} {2021})}\BibitemShut {NoStop}%
\bibitem [{\citenamefont {Zeng}\ \emph {et~al.}(2018)\citenamefont {Zeng},
  \citenamefont {Deng}, \citenamefont {Russo},\ and\ \citenamefont
  {Barnes}}]{Zeng.NJP.2018}%
  \BibitemOpen
  \bibfield  {author} {\bibinfo {author} {\bibfnamefont {Junkai}\ \bibnamefont
  {Zeng}}, \bibinfo {author} {\bibfnamefont {Xiu-Hao}\ \bibnamefont {Deng}},
  \bibinfo {author} {\bibfnamefont {Antonio}\ \bibnamefont {Russo}}, \ and\
  \bibinfo {author} {\bibfnamefont {Edwin}\ \bibnamefont {Barnes}},\ }\bibfield
   {title} {\enquote {\bibinfo {title} {General solution to inhomogeneous
  dephasing and smooth pulse dynamical decoupling},}\ }\href {\doibase
  10.1088/1367-2630/aaafe9} {\bibfield  {journal} {\bibinfo  {journal} {New
  Journal of Physics}\ }\textbf {\bibinfo {volume} {20}},\ \bibinfo {pages}
  {033011} (\bibinfo {year} {2018})}\BibitemShut {NoStop}%
\bibitem [{\citenamefont {Zeng}\ and\ \citenamefont
  {Barnes}(2018)}]{ZengPRA2018}%
  \BibitemOpen
  \bibfield  {author} {\bibinfo {author} {\bibfnamefont {Junkai}\ \bibnamefont
  {Zeng}}\ and\ \bibinfo {author} {\bibfnamefont {Edwin}\ \bibnamefont
  {Barnes}},\ }\bibfield  {title} {\enquote {\bibinfo {title} {Fastest pulses
  that implement dynamically corrected single-qubit phase gates},}\ }\href
  {\doibase 10.1103/PhysRevA.98.012301} {\bibfield  {journal} {\bibinfo
  {journal} {Phys. Rev. A}\ }\textbf {\bibinfo {volume} {98}},\ \bibinfo
  {pages} {012301} (\bibinfo {year} {2018})}\BibitemShut {NoStop}%
\bibitem [{\citenamefont {Buterakos}\ \emph {et~al.}(2021)\citenamefont
  {Buterakos}, \citenamefont {Das~Sarma},\ and\ \citenamefont
  {Barnes}}]{Buterakos.PRXQ.2021}%
  \BibitemOpen
  \bibfield  {author} {\bibinfo {author} {\bibfnamefont {Donovan}\ \bibnamefont
  {Buterakos}}, \bibinfo {author} {\bibfnamefont {Sankar}\ \bibnamefont
  {Das~Sarma}}, \ and\ \bibinfo {author} {\bibfnamefont {Edwin}\ \bibnamefont
  {Barnes}},\ }\bibfield  {title} {\enquote {\bibinfo {title} {Geometrical
  formalism for dynamically corrected gates in multiqubit systems},}\ }\href
  {\doibase 10.1103/PRXQuantum.2.010341} {\bibfield  {journal} {\bibinfo
  {journal} {PRX Quantum}\ }\textbf {\bibinfo {volume} {2}},\ \bibinfo {pages}
  {010341} (\bibinfo {year} {2021})}\BibitemShut {NoStop}%
\bibitem [{\citenamefont {Throckmorton}\ and\ \citenamefont
  {Das~Sarma}(2019)}]{Throckmorton2019}%
  \BibitemOpen
  \bibfield  {author} {\bibinfo {author} {\bibfnamefont {Robert~E.}\
  \bibnamefont {Throckmorton}}\ and\ \bibinfo {author} {\bibfnamefont
  {S.}~\bibnamefont {Das~Sarma}},\ }\bibfield  {title} {\enquote {\bibinfo
  {title} {Conditions allowing error correction in driven qubits},}\ }\href
  {\doibase 10.1103/PhysRevB.99.045422} {\bibfield  {journal} {\bibinfo
  {journal} {Phys. Rev. B}\ }\textbf {\bibinfo {volume} {99}},\ \bibinfo
  {pages} {045422} (\bibinfo {year} {2019})}\BibitemShut {NoStop}%
\bibitem [{\citenamefont {G\"ung\"ord\"u}\ and\ \citenamefont
  {Kestner}(2019)}]{Gungordu2019}%
  \BibitemOpen
  \bibfield  {author} {\bibinfo {author} {\bibfnamefont {Utkan}\ \bibnamefont
  {G\"ung\"ord\"u}}\ and\ \bibinfo {author} {\bibfnamefont {J.~P.}\
  \bibnamefont {Kestner}},\ }\bibfield  {title} {\enquote {\bibinfo {title}
  {Analytically parametrized solutions for robust quantum control using smooth
  pulses},}\ }\href {\doibase 10.1103/PhysRevA.100.062310} {\bibfield
  {journal} {\bibinfo  {journal} {Phys. Rev. A}\ }\textbf {\bibinfo {volume}
  {100}},\ \bibinfo {pages} {062310} (\bibinfo {year} {2019})}\BibitemShut
  {NoStop}%
\bibitem [{\citenamefont {Samuel}\ and\ \citenamefont
  {Bhandari}(1988)}]{Samuel.Bhandari.PRL.1988}%
  \BibitemOpen
  \bibfield  {author} {\bibinfo {author} {\bibfnamefont {Joseph}\ \bibnamefont
  {Samuel}}\ and\ \bibinfo {author} {\bibfnamefont {Rajendra}\ \bibnamefont
  {Bhandari}},\ }\bibfield  {title} {\enquote {\bibinfo {title} {General
  setting for berry's phase},}\ }\href {\doibase 10.1103/PhysRevLett.60.2339}
  {\bibfield  {journal} {\bibinfo  {journal} {Phys. Rev. Lett.}\ }\textbf
  {\bibinfo {volume} {60}},\ \bibinfo {pages} {2339--2342} (\bibinfo {year}
  {1988})}\BibitemShut {NoStop}%
\bibitem [{\citenamefont {Pachos}\ and\ \citenamefont
  {Zanardi}(2001)}]{Pachos.IJMPB.2001}%
  \BibitemOpen
  \bibfield  {author} {\bibinfo {author} {\bibfnamefont {Jiannis}\ \bibnamefont
  {Pachos}}\ and\ \bibinfo {author} {\bibfnamefont {Paolo}\ \bibnamefont
  {Zanardi}},\ }\bibfield  {title} {\enquote {\bibinfo {title} {Quantum
  holonomies for quantum computing},}\ }\href {\doibase
  10.1142/S0217979201004836} {\bibfield  {journal} {\bibinfo  {journal}
  {International Journal of Modern Physics B}\ }\textbf {\bibinfo {volume}
  {15}},\ \bibinfo {pages} {1257--1285} (\bibinfo {year} {2001})}\BibitemShut
  {NoStop}%
\bibitem [{\citenamefont {Zhu}\ and\ \citenamefont
  {Zanardi}(2005)}]{ZhuShiLiang.PRA.2005}%
  \BibitemOpen
  \bibfield  {author} {\bibinfo {author} {\bibfnamefont {Shi-Liang}\
  \bibnamefont {Zhu}}\ and\ \bibinfo {author} {\bibfnamefont {Paolo}\
  \bibnamefont {Zanardi}},\ }\bibfield  {title} {\enquote {\bibinfo {title}
  {Geometric quantum gates that are robust against stochastic control
  errors},}\ }\href {\doibase 10.1103/PhysRevA.72.020301} {\bibfield  {journal}
  {\bibinfo  {journal} {Phys. Rev. A}\ }\textbf {\bibinfo {volume} {72}},\
  \bibinfo {pages} {020301} (\bibinfo {year} {2005})}\BibitemShut {NoStop}%
\bibitem [{\citenamefont {Zhao}\ \emph {et~al.}(2020)\citenamefont {Zhao},
  \citenamefont {Li}, \citenamefont {Xu},\ and\ \citenamefont
  {Tong}}]{Zhao.PZ.PRA.2020}%
  \BibitemOpen
  \bibfield  {author} {\bibinfo {author} {\bibfnamefont {P.~Z.}\ \bibnamefont
  {Zhao}}, \bibinfo {author} {\bibfnamefont {K.~Z.}\ \bibnamefont {Li}},
  \bibinfo {author} {\bibfnamefont {G.~F.}\ \bibnamefont {Xu}}, \ and\ \bibinfo
  {author} {\bibfnamefont {D.~M.}\ \bibnamefont {Tong}},\ }\bibfield  {title}
  {\enquote {\bibinfo {title} {General approach for constructing hamiltonians
  for nonadiabatic holonomic quantum computation},}\ }\href {\doibase
  10.1103/PhysRevA.101.062306} {\bibfield  {journal} {\bibinfo  {journal}
  {Phys. Rev. A}\ }\textbf {\bibinfo {volume} {101}},\ \bibinfo {pages}
  {062306} (\bibinfo {year} {2020})}\BibitemShut {NoStop}%
\bibitem [{\citenamefont {Li}\ \emph {et~al.}(2020)\citenamefont {Li},
  \citenamefont {Zhao},\ and\ \citenamefont {Tong}}]{LiKZ.PRR.2020}%
  \BibitemOpen
  \bibfield  {author} {\bibinfo {author} {\bibfnamefont {K.~Z.}\ \bibnamefont
  {Li}}, \bibinfo {author} {\bibfnamefont {P.~Z.}\ \bibnamefont {Zhao}}, \ and\
  \bibinfo {author} {\bibfnamefont {D.~M.}\ \bibnamefont {Tong}},\ }\bibfield
  {title} {\enquote {\bibinfo {title} {Approach to realizing nonadiabatic
  geometric gates with prescribed evolution paths},}\ }\href {\doibase
  10.1103/PhysRevResearch.2.023295} {\bibfield  {journal} {\bibinfo  {journal}
  {Phys. Rev. Research}\ }\textbf {\bibinfo {volume} {2}},\ \bibinfo {pages}
  {023295} (\bibinfo {year} {2020})}\BibitemShut {NoStop}%
\bibitem [{\citenamefont {Kwiat}\ and\ \citenamefont
  {Chiao}(1991)}]{KwiatPRL1991}%
  \BibitemOpen
  \bibfield  {author} {\bibinfo {author} {\bibfnamefont {Paul~G.}\ \bibnamefont
  {Kwiat}}\ and\ \bibinfo {author} {\bibfnamefont {Raymond~Y.}\ \bibnamefont
  {Chiao}},\ }\bibfield  {title} {\enquote {\bibinfo {title} {Observation of a
  nonclassical berry's phase for the photon},}\ }\href {\doibase
  10.1103/PhysRevLett.66.588} {\bibfield  {journal} {\bibinfo  {journal} {Phys.
  Rev. Lett.}\ }\textbf {\bibinfo {volume} {66}},\ \bibinfo {pages} {588--591}
  (\bibinfo {year} {1991})}\BibitemShut {NoStop}%
\bibitem [{\citenamefont {Mousolou}\ and\ \citenamefont
  {Sj\"oqvist}(2014)}]{Mousolou.PRA.2014}%
  \BibitemOpen
  \bibfield  {author} {\bibinfo {author} {\bibfnamefont {Vahid~Azimi}\
  \bibnamefont {Mousolou}}\ and\ \bibinfo {author} {\bibfnamefont {Erik}\
  \bibnamefont {Sj\"oqvist}},\ }\bibfield  {title} {\enquote {\bibinfo {title}
  {Non-abelian geometric phases in a system of coupled quantum bits},}\ }\href
  {\doibase 10.1103/PhysRevA.89.022117} {\bibfield  {journal} {\bibinfo
  {journal} {Phys. Rev. A}\ }\textbf {\bibinfo {volume} {89}},\ \bibinfo
  {pages} {022117} (\bibinfo {year} {2014})}\BibitemShut {NoStop}%
\bibitem [{\citenamefont {Bluhm}\ \emph {et~al.}(2010)\citenamefont {Bluhm},
  \citenamefont {Foletti}, \citenamefont {Mahalu}, \citenamefont {Umansky},\
  and\ \citenamefont {Yacoby}}]{BluhmPRL2010}%
  \BibitemOpen
  \bibfield  {author} {\bibinfo {author} {\bibfnamefont {Hendrik}\ \bibnamefont
  {Bluhm}}, \bibinfo {author} {\bibfnamefont {Sandra}\ \bibnamefont {Foletti}},
  \bibinfo {author} {\bibfnamefont {Diana}\ \bibnamefont {Mahalu}}, \bibinfo
  {author} {\bibfnamefont {Vladimir}\ \bibnamefont {Umansky}}, \ and\ \bibinfo
  {author} {\bibfnamefont {Amir}\ \bibnamefont {Yacoby}},\ }\bibfield  {title}
  {\enquote {\bibinfo {title} {Enhancing the coherence of a spin qubit by
  operating it as a feedback loop that controls its nuclear spin bath},}\
  }\href {\doibase 10.1103/PhysRevLett.105.216803} {\bibfield  {journal}
  {\bibinfo  {journal} {Phys. Rev. Lett.}\ }\textbf {\bibinfo {volume} {105}},\
  \bibinfo {pages} {216803} (\bibinfo {year} {2010})}\BibitemShut {NoStop}%
\bibitem [{\citenamefont {Bylander}\ \emph {et~al.}(2011)\citenamefont
  {Bylander}, \citenamefont {Gustavsson}, \citenamefont {Yan}, \citenamefont
  {Yoshihara}, \citenamefont {Harrabi}, \citenamefont {Fitch}, \citenamefont
  {Cory}, \citenamefont {Nakamura}, \citenamefont {Tsai},\ and\ \citenamefont
  {Oliver}}]{BylanderNaturePhysics2011}%
  \BibitemOpen
  \bibfield  {author} {\bibinfo {author} {\bibfnamefont {Jonas}\ \bibnamefont
  {Bylander}}, \bibinfo {author} {\bibfnamefont {Simon}\ \bibnamefont
  {Gustavsson}}, \bibinfo {author} {\bibfnamefont {Fei}\ \bibnamefont {Yan}},
  \bibinfo {author} {\bibfnamefont {Fumiki}\ \bibnamefont {Yoshihara}},
  \bibinfo {author} {\bibfnamefont {Khalil}\ \bibnamefont {Harrabi}}, \bibinfo
  {author} {\bibfnamefont {George}\ \bibnamefont {Fitch}}, \bibinfo {author}
  {\bibfnamefont {David~G.}\ \bibnamefont {Cory}}, \bibinfo {author}
  {\bibfnamefont {Yasunobu}\ \bibnamefont {Nakamura}}, \bibinfo {author}
  {\bibfnamefont {Jaw-Shen}\ \bibnamefont {Tsai}}, \ and\ \bibinfo {author}
  {\bibfnamefont {William~D.}\ \bibnamefont {Oliver}},\ }\bibfield  {title}
  {\enquote {\bibinfo {title} {Noise spectroscopy through dynamical decoupling
  with a superconducting flux qubit},}\ }\href {\doibase 10.1038/nphys1994}
  {\bibfield  {journal} {\bibinfo  {journal} {Nature Physics}\ }\textbf
  {\bibinfo {volume} {7}},\ \bibinfo {pages} {565--570} (\bibinfo {year}
  {2011})}\BibitemShut {NoStop}%
\bibitem [{\citenamefont {Martins}\ \emph {et~al.}(2016)\citenamefont
  {Martins}, \citenamefont {Malinowski}, \citenamefont {Nissen}, \citenamefont
  {Barnes}, \citenamefont {Fallahi}, \citenamefont {Gardner}, \citenamefont
  {Manfra}, \citenamefont {Marcus},\ and\ \citenamefont
  {Kuemmeth}}]{MartinsPRL2016}%
  \BibitemOpen
  \bibfield  {author} {\bibinfo {author} {\bibfnamefont {Frederico}\
  \bibnamefont {Martins}}, \bibinfo {author} {\bibfnamefont {Filip~K.}\
  \bibnamefont {Malinowski}}, \bibinfo {author} {\bibfnamefont {Peter~D.}\
  \bibnamefont {Nissen}}, \bibinfo {author} {\bibfnamefont {Edwin}\
  \bibnamefont {Barnes}}, \bibinfo {author} {\bibfnamefont {Saeed}\
  \bibnamefont {Fallahi}}, \bibinfo {author} {\bibfnamefont {Geoffrey~C.}\
  \bibnamefont {Gardner}}, \bibinfo {author} {\bibfnamefont {Michael~J.}\
  \bibnamefont {Manfra}}, \bibinfo {author} {\bibfnamefont {Charles~M.}\
  \bibnamefont {Marcus}}, \ and\ \bibinfo {author} {\bibfnamefont {Ferdinand}\
  \bibnamefont {Kuemmeth}},\ }\bibfield  {title} {\enquote {\bibinfo {title}
  {Noise suppression using symmetric exchange gates in spin qubits},}\ }\href
  {\doibase 10.1103/PhysRevLett.116.116801} {\bibfield  {journal} {\bibinfo
  {journal} {Phys. Rev. Lett.}\ }\textbf {\bibinfo {volume} {116}},\ \bibinfo
  {pages} {116801} (\bibinfo {year} {2016})}\BibitemShut {NoStop}%
\bibitem [{\citenamefont {Klimov}\ \emph {et~al.}(2018)\citenamefont {Klimov},
  \citenamefont {Kelly}, \citenamefont {Chen}, \citenamefont {Neeley},
  \citenamefont {Megrant}, \citenamefont {Burkett}, \citenamefont {Barends},
  \citenamefont {Arya}, \citenamefont {Chiaro}, \citenamefont {Chen},
  \citenamefont {Dunsworth}, \citenamefont {Fowler}, \citenamefont {Foxen},
  \citenamefont {Gidney}, \citenamefont {Giustina}, \citenamefont {Graff},
  \citenamefont {Huang}, \citenamefont {Jeffrey}, \citenamefont {Lucero},
  \citenamefont {Mutus}, \citenamefont {Naaman}, \citenamefont {Neill},
  \citenamefont {Quintana}, \citenamefont {Roushan}, \citenamefont {Sank},
  \citenamefont {Vainsencher}, \citenamefont {Wenner}, \citenamefont {White},
  \citenamefont {Boixo}, \citenamefont {Babbush}, \citenamefont {Smelyanskiy},
  \citenamefont {Neven},\ and\ \citenamefont {Martinis}}]{KlimovPRL2018}%
  \BibitemOpen
  \bibfield  {author} {\bibinfo {author} {\bibfnamefont {P.~V.}\ \bibnamefont
  {Klimov}}, \bibinfo {author} {\bibfnamefont {J.}~\bibnamefont {Kelly}},
  \bibinfo {author} {\bibfnamefont {Z.}~\bibnamefont {Chen}}, \bibinfo {author}
  {\bibfnamefont {M.}~\bibnamefont {Neeley}}, \bibinfo {author} {\bibfnamefont
  {A.}~\bibnamefont {Megrant}}, \bibinfo {author} {\bibfnamefont
  {B.}~\bibnamefont {Burkett}}, \bibinfo {author} {\bibfnamefont
  {R.}~\bibnamefont {Barends}}, \bibinfo {author} {\bibfnamefont
  {K.}~\bibnamefont {Arya}}, \bibinfo {author} {\bibfnamefont {B.}~\bibnamefont
  {Chiaro}}, \bibinfo {author} {\bibfnamefont {Yu}~\bibnamefont {Chen}},
  \bibinfo {author} {\bibfnamefont {A.}~\bibnamefont {Dunsworth}}, \bibinfo
  {author} {\bibfnamefont {A.}~\bibnamefont {Fowler}}, \bibinfo {author}
  {\bibfnamefont {B.}~\bibnamefont {Foxen}}, \bibinfo {author} {\bibfnamefont
  {C.}~\bibnamefont {Gidney}}, \bibinfo {author} {\bibfnamefont
  {M.}~\bibnamefont {Giustina}}, \bibinfo {author} {\bibfnamefont
  {R.}~\bibnamefont {Graff}}, \bibinfo {author} {\bibfnamefont
  {T.}~\bibnamefont {Huang}}, \bibinfo {author} {\bibfnamefont
  {E.}~\bibnamefont {Jeffrey}}, \bibinfo {author} {\bibfnamefont {Erik}\
  \bibnamefont {Lucero}}, \bibinfo {author} {\bibfnamefont {J.~Y.}\
  \bibnamefont {Mutus}}, \bibinfo {author} {\bibfnamefont {O.}~\bibnamefont
  {Naaman}}, \bibinfo {author} {\bibfnamefont {C.}~\bibnamefont {Neill}},
  \bibinfo {author} {\bibfnamefont {C.}~\bibnamefont {Quintana}}, \bibinfo
  {author} {\bibfnamefont {P.}~\bibnamefont {Roushan}}, \bibinfo {author}
  {\bibfnamefont {Daniel}\ \bibnamefont {Sank}}, \bibinfo {author}
  {\bibfnamefont {A.}~\bibnamefont {Vainsencher}}, \bibinfo {author}
  {\bibfnamefont {J.}~\bibnamefont {Wenner}}, \bibinfo {author} {\bibfnamefont
  {T.~C.}\ \bibnamefont {White}}, \bibinfo {author} {\bibfnamefont
  {S.}~\bibnamefont {Boixo}}, \bibinfo {author} {\bibfnamefont
  {R.}~\bibnamefont {Babbush}}, \bibinfo {author} {\bibfnamefont {V.~N.}\
  \bibnamefont {Smelyanskiy}}, \bibinfo {author} {\bibfnamefont
  {H.}~\bibnamefont {Neven}}, \ and\ \bibinfo {author} {\bibfnamefont
  {John~M.}\ \bibnamefont {Martinis}},\ }\bibfield  {title} {\enquote {\bibinfo
  {title} {Fluctuations of energy-relaxation times in superconducting
  qubits},}\ }\href {\doibase 10.1103/PhysRevLett.121.090502} {\bibfield
  {journal} {\bibinfo  {journal} {Phys. Rev. Lett.}\ }\textbf {\bibinfo
  {volume} {121}},\ \bibinfo {pages} {090502} (\bibinfo {year}
  {2018})}\BibitemShut {NoStop}%
\bibitem [{\citenamefont {Burnett}\ \emph
  {et~al.}(2019{\natexlab{a}})\citenamefont {Burnett}, \citenamefont
  {Bengtsson}, \citenamefont {Scigliuzzo}, \citenamefont {Niepce},
  \citenamefont {Kudra}, \citenamefont {Delsing},\ and\ \citenamefont
  {Bylander}}]{Burnett.npjQI.2019}%
  \BibitemOpen
  \bibfield  {author} {\bibinfo {author} {\bibfnamefont {Jonathan~J}\
  \bibnamefont {Burnett}}, \bibinfo {author} {\bibfnamefont {Andreas}\
  \bibnamefont {Bengtsson}}, \bibinfo {author} {\bibfnamefont {Marco}\
  \bibnamefont {Scigliuzzo}}, \bibinfo {author} {\bibfnamefont {David}\
  \bibnamefont {Niepce}}, \bibinfo {author} {\bibfnamefont {Marina}\
  \bibnamefont {Kudra}}, \bibinfo {author} {\bibfnamefont {Per}\ \bibnamefont
  {Delsing}}, \ and\ \bibinfo {author} {\bibfnamefont {Jonas}\ \bibnamefont
  {Bylander}},\ }\bibfield  {title} {\enquote {\bibinfo {title} {Decoherence
  benchmarking of superconducting qubits},}\ }\href
  {https://www.nature.com/articles/s41534-019-0168-5} {\bibfield  {journal}
  {\bibinfo  {journal} {npj Quantum Information}\ }\textbf {\bibinfo {volume}
  {5}},\ \bibinfo {pages} {1--8} (\bibinfo {year}
  {2019}{\natexlab{a}})}\BibitemShut {NoStop}%
\bibitem [{\citenamefont {Yang}\ \emph
  {et~al.}(2019{\natexlab{a}})\citenamefont {Yang}, \citenamefont {Chan},
  \citenamefont {Harper}, \citenamefont {Huang}, \citenamefont {Evans},
  \citenamefont {Hwang}, \citenamefont {Hensen}, \citenamefont {Laucht},
  \citenamefont {Tanttu}, \citenamefont {Hudson} \emph
  {et~al.}}]{CHYang.NatEle.2019}%
  \BibitemOpen
  \bibfield  {author} {\bibinfo {author} {\bibfnamefont {CH}~\bibnamefont
  {Yang}}, \bibinfo {author} {\bibfnamefont {KW}~\bibnamefont {Chan}}, \bibinfo
  {author} {\bibfnamefont {R}~\bibnamefont {Harper}}, \bibinfo {author}
  {\bibfnamefont {W}~\bibnamefont {Huang}}, \bibinfo {author} {\bibfnamefont
  {T}~\bibnamefont {Evans}}, \bibinfo {author} {\bibfnamefont {JCC}\
  \bibnamefont {Hwang}}, \bibinfo {author} {\bibfnamefont {B}~\bibnamefont
  {Hensen}}, \bibinfo {author} {\bibfnamefont {A}~\bibnamefont {Laucht}},
  \bibinfo {author} {\bibfnamefont {T}~\bibnamefont {Tanttu}}, \bibinfo
  {author} {\bibfnamefont {FE}~\bibnamefont {Hudson}},  \emph {et~al.},\
  }\bibfield  {title} {\enquote {\bibinfo {title} {Silicon qubit fidelities
  approaching incoherent noise limits via pulse engineering},}\ }\href
  {https://www.nature.com/articles/s41928-019-0234-1} {\bibfield  {journal}
  {\bibinfo  {journal} {Nature Electronics}\ }\textbf {\bibinfo {volume} {2}},\
  \bibinfo {pages} {151--158} (\bibinfo {year}
  {2019}{\natexlab{a}})}\BibitemShut {NoStop}%
\bibitem [{\citenamefont {Burnett}\ \emph
  {et~al.}(2019{\natexlab{b}})\citenamefont {Burnett}, \citenamefont
  {Bengtsson}, \citenamefont {Scigliuzzo}, \citenamefont {Niepce},
  \citenamefont {Kudra}, \citenamefont {Delsing},\ and\ \citenamefont
  {Bylander}}]{BurnettnpjQI2019}%
  \BibitemOpen
  \bibfield  {author} {\bibinfo {author} {\bibfnamefont {Jonathan~J.}\
  \bibnamefont {Burnett}}, \bibinfo {author} {\bibfnamefont {Andreas}\
  \bibnamefont {Bengtsson}}, \bibinfo {author} {\bibfnamefont {Marco}\
  \bibnamefont {Scigliuzzo}}, \bibinfo {author} {\bibfnamefont {David}\
  \bibnamefont {Niepce}}, \bibinfo {author} {\bibfnamefont {Marina}\
  \bibnamefont {Kudra}}, \bibinfo {author} {\bibfnamefont {Per}\ \bibnamefont
  {Delsing}}, \ and\ \bibinfo {author} {\bibfnamefont {Jonas}\ \bibnamefont
  {Bylander}},\ }\bibfield  {title} {\enquote {\bibinfo {title} {Decoherence
  benchmarking of superconducting qubits},}\ }\href {\doibase
  10.1038/s41534-019-0168-5} {\bibfield  {journal} {\bibinfo  {journal} {npj
  Quantum Information}\ }\textbf {\bibinfo {volume} {5}},\ \bibinfo {pages}
  {54} (\bibinfo {year} {2019}{\natexlab{b}})}\BibitemShut {NoStop}%
\bibitem [{\citenamefont {Yang}\ \emph
  {et~al.}(2019{\natexlab{b}})\citenamefont {Yang}, \citenamefont {Chan},
  \citenamefont {Harper}, \citenamefont {Huang}, \citenamefont {Evans},
  \citenamefont {Hwang}, \citenamefont {Hensen}, \citenamefont {Laucht},
  \citenamefont {Tanttu}, \citenamefont {Hudson}, \citenamefont {Flammia},
  \citenamefont {Itoh}, \citenamefont {Morello}, \citenamefont {Bartlett},\
  and\ \citenamefont {Dzurak}}]{YangNatureElectronics2019}%
  \BibitemOpen
  \bibfield  {author} {\bibinfo {author} {\bibfnamefont {C.~H.}\ \bibnamefont
  {Yang}}, \bibinfo {author} {\bibfnamefont {K.~W.}\ \bibnamefont {Chan}},
  \bibinfo {author} {\bibfnamefont {R.}~\bibnamefont {Harper}}, \bibinfo
  {author} {\bibfnamefont {W.}~\bibnamefont {Huang}}, \bibinfo {author}
  {\bibfnamefont {T.}~\bibnamefont {Evans}}, \bibinfo {author} {\bibfnamefont
  {J.~C.~C.}\ \bibnamefont {Hwang}}, \bibinfo {author} {\bibfnamefont
  {B.}~\bibnamefont {Hensen}}, \bibinfo {author} {\bibfnamefont
  {A.}~\bibnamefont {Laucht}}, \bibinfo {author} {\bibfnamefont
  {T.}~\bibnamefont {Tanttu}}, \bibinfo {author} {\bibfnamefont {F.~E.}\
  \bibnamefont {Hudson}}, \bibinfo {author} {\bibfnamefont {S.~T.}\
  \bibnamefont {Flammia}}, \bibinfo {author} {\bibfnamefont {K.~M.}\
  \bibnamefont {Itoh}}, \bibinfo {author} {\bibfnamefont {A.}~\bibnamefont
  {Morello}}, \bibinfo {author} {\bibfnamefont {S.~D.}\ \bibnamefont
  {Bartlett}}, \ and\ \bibinfo {author} {\bibfnamefont {A.~S.}\ \bibnamefont
  {Dzurak}},\ }\bibfield  {title} {\enquote {\bibinfo {title} {Silicon qubit
  fidelities approaching incoherent noise limits via pulse engineering},}\
  }\href {\doibase 10.1038/s41928-019-0234-1} {\bibfield  {journal} {\bibinfo
  {journal} {Nature Electronics}\ }\textbf {\bibinfo {volume} {2}},\ \bibinfo
  {pages} {151--158} (\bibinfo {year} {2019}{\natexlab{b}})}\BibitemShut
  {NoStop}%
\bibitem [{\citenamefont {Economou}\ \emph {et~al.}(2006)\citenamefont
  {Economou}, \citenamefont {Sham}, \citenamefont {Wu},\ and\ \citenamefont
  {Steel}}]{PhysRevB.74.205415}%
  \BibitemOpen
  \bibfield  {author} {\bibinfo {author} {\bibfnamefont {Sophia~E.}\
  \bibnamefont {Economou}}, \bibinfo {author} {\bibfnamefont {L.~J.}\
  \bibnamefont {Sham}}, \bibinfo {author} {\bibfnamefont {Yanwen}\ \bibnamefont
  {Wu}}, \ and\ \bibinfo {author} {\bibfnamefont {D.~G.}\ \bibnamefont
  {Steel}},\ }\bibfield  {title} {\enquote {\bibinfo {title} {Proposal for
  optical u(1) rotations of electron spin trapped in a quantum dot},}\ }\href
  {\doibase 10.1103/PhysRevB.74.205415} {\bibfield  {journal} {\bibinfo
  {journal} {Phys. Rev. B}\ }\textbf {\bibinfo {volume} {74}},\ \bibinfo
  {pages} {205415} (\bibinfo {year} {2006})}\BibitemShut {NoStop}%
\bibitem [{\citenamefont {Economou}\ and\ \citenamefont
  {Reinecke}(2007)}]{EconomouPRL2007}%
  \BibitemOpen
  \bibfield  {author} {\bibinfo {author} {\bibfnamefont {Sophia~E.}\
  \bibnamefont {Economou}}\ and\ \bibinfo {author} {\bibfnamefont {T.~L.}\
  \bibnamefont {Reinecke}},\ }\bibfield  {title} {\enquote {\bibinfo {title}
  {Theory of fast optical spin rotation in a quantum dot based on geometric
  phases and trapped states},}\ }\href {\doibase 10.1103/PhysRevLett.99.217401}
  {\bibfield  {journal} {\bibinfo  {journal} {Phys. Rev. Lett.}\ }\textbf
  {\bibinfo {volume} {99}},\ \bibinfo {pages} {217401} (\bibinfo {year}
  {2007})}\BibitemShut {NoStop}%
\bibitem [{\citenamefont {Economou}(2012)}]{EconomouPRB2012}%
  \BibitemOpen
  \bibfield  {author} {\bibinfo {author} {\bibfnamefont {Sophia~E.}\
  \bibnamefont {Economou}},\ }\bibfield  {title} {\enquote {\bibinfo {title}
  {High-fidelity quantum gates via analytically solvable pulses},}\ }\href
  {\doibase 10.1103/PhysRevB.85.241401} {\bibfield  {journal} {\bibinfo
  {journal} {Phys. Rev. B}\ }\textbf {\bibinfo {volume} {85}},\ \bibinfo
  {pages} {241401} (\bibinfo {year} {2012})}\BibitemShut {NoStop}%
\bibitem [{\citenamefont {Greilich}\ \emph {et~al.}(2009)\citenamefont
  {Greilich}, \citenamefont {Economou}, \citenamefont {Spatzek}, \citenamefont
  {Yakovlev}, \citenamefont {Reuter}, \citenamefont {Wieck}, \citenamefont
  {Reinecke},\ and\ \citenamefont {Bayer}}]{Greilich2009}%
  \BibitemOpen
  \bibfield  {author} {\bibinfo {author} {\bibfnamefont {A.}~\bibnamefont
  {Greilich}}, \bibinfo {author} {\bibfnamefont {Sophia~E.}\ \bibnamefont
  {Economou}}, \bibinfo {author} {\bibfnamefont {S.}~\bibnamefont {Spatzek}},
  \bibinfo {author} {\bibfnamefont {D.~R.}\ \bibnamefont {Yakovlev}}, \bibinfo
  {author} {\bibfnamefont {D.}~\bibnamefont {Reuter}}, \bibinfo {author}
  {\bibfnamefont {A.~D.}\ \bibnamefont {Wieck}}, \bibinfo {author}
  {\bibfnamefont {T.~L.}\ \bibnamefont {Reinecke}}, \ and\ \bibinfo {author}
  {\bibfnamefont {M.}~\bibnamefont {Bayer}},\ }\bibfield  {title} {\enquote
  {\bibinfo {title} {Ultrafast optical rotations of electron spins in quantum
  dots},}\ }\href {\doibase 10.1038/nphys1226} {\bibfield  {journal} {\bibinfo
  {journal} {Nature Physics}\ }\textbf {\bibinfo {volume} {5}},\ \bibinfo
  {pages} {262--266} (\bibinfo {year} {2009})}\BibitemShut {NoStop}%
\bibitem [{\citenamefont {Ku}\ \emph {et~al.}(2017)\citenamefont {Ku},
  \citenamefont {Long}, \citenamefont {Wu}, \citenamefont {Bal}, \citenamefont
  {Lake}, \citenamefont {Barnes}, \citenamefont {Economou},\ and\ \citenamefont
  {Pappas}}]{KuPRA2017}%
  \BibitemOpen
  \bibfield  {author} {\bibinfo {author} {\bibfnamefont {H.~S.}\ \bibnamefont
  {Ku}}, \bibinfo {author} {\bibfnamefont {J.~L.}\ \bibnamefont {Long}},
  \bibinfo {author} {\bibfnamefont {X.}~\bibnamefont {Wu}}, \bibinfo {author}
  {\bibfnamefont {M.}~\bibnamefont {Bal}}, \bibinfo {author} {\bibfnamefont
  {R.~E.}\ \bibnamefont {Lake}}, \bibinfo {author} {\bibfnamefont {Edwin}\
  \bibnamefont {Barnes}}, \bibinfo {author} {\bibfnamefont {Sophia~E.}\
  \bibnamefont {Economou}}, \ and\ \bibinfo {author} {\bibfnamefont {D.~P.}\
  \bibnamefont {Pappas}},\ }\bibfield  {title} {\enquote {\bibinfo {title}
  {Single qubit operations using microwave hyperbolic secant pulses},}\ }\href
  {\doibase 10.1103/PhysRevA.96.042339} {\bibfield  {journal} {\bibinfo
  {journal} {Phys. Rev. A}\ }\textbf {\bibinfo {volume} {96}},\ \bibinfo
  {pages} {042339} (\bibinfo {year} {2017})}\BibitemShut {NoStop}%
\bibitem [{\citenamefont {Pedersen}\ \emph {et~al.}(2007)\citenamefont
  {Pedersen}, \citenamefont {Møller},\ and\ \citenamefont
  {Mølmer}}]{PEDERSEN200747}%
  \BibitemOpen
  \bibfield  {author} {\bibinfo {author} {\bibfnamefont {Line~Hjortshøj}\
  \bibnamefont {Pedersen}}, \bibinfo {author} {\bibfnamefont {Niels~Martin}\
  \bibnamefont {Møller}}, \ and\ \bibinfo {author} {\bibfnamefont {Klaus}\
  \bibnamefont {Mølmer}},\ }\bibfield  {title} {\enquote {\bibinfo {title}
  {Fidelity of quantum operations},}\ }\href {\doibase
  https://doi.org/10.1016/j.physleta.2007.02.069} {\bibfield  {journal}
  {\bibinfo  {journal} {Physics Letters A}\ }\textbf {\bibinfo {volume}
  {367}},\ \bibinfo {pages} {47 -- 51} (\bibinfo {year} {2007})}\BibitemShut
  {NoStop}%
\bibitem [{\citenamefont {Margolus}\ and\ \citenamefont
  {Levitin}(1998)}]{Margolus.PhysicaD.1998}%
  \BibitemOpen
  \bibfield  {author} {\bibinfo {author} {\bibfnamefont {Norman}\ \bibnamefont
  {Margolus}}\ and\ \bibinfo {author} {\bibfnamefont {Lev~B}\ \bibnamefont
  {Levitin}},\ }\bibfield  {title} {\enquote {\bibinfo {title} {The maximum
  speed of dynamical evolution},}\ }\href
  {https://www.sciencedirect.com/science/article/abs/pii/S0167278998000542}
  {\bibfield  {journal} {\bibinfo  {journal} {Physica D: Nonlinear Phenomena}\
  }\textbf {\bibinfo {volume} {120}},\ \bibinfo {pages} {188--195} (\bibinfo
  {year} {1998})}\BibitemShut {NoStop}%
\bibitem [{\citenamefont {Deffner}\ and\ \citenamefont
  {Lutz}(2013)}]{Deffner.PRL.2013}%
  \BibitemOpen
  \bibfield  {author} {\bibinfo {author} {\bibfnamefont {Sebastian}\
  \bibnamefont {Deffner}}\ and\ \bibinfo {author} {\bibfnamefont {Eric}\
  \bibnamefont {Lutz}},\ }\bibfield  {title} {\enquote {\bibinfo {title}
  {Quantum speed limit for non-markovian dynamics},}\ }\href {\doibase
  10.1103/PhysRevLett.111.010402} {\bibfield  {journal} {\bibinfo  {journal}
  {Phys. Rev. Lett.}\ }\textbf {\bibinfo {volume} {111}},\ \bibinfo {pages}
  {010402} (\bibinfo {year} {2013})}\BibitemShut {NoStop}%
\bibitem [{\citenamefont {Deffner}\ and\ \citenamefont
  {Campbell}(2017)}]{Deffner.JPA.2017}%
  \BibitemOpen
  \bibfield  {author} {\bibinfo {author} {\bibfnamefont {Sebastian}\
  \bibnamefont {Deffner}}\ and\ \bibinfo {author} {\bibfnamefont {Steve}\
  \bibnamefont {Campbell}},\ }\bibfield  {title} {\enquote {\bibinfo {title}
  {Quantum speed limits: from {Heisenberg}’s uncertainty principle to optimal
  quantum control},}\ }\href
  {https://iopscience.iop.org/article/10.1088/1751-8121/aa86c6} {\bibfield
  {journal} {\bibinfo  {journal} {Journal of Physics A: Mathematical and
  Theoretical}\ }\textbf {\bibinfo {volume} {50}},\ \bibinfo {pages} {453001}
  (\bibinfo {year} {2017})}\BibitemShut {NoStop}%
\bibitem [{\citenamefont {Krantz}\ \emph {et~al.}(2019)\citenamefont {Krantz},
  \citenamefont {Kjaergaard}, \citenamefont {Yan}, \citenamefont {Orlando},
  \citenamefont {Gustavsson},\ and\ \citenamefont {Oliver}}]{Krantz.APR.2019}%
  \BibitemOpen
  \bibfield  {author} {\bibinfo {author} {\bibfnamefont {Philip}\ \bibnamefont
  {Krantz}}, \bibinfo {author} {\bibfnamefont {Morten}\ \bibnamefont
  {Kjaergaard}}, \bibinfo {author} {\bibfnamefont {Fei}\ \bibnamefont {Yan}},
  \bibinfo {author} {\bibfnamefont {Terry~P}\ \bibnamefont {Orlando}}, \bibinfo
  {author} {\bibfnamefont {Simon}\ \bibnamefont {Gustavsson}}, \ and\ \bibinfo
  {author} {\bibfnamefont {William~D}\ \bibnamefont {Oliver}},\ }\bibfield
  {title} {\enquote {\bibinfo {title} {A quantum engineer's guide to
  superconducting qubits},}\ }\href
  {https://aip.scitation.org/doi/abs/10.1063/1.5089550} {\bibfield  {journal}
  {\bibinfo  {journal} {Applied Physics Reviews}\ }\textbf {\bibinfo {volume}
  {6}},\ \bibinfo {pages} {021318} (\bibinfo {year} {2019})}\BibitemShut
  {NoStop}%
\bibitem [{\citenamefont {Bruzewicz}\ \emph {et~al.}(2019)\citenamefont
  {Bruzewicz}, \citenamefont {Chiaverini}, \citenamefont {McConnell},\ and\
  \citenamefont {Sage}}]{Bruzewicz.APR.2019}%
  \BibitemOpen
  \bibfield  {author} {\bibinfo {author} {\bibfnamefont {Colin~D}\ \bibnamefont
  {Bruzewicz}}, \bibinfo {author} {\bibfnamefont {John}\ \bibnamefont
  {Chiaverini}}, \bibinfo {author} {\bibfnamefont {Robert}\ \bibnamefont
  {McConnell}}, \ and\ \bibinfo {author} {\bibfnamefont {Jeremy~M}\
  \bibnamefont {Sage}},\ }\bibfield  {title} {\enquote {\bibinfo {title}
  {Trapped-ion quantum computing: Progress and challenges},}\ }\href
  {https://aip.scitation.org/doi/10.1063/1.5088164} {\bibfield  {journal}
  {\bibinfo  {journal} {Applied Physics Reviews}\ }\textbf {\bibinfo {volume}
  {6}},\ \bibinfo {pages} {021314} (\bibinfo {year} {2019})}\BibitemShut
  {NoStop}%
\bibitem [{\citenamefont {Milne}\ \emph {et~al.}(2020)\citenamefont {Milne},
  \citenamefont {Edmunds}, \citenamefont {Hempel}, \citenamefont {Roy},
  \citenamefont {Mavadia},\ and\ \citenamefont {Biercuk}}]{Milne.PRApp.2020}%
  \BibitemOpen
  \bibfield  {author} {\bibinfo {author} {\bibfnamefont {Alistair~R.}\
  \bibnamefont {Milne}}, \bibinfo {author} {\bibfnamefont {Claire~L.}\
  \bibnamefont {Edmunds}}, \bibinfo {author} {\bibfnamefont {Cornelius}\
  \bibnamefont {Hempel}}, \bibinfo {author} {\bibfnamefont {Federico}\
  \bibnamefont {Roy}}, \bibinfo {author} {\bibfnamefont {Sandeep}\ \bibnamefont
  {Mavadia}}, \ and\ \bibinfo {author} {\bibfnamefont {Michael~J.}\
  \bibnamefont {Biercuk}},\ }\bibfield  {title} {\enquote {\bibinfo {title}
  {Phase-modulated entangling gates robust to static and time-varying
  errors},}\ }\href {\doibase 10.1103/PhysRevApplied.13.024022} {\bibfield
  {journal} {\bibinfo  {journal} {Phys. Rev. Applied}\ }\textbf {\bibinfo
  {volume} {13}},\ \bibinfo {pages} {024022} (\bibinfo {year}
  {2020})}\BibitemShut {NoStop}%
\bibitem [{\citenamefont {Rong}\ \emph {et~al.}(2015)\citenamefont {Rong},
  \citenamefont {Geng}, \citenamefont {Shi}, \citenamefont {Liu}, \citenamefont
  {Xu}, \citenamefont {Ma}, \citenamefont {Kong}, \citenamefont {Jiang},
  \citenamefont {Wu},\ and\ \citenamefont {Du}}]{Rong.NatCom.2015}%
  \BibitemOpen
  \bibfield  {author} {\bibinfo {author} {\bibfnamefont {Xing}\ \bibnamefont
  {Rong}}, \bibinfo {author} {\bibfnamefont {Jianpei}\ \bibnamefont {Geng}},
  \bibinfo {author} {\bibfnamefont {Fazhan}\ \bibnamefont {Shi}}, \bibinfo
  {author} {\bibfnamefont {Ying}\ \bibnamefont {Liu}}, \bibinfo {author}
  {\bibfnamefont {Kebiao}\ \bibnamefont {Xu}}, \bibinfo {author} {\bibfnamefont
  {Wenchao}\ \bibnamefont {Ma}}, \bibinfo {author} {\bibfnamefont {Fei}\
  \bibnamefont {Kong}}, \bibinfo {author} {\bibfnamefont {Zhen}\ \bibnamefont
  {Jiang}}, \bibinfo {author} {\bibfnamefont {Yang}\ \bibnamefont {Wu}}, \ and\
  \bibinfo {author} {\bibfnamefont {Jiangfeng}\ \bibnamefont {Du}},\ }\bibfield
   {title} {\enquote {\bibinfo {title} {Experimental fault-tolerant universal
  quantum gates with solid-state spins under ambient conditions},}\ }\href
  {https://www.nature.com/articles/ncomms9748} {\bibfield  {journal} {\bibinfo
  {journal} {Nature communications}\ }\textbf {\bibinfo {volume} {6}},\
  \bibinfo {pages} {1--7} (\bibinfo {year} {2015})}\BibitemShut {NoStop}%
\bibitem [{CZ.()}]{CZ.equiv}%
  \BibitemOpen
  \href@noop {} {}\bibinfo {note} {The DoG gate $\text{diag}\{e^{3 i \pi/
  4},e^{-3 i \pi/ 4},e^{-3 i \pi/ 4},e^{3 i \pi/ 4}\}$ is equivalent to CZ gate
  up to a single-qubit gate $e^{-i3\pi/4 (\sigma^{(1)}_z+\sigma^{(2)}_z) }$ and
  a phase factor $\frac{i-1}{\sqrt{2}}$}\BibitemShut {NoStop}%
\bibitem [{don()}]{donovan.curves.comment}%
  \BibitemOpen
  \href@noop {} {}\bibinfo {note} {We do not provide the DoG solution based on
  two curves with constant and positive torsions in \cite{Buterakos.PRXQ.2021}
  because the corresponding DoG control fields are not experimentally
  friendly.}\BibitemShut {Stop}%
\bibitem [{\citenamefont {Frey}\ \emph {et~al.}(2020)\citenamefont {Frey},
  \citenamefont {Norris}, \citenamefont {Viola},\ and\ \citenamefont
  {Biercuk}}]{Frey.PRApp.2020}%
  \BibitemOpen
  \bibfield  {author} {\bibinfo {author} {\bibfnamefont {Virginia}\
  \bibnamefont {Frey}}, \bibinfo {author} {\bibfnamefont {Leigh~M.}\
  \bibnamefont {Norris}}, \bibinfo {author} {\bibfnamefont {Lorenza}\
  \bibnamefont {Viola}}, \ and\ \bibinfo {author} {\bibfnamefont {Michael~J.}\
  \bibnamefont {Biercuk}},\ }\bibfield  {title} {\enquote {\bibinfo {title}
  {Simultaneous spectral estimation of dephasing and amplitude noise on a qubit
  sensor via optimally band-limited control},}\ }\href {\doibase
  10.1103/PhysRevApplied.14.024021} {\bibfield  {journal} {\bibinfo  {journal}
  {Phys. Rev. Applied}\ }\textbf {\bibinfo {volume} {14}},\ \bibinfo {pages}
  {024021} (\bibinfo {year} {2020})}\BibitemShut {NoStop}%
\bibitem [{\citenamefont {Chalermpusitarak}\ \emph {et~al.}(2020)\citenamefont
  {Chalermpusitarak}, \citenamefont {Tonekaboni}, \citenamefont {Wang},
  \citenamefont {Norris}, \citenamefont {Viola},\ and\ \citenamefont
  {Paz-Silva}}]{chalermpusitarak2020framebased}%
  \BibitemOpen
  \bibfield  {author} {\bibinfo {author} {\bibfnamefont {Teerawat}\
  \bibnamefont {Chalermpusitarak}}, \bibinfo {author} {\bibfnamefont {Behnam}\
  \bibnamefont {Tonekaboni}}, \bibinfo {author} {\bibfnamefont {Yuanlong}\
  \bibnamefont {Wang}}, \bibinfo {author} {\bibfnamefont {Leigh~M.}\
  \bibnamefont {Norris}}, \bibinfo {author} {\bibfnamefont {Lorenza}\
  \bibnamefont {Viola}}, \ and\ \bibinfo {author} {\bibfnamefont {Gerardo~A.}\
  \bibnamefont {Paz-Silva}},\ }\href@noop {} {\enquote {\bibinfo {title}
  {Frame-based filter-function formalism for quantum characterization and
  control},}\ } (\bibinfo {year} {2020}),\ \Eprint
  {http://arxiv.org/abs/2008.13216} {arXiv:2008.13216 [quant-ph]} \BibitemShut
  {NoStop}%
\bibitem [{\citenamefont {Zheng}\ \emph {et~al.}(2016)\citenamefont {Zheng},
  \citenamefont {Yang},\ and\ \citenamefont {Nori}}]{ZhengShibiao.PRA.2016}%
  \BibitemOpen
  \bibfield  {author} {\bibinfo {author} {\bibfnamefont {Shi-Biao}\
  \bibnamefont {Zheng}}, \bibinfo {author} {\bibfnamefont {Chui-Ping}\
  \bibnamefont {Yang}}, \ and\ \bibinfo {author} {\bibfnamefont {Franco}\
  \bibnamefont {Nori}},\ }\bibfield  {title} {\enquote {\bibinfo {title}
  {Comparison of the sensitivity to systematic errors between nonadiabatic
  non-abelian geometric gates and their dynamical counterparts},}\ }\href
  {\doibase 10.1103/PhysRevA.93.032313} {\bibfield  {journal} {\bibinfo
  {journal} {Phys. Rev. A}\ }\textbf {\bibinfo {volume} {93}},\ \bibinfo
  {pages} {032313} (\bibinfo {year} {2016})}\BibitemShut {NoStop}%
\bibitem [{\citenamefont {Ribeiro}\ \emph {et~al.}(2017)\citenamefont
  {Ribeiro}, \citenamefont {Baksic},\ and\ \citenamefont
  {Clerk}}]{Ribeiro.PRX.2017}%
  \BibitemOpen
  \bibfield  {author} {\bibinfo {author} {\bibfnamefont {Hugo}\ \bibnamefont
  {Ribeiro}}, \bibinfo {author} {\bibfnamefont {Alexandre}\ \bibnamefont
  {Baksic}}, \ and\ \bibinfo {author} {\bibfnamefont {Aashish~A.}\ \bibnamefont
  {Clerk}},\ }\bibfield  {title} {\enquote {\bibinfo {title} {Systematic
  magnus-based approach for suppressing leakage and nonadiabatic errors in
  quantum dynamics},}\ }\href {\doibase 10.1103/PhysRevX.7.011021} {\bibfield
  {journal} {\bibinfo  {journal} {Phys. Rev. X}\ }\textbf {\bibinfo {volume}
  {7}},\ \bibinfo {pages} {011021} (\bibinfo {year} {2017})}\BibitemShut
  {NoStop}%
\bibitem [{\citenamefont {S\o{}rensen}\ and\ \citenamefont
  {M\o{}lmer}(1999)}]{Sorrensen.Molmer.PRL.1999}%
  \BibitemOpen
  \bibfield  {author} {\bibinfo {author} {\bibfnamefont {Anders}\ \bibnamefont
  {S\o{}rensen}}\ and\ \bibinfo {author} {\bibfnamefont {Klaus}\ \bibnamefont
  {M\o{}lmer}},\ }\bibfield  {title} {\enquote {\bibinfo {title} {Quantum
  computation with ions in thermal motion},}\ }\href {\doibase
  10.1103/PhysRevLett.82.1971} {\bibfield  {journal} {\bibinfo  {journal}
  {Phys. Rev. Lett.}\ }\textbf {\bibinfo {volume} {82}},\ \bibinfo {pages}
  {1971--1974} (\bibinfo {year} {1999})}\BibitemShut {NoStop}%
\bibitem [{\citenamefont {S\o{}rensen}\ and\ \citenamefont
  {M\o{}lmer}(2000)}]{Sorrensen.Molmer.PRA.2000}%
  \BibitemOpen
  \bibfield  {author} {\bibinfo {author} {\bibfnamefont {Anders}\ \bibnamefont
  {S\o{}rensen}}\ and\ \bibinfo {author} {\bibfnamefont {Klaus}\ \bibnamefont
  {M\o{}lmer}},\ }\bibfield  {title} {\enquote {\bibinfo {title} {Entanglement
  and quantum computation with ions in thermal motion},}\ }\href {\doibase
  10.1103/PhysRevA.62.022311} {\bibfield  {journal} {\bibinfo  {journal} {Phys.
  Rev. A}\ }\textbf {\bibinfo {volume} {62}},\ \bibinfo {pages} {022311}
  (\bibinfo {year} {2000})}\BibitemShut {NoStop}%
\end{thebibliography}
%

\end{document}